\documentclass[12pt,preprint]{aastex}
\newcommand{\HI}{H~{\sc i}} 

\newcommand{\kms}{${\rm km~s^{-1}}$}

\slugcomment{Accepted to ApJS 10 Jan 2012}
\shortauthors{McCLURE-GRIFFITHS ET AL.} 
\shorttitle{ATCA HI Survey of the Galactic Center}

\begin{document} 

\title{The ATCA H~{\sc i} Survey of the Galactic Center}

\author{N.\ M.\ McClure-Griffiths,\altaffilmark{1} J.\ M.\
  Dickey,\altaffilmark{2} B.\ M.\ Gaensler,\altaffilmark{3} A.\ J.\
  Green,\altaffilmark{3} J.\ A.\
  Green\altaffilmark{1} , \& M.\ Haverkorn\altaffilmark{4,5}} 

\altaffiltext{1}{Australia Telescope National Facility, CSIRO
  Astronomy \& Space Science,  Marsfield NSW 2122, Australia;
  naomi.mcclure-griffiths@csiro.au, james.green@csiro.au}
\altaffiltext{2}{School of Physics and Mathematics, University of
  Tasmania, TAS 7001, Australia; john.dickey@utas.edu.au}
\altaffiltext{3}{Sydney Institute for Astronomy, School of Physics,
  The University of Sydney, NSW 2006, Australia;
  bryan.gaensler@sydney.edu.au; anne.green@sydney.edu.au}
\altaffiltext{4}{Department of Astrophysics/IMAPP, Radboud University
  Nijmegen, 6500 GL Nijmegen, The Netherlands; m.haverkorn@astro.ru.nl}
\altaffiltext{5}{Leiden Observatory, Leiden University, 2300 RA Leiden, 
The Netherlands}
\begin{abstract}
  We present a survey of atomic hydrogen (\HI) emission in the
  direction of the Galactic Center conducted with the CSIRO Australia
  Telescope Compact Array (ATCA).  The survey covers the area $-5
  \arcdeg \leq l \leq +5\arcdeg$, $-5 \arcdeg \leq b \leq +5\arcdeg$
  over the velocity range $-309 \leq v_{LSR} \leq 349$ \kms\ with a
  velocity resolution of 1 \kms.  The ATCA data are supplemented with
  data from the Parkes Radio Telescope for sensitivity to all angular
  scales larger than the $145\arcsec$ angular resolution of the survey.
  The mean rms brightness temperature across the field is $0.7$ K,
  except near $(l,b)=0\arcdeg,0\arcdeg$ where it increases to $\sim 2$
  K.  This survey complements the Southern Galactic Plane Survey to
  complete the continuous coverage of the inner Galactic plane in \HI\
  at $\sim 2\arcmin$ resolution.  Here we describe the observations
  and analysis of this Galactic Center survey and present the final
  data product.  Features such as Bania's Clump~2, the far 3
  kiloparsec arm and small high velocity clumps are briefly described.
\end{abstract}

\section{Introduction}
\label{sec:intro}
The central 3 kiloparsecs of the Milky Way contain information on a
wealth of astrophysical processes from Galactic dynamics to Galactic
outflows. The extended environment of the Galactic Center (GC)
provides an outstanding opportunity to study the dynamics of gas flow
in a barred Galaxy and the relationship of molecular and atomic gas in
the unique GC environment. Atomic hydrogen (\HI) observations of this
region offer a clear view of the gas dynamics at the center of our
Galaxy, probing the transition between orbits associated with the bar,
circular orbits, and the beginnings of the spiral arms.  

The spatial distribution of interstellar matter in the innermost 3 kpc
has recently been reviewed by \citet{morris96} and
\citet{ferriere07}. The main structural features that make up the GC
region have been discussed extensively in the literature, including
reviews by \citet{blitz93} and \citet{genzel94}. Amongst the main
structural features inside the GC are: both a long, thin bar
\citep{hammersley00,benjamin05,cabrera-lavers08} and a short,
boxy-bulge bar \citep[e.g.][]{blitz91,dwek95, babusiaux05}; the near 3
kpc arm \citep{vanwoerden57,oort58}; and the Central Molecular Zone
(CMZ) covering the inner 150 - 250 pc \citep{morris96}.  More recently
discovered features include the far-side 3 kpc arm \citep{dame08}, and
a thin twisted 100 pc ring
\citep{sawada04,martin04,liszt08,molinari11}.

Studies of the dynamics of gas, including high resolution CO
observations \citep[e.g][]{liszt78,burton92,oka98} and high
sensitivity, low resolution \HI\ \citep{liszt80}, detail the
distribution and kinematics of molecular gas in the CMZ. These data
have constrained extensive models of the dynamics of the GC, including
those incorporating the effects of the bar(s)
\citep[e.g.][]{peters75,binney91,fux99,rodriguez-fernandez06,
  romero-gomez11}.

In addition to Galactic dynamics, \HI\ observations of the GC region
may reveal any putative gas outflows from the GC. Much of the early
analysis of the GC \HI\ invoked large-scale gas expulsive events
\citep{vanderkruit70,vanderkruit71,sanders72} to explain the highly
non-circular motions observed in the \HI\ emission envelope. Since
\citet{liszt80} showed that the forbidden velocity gas and its
prevalence off the Galactic mid-plane could be explained by a
tilted-bar inducing elliptical gas streamlines the idea of large-scale
neutral gas outflows has largely disappeared. However, due to lack of
high-resolution data probing small physical scales there is very
little information about any small-scale outflows associated with star
formation at the GC. For example, we know nothing about whether there
is \HI\ associated with the Galactic wind suggested by
\citet{bland-hawthorn03}, despite evidence for cool and warm gas in
the wind.

Large-scale surveys of the distribution of atomic hydrogen around the
GC have a long history and include observations with the Dwingeloo 26m
\citep{vanwoerden57,vanderkruit70}, the Parkes 18m \citep{kerr67}, an
extensive Jodrell Bank survey of $355\arcdeg \leq l \leq 10\arcdeg$,
$|b| \leq 5\arcdeg$, $|v|\leq 530$ \kms\ \citep{cohen75}, the NRAO
140-ft survey of $349\arcdeg \leq l \leq 13\arcdeg$, $|b| \leq
10\arcdeg$, $|v|\leq 350$ \kms\ \citep{burton78a,liszt80,burton83} and
the recent all-sky \HI\ surveys: the Leiden-Argentine-Bonn survey
\citep{kalberla05} and the Galactic All-Sky Survey
\citep{mcgriff09,kalberla10}.  Although the \HI\ in the GC has been
well-studied at low resolution, the large-scale environment is
relatively untapped at angular resolutions better than $>0.25\arcdeg$.
\citet{braunfurth81} conducted a slightly higher resolution, limited
area, survey of $|l| \leq 1.5\arcdeg$, $|b| \leq 1.5\arcdeg$ with the
Effelsberg 100m telescope, with a sampling grid and beamsize of
$0.15\arcdeg$.  More recently \citet{lang10} produced a high angular
resolution survey of \HI\ absorption, but only in the central
$100\arcmin \times 50\arcmin$.  Hence, there are currently no
high-resolution \HI\ datasets covering the entire GC region to compare
with detailed dynamical models and theories of molecular transition
(about 180--250 pc from the GC, outside the CMZ, the nuclear disk has
been thought to transition from molecular hydrogen to \HI\
(\citealt{liszt78,bitran97,morris96})). Furthermore, the angular
resolution of new infrared data on the GC from {\em Spitzer} \citep{arendt08}
and {\em Herschel} \citep{molinari10,molinari11} exceeds the
resolution of existing \HI\ data of the inner 3 kpc by more than an
order of magnitude.

In this paper we describe the atomic hydrogen (\HI) component of a
survey of the Galactic Center conducted with the Australia Telescope
Compact Array and incorporating single dish data from Parkes.  Here we
take an intermediate approach, covering an area of
$10^{\circ}\times10^{\circ}$ around the Galactic Center with an
angular resolution of $145$ arcseconds in atomic hydrogen (\HI) and in
20 cm continuum.  At a distance of 8.4 kpc \citep{ghez08,reid09} the
survey covers about 1500 pc on all sides of the Galactic Center with a
spatial resolution of about 6 pc.  This survey builds on the Southern
Galactic Plane Survey \citep[SGPS;][]{mcgriff05}, VLA Galactic Plane
Survey \citep[VGPS;][]{stil06} and Canadian Galactic Plane Survey
\citep[CGPS;][]{taylor03} to complete the International Galactic Plane
Survey of the first, second and fourth Galactic quadrants.  We discuss
the observations and data reduction of our Galactic Center \HI\ survey
(\S \ref{sec:obs} \& \S \ref{sec:imaging}), present the data cubes (\S
\ref{sec:data}) and discuss noteworthy features (\S
\ref{subsec:features}).  The data have been made publicly available
at: \url{http://www.atnf.csiro.au/research/HI/sgps/GalacticCenter}.

\section{ATCA Observations}
\label{sec:obs}
The Galactic Center survey data were obtained with the Australia
Telescope Compact Array (ATCA) and the Parkes Radio Telescope.  The
Parkes data are from the Galactic All-Sky Survey
\citep{mcgriff09,kalberla10} and are described therein.  The ATCA
observations were conducted between December 2002 and March 2004.  The
ATCA consists of five moveable 22-m antennas on a 3-km east-west track
and an additional 22-m antenna at a fixed position 3-km west of the
end of the track.  Observations were conducted in six different array
configurations: EW352, EW367, 750A, 750B, 750C, and 750D.  The array
configurations were chosen to give optimum coverage of the inner {\em
  u-v} plane out to a baseline of 750 m.  Data were recorded from the
sixth antenna, with a maximum baseline of 6 km, but these are not used
in the image cubes presented here.  The shortest baseline is 31 m and
every baseline, in intervals of 15.5m, to 245 m is sampled

\begin{figure}
\centering
\includegraphics[width=6in]{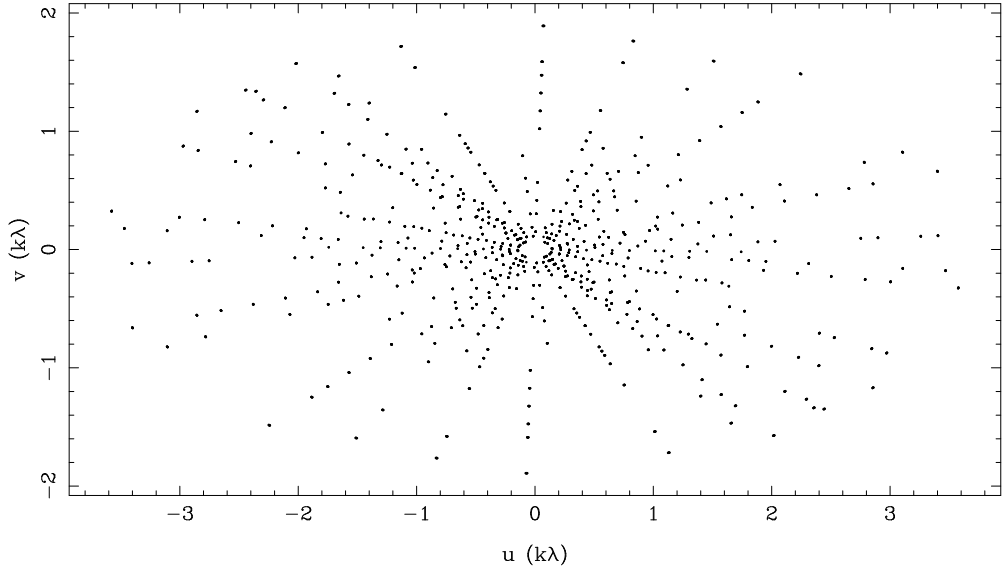}
\caption[]{U-V coverage of a typical pointing in the mosaic.
\label{fig:uvcover}}
\end{figure}

To image the full $10^{\circ} \times 10^{\circ} $ area the survey was
conducted as a mosaic of many pointings.  The primary beam of the ATCA
antennas, $\lambda/D = 33^{\prime}$, determines the
optimum spacing between pointings, where $\lambda = 21$ cm is the
observing wavelength and $D=22~{\rm m}$ is the ATCA antenna diameter.
The mosaic pointings were arranged on a hexagonal grid with a
separation of 20$^\prime$.  Although this is slightly wider than the
optimum hexagonal spacing of $(2/\sqrt{3})(\lambda/2D)\sim
19\arcmin$, the theoretical sensitivity varies by less than 2\% across
the field.  A total of 1031 pointings were required to cover the full
area; 64 of these were previously observed as part of the Southern
Galactic Plane Survey \citep{mcgriff05}.  The 64 SGPS
pointings were re-observed several times with the 750A and 750D arrays
as part of the Galactic Center project to help improve the {\em u-v}
sampling of the longer baselines.  The overall observing strategy was
similar to that used for the SGPS and described by \citet{mcgriff05}.
The 967 new pointings were observed in 60s integrations approximately
25 times at widely spread hour angles.  We chose 60s integrations to
mitigate against the effects of fluctuating sampler statistics caused
by mosaicing very strong continuum sources.  In practice we found the
first 10s sample was occasionally flagged as bad because of sample
statistics when the mosaic centre was moved to some of the pointings
closest to $(l,b)\approx (0\arcdeg, 0\arcdeg)$; samplers did not cause
problems for the majority of the pointings.  The 967 pointings were
divided into subfields of 42 or 36 pointings that were observed as
continuous blocks.  Within each of these blocks the {\em u-v} coverage
and integration time was essentially the same for all pointings.
Multiple blocks were observed during the course of each observing day
and care was taken to ensure that the hour angle coverage for each
block was as uniform as possible.  An example of the {\em u-v}
coverage for a pointing in the north-western corner of the field is
shown in Figure~\ref{fig:uvcover}.

The ATCA has linear feeds, recording two orthogonal linear
polarizations, $X$ and $Y$.  These data were obtained prior to the
Compact Array Broadband Backend upgrade \citep{wilson11}, and
therefore were recorded simultaneously in a narrowband spectral line
mode with 1024 channels across a 4 MHz bandwidth centered on 1420 MHz
and a broadband mode recording 32 channels across a 128 MHz bandwidth
centered on 1384 MHz.  For the narrowband mode that is the focus of
this paper, only auto correlations, $XX$ and $YY$ were recorded.  All
four polarization products, $XX$, $YY$, $XY$, and $YX$ were recorded
in the broadband mode.  The broadband data will be the subject of a
future paper.

\section{Calibration and Imaging}
\label{sec:imaging}
Data editing, calibration and imaging were performed in the Miriad
package using standard techniques \citep{sault95}.  The ATCA's primary
flux calibrator, PKS B1934-638, was observed at least once per day.
PKS B1934-638 was used for bandpass calibration and primary flux
calibration assuming a flux of 14.86 Jy at 1420 MHz \citep{reynolds94}.
In addition, a secondary calibrator, PKS B1827-360, was observed for 2
minutes every $\sim 1$ hr to solve for time varying complex gains.
Gain solutions determined on PKS 1827-360 were copied across to the
Galactic Center pointings.

Data-editing, or {\em flagging}, was carried out by hand within
Miriad.  Although the 4 MHz band around the \HI\ line is generally
free from significant radio-frequency interference (RFI) there was
some intermittent locally generated RFI at 1420.0 MHz, apparent on the shortest
baselines.  This was flagged by hand and after Doppler correction the
flagged frequency channels were spread around many velocity channels.
However, in some cases the 1420.0 MHz RFI constituted a significant
amount of the short baseline data for channels in the velocity range
$95~{\rm km~s^{-1}} \lesssim v \lesssim105~{\rm km~s^{-1}}$.  For
these channels the noise in the final images is slightly higher and
the image fidelity is generally worse than in other parts of the
dataset due to poor {\em u-v} sampling.

Continuum emission was subtracted from the data in the {\em u-v} plane
using a linear fit to ranges of channels on either side of the main
\HI\ line \citep{sault94}.  The ATCA does not Doppler track so Doppler
corrections were applied to the data after calibration and continuum
subtraction and before imaging.  The data were shifted to the IAU
defined Local Standard of Rest (LSR) assuming a solar velocity
$V_{sun} = 20.0$ \kms\ in the direction of $RA=18^{h} 07^{m} 50.3^s$,
$\delta=+30^{\circ} 00^{\prime} 52^{\prime\prime}$, J2000.0.  The
continuum subtracted data of the individual 1031 pointings were
linearly combined to form a dirty image of the entire field.  This
approach to mosaicing, rather than the approach of combining the
pointings after deconvolution, is most effective at recovering
information on the largest angular scales.  Robust weighting, with a
robust factor of +0.7, was used in the imaging.  This robustness was
found to optimize the sidelobe levels, synthesized beam size and
overall noise for these images.

The Galactic Center region contains several very strong continuum
sources which, when seen in absorption, appear as negative sources in
the continuum subtracted \HI\ line cubes.  For the sources
near $(l,b)=(0.0^{\circ},0.0^{\circ})$ (Sgr A), $(l,b)=(0.86^{\circ},0.07^{\circ})$ (Sgr B2) and
$(l,b)=(0.6^{\circ},-0.05^{\circ})$ (Sgr B)  the continuum flux is sufficient
that absorption is observed in almost every channel where there is
\HI\ emission.  The maximum entropy deconvolution techniques
traditionally used for large Galactic \HI\ mosaics are not sufficient
to deconvolve diffuse emission in the presence of strong negative
sources.  The standard maximum entropy technique has a positivity
constraint, which means that rather than deconvolving the strongly
negative sources, the sidelobes for these would be treated as structure
in the \HI\ emission and the algorithm would attempt to deconvolve
these structures.

In order to deconvolve the \HI\ cube we used a combined approach of
maximum entropy and traditional CLEAN \citep{hogbom74}.  Before
deconvolution of the diffuse \HI\ emission we deconvolved the
strongest negative sources using the clean algorithm in Miriad's
MOSSDI.  Differing from the technique employed by \citet{stil06}, we
did not filter out small {\em u-v} distances because the continuum
structure near ($l,b$) = ($0.0\arcdeg,0.0\arcdeg$) is extended and the
absorbed flux exceeds that of the diffuse emission by a factor of $\sim 20$.
After deconvolving the absorbed emission, and taking care to not
iterate so far as to start cleaning the diffuse emission, the
clean-component model was subtracted from the {\em u-v} data.  The
resulting data were then re-imaged and the \HI\ emission deconvolved
using Miriad's maximum entropy algorithm, MOSMEM.  To construct the
final image cube we first created a 'residual' cube.  The residual is
the difference of the clean-component model of the {\em u-v}
subtracted dirty cube and the maximum entropy model convolved with
the dirty beam.  This residual image cube was then combined with the
clean and maximum entropy models convolved with a circular Gaussian of
145$^{\prime\prime}$.  This combined maximum entropy and traditional
clean technique proved quite effective, removing most of the sidelobe
structure around the strongest continuum sources near
$-0.2^{\circ}\leq l \leq 0.2^{\circ}$, $-0.2^{\circ}\leq b \leq
0.2^{\circ}$, plus toward $(l,b)=(0.86^{\circ},0.07^{\circ})$ and
$(l,b)=(0.6^{\circ},-0.05^{\circ})$.

\subsection{Single-dish and Interferometer Combination}
Although mosaicing recovers information on angular scales up to
$\lambda/(d_{min}-D/2)\sim 36^{\prime}$, where $d_{min}=31~{\rm m}$ is
the minimum baseline measured and $D=22~{\rm m}$ for the ATCA,
information on larger angular scales is lost to the interferometer.
To recover information on larger angular scales it has become common
practice to combine the interferometric data with images from a
single-dish telescope.  For this survey we use \HI\ data cubes from
the Parkes Galactic All-Sky Survey \citep[GASS;][]{mcgriff09,kalberla10}, which
have an angular resolution of 16\arcmin\ and the same velocity
resolution as the ATCA data (1 \kms). The GASS data are significantly
more sensitive than the ATCA data, with an rms brightness temperature
of $\sim 55$ mK, and therefore do not contribute significantly to the
final noise of the survey presented here.

There are a number of techniques for combining single dish and
interferometric data, including combination prior to deconvolution,
during deconvolution and after deconvolution.  The techniques and
the merits of each are described in detail by
\citet{stanimirovic02}.  For ATCA data with moderate {\em u-v}
sampling, experience has shown that the most robust technique is to
combine the images in the Fourier domain after deconvolution.  As
implemented in Miriad's IMMERGE, the single dish and interferometric
data are deconvolved separately, then Fourier transformed, reweighted,
linearly combined and then inverse Fourier transformed.
Mathematically, the Fourier transform of the combined image,
$V_{comb}$ can be expressed as:
\begin{equation}
  V_{comb} (k) = \omega^{\prime}(k) V_{int} (k) +
  f_{cal}\omega^{\prime\prime}(k) V_{sd} (k), 
\end{equation}
where $V_{int} (k)$ is the Fourier transform of the deconvolved ATCA
mosaic and $V_{sd}(k)$ is the Fourier transform of the deconvolved
Parkes image.  The weighting functions, $\omega^{\prime}(k)$ and
$\omega^{\prime\prime}(k)$, are defined such that $\omega^{\prime}(k)
+ \omega^{\prime\prime}(k)$ is a Gaussian whose full width at
half-maximum is the same as the synthesized ATCA beam. In this way the
ATCA data are down-weighted at the large angular scales and the Parkes data are
down-weighted at the small angular scales. The calibration factor,
$f_{cal}$, scales the Parkes data to match the flux scale of the ATCA
data.  This factor is determined by comparing the ATCA and Parkes
images at every pixel and frequency in the range of overlapping
spatial frequencies, which for an ATCA mosaic and a Parkes image is
$120\lambda$ to $190\lambda$, for $\lambda=21$ cm.  For GASS data
and our ATCA image we found a calibration factor $f_{cal}=1.2$.

\section{Data product}
\label{sec:data}
The final combined \HI\ cube was regridded to Galactic coordinates and
converted to Kelvins of brightness temperature, assuming filled
emission on the scale of the ATCA synthesized beam.  The cube covers
the area $-5.1^{\circ}\leq l \leq +5.1^{\circ}$, $-5.1^{\circ}\leq b
\leq 5.1^{\circ}$ with a pixel size of 35\arcsec\ and a beam size of
145\arcsec.  The velocity coverage is $-309~{\rm km~s^{-1}} \leq v
\leq 349~{\rm km~s^{-1}}$ with a channel spacing of $0.82$ \kms\ and
an effective velocity resolution of 1 \kms.

\subsection{Data Quality and Artifacts}
\begin{figure}
\centering
\includegraphics[width=6in]{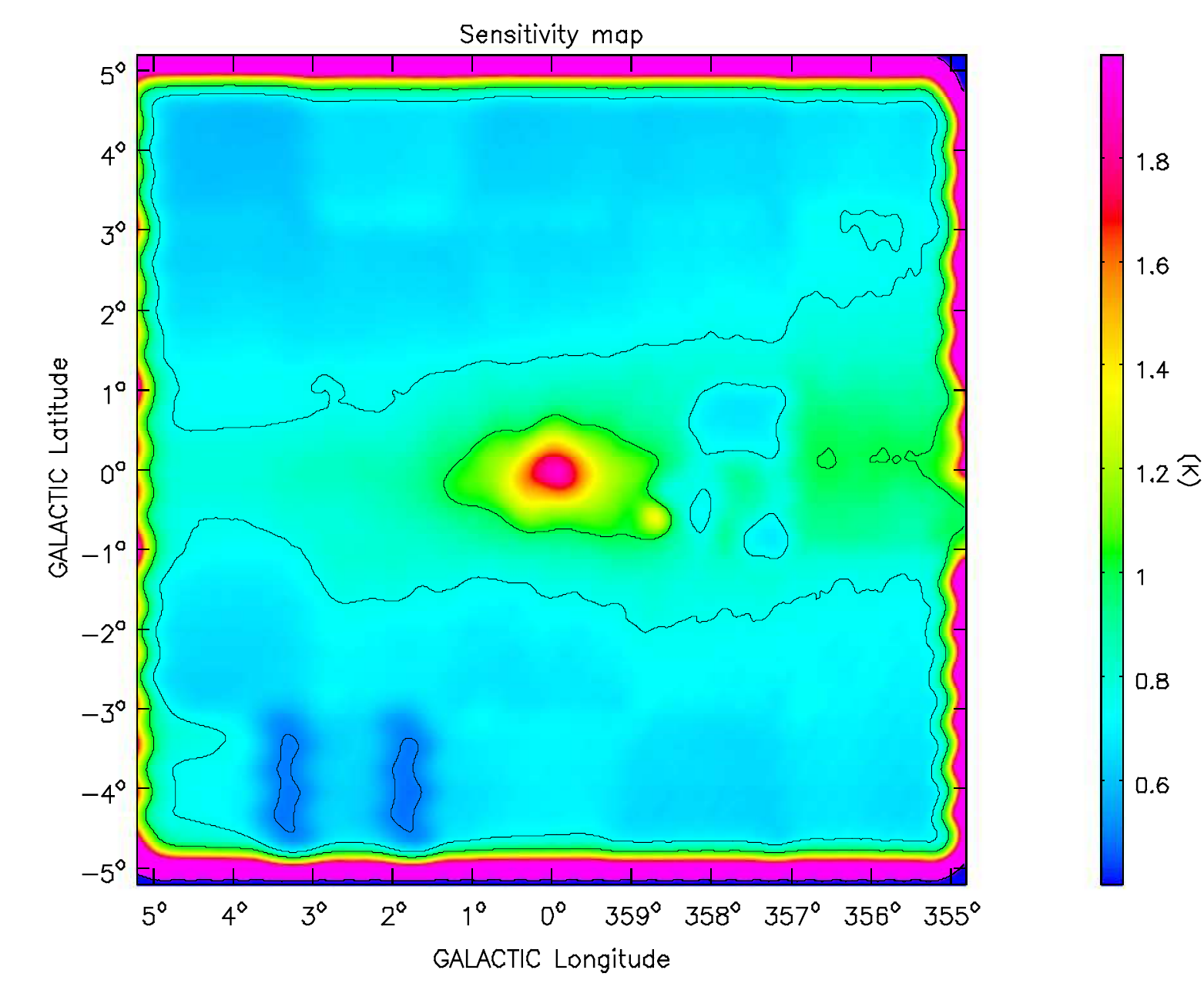}
\caption[]{Sensitivity map of the survey region.  The color scale is
  shown in the wedge at the right and varies linearly from 0.6 - 2.5
  K.  The contours are at 0.5 - 2.0 K in increments of 0.25 K.  The sensitivity is
  not completely uniform due to: (1) slightly uneven {\em u-v} coverage
  of the individual fields that make up the mosaic and (2) bright
  continuum sources, which increase the system temperature and (3)
  broad \HI\ linewidths toward $b=0\arcdeg$, which occupy a large
  fraction of the observing bandwidth and increase the system temperature.
\label{fig:sens_map}}
\end{figure}

The sensitivity in the data cube varies slightly across the field because of observing
strategy used to cover the subregions.  Figure~\ref{fig:sens_map} is
an image of the rms brightness temperature sensitivity per channel
created from the expected sensitivity based on the per pointing
integration time, scaled to the actual sensitivity as measured from
the rms in off-line regions of the cube.  Variations in rectangular
patches are due to differences in the total number of snapshots on
each sub-field.  The remaining variations, in particular the increased
rms towards the Galactic Center and around the Galactic plane, in
general are due to true increases in system temperature from the
contributions of the strong continuum emission and bright \HI, which
fills most of the velocity band.  The mean rms, $\sigma_{tb}$, per
channel away from the Galactic plane is $\sim 0.7$ K, increasing to
$\sim 1.0$ K at $b\sim 0\arcdeg$ and $\sim 2.0$ K towards $(l,b) =
(0.0\arcdeg, 0.0\arcdeg)$.

In general the mean rms measured in individual off-line channels is
comparable to the channel-to-channel rms measured in individual
spectra.  However, the spatial distribution of the noise is not entirely
constant with velocity.  The velocity coverage differs slightly from
field to field and particularly for the data incorporated from the
SGPS, which covered a slightly smaller range.  As a result the noise
increases for $v<-225$ \kms\ and $v>267$ \kms\ in the region that was
originally covered by the SGPS, namely $l<-3\arcdeg$,
$|b|<1.2\arcdeg$.

\begin{figure}
\centering
\includegraphics[width=4in]{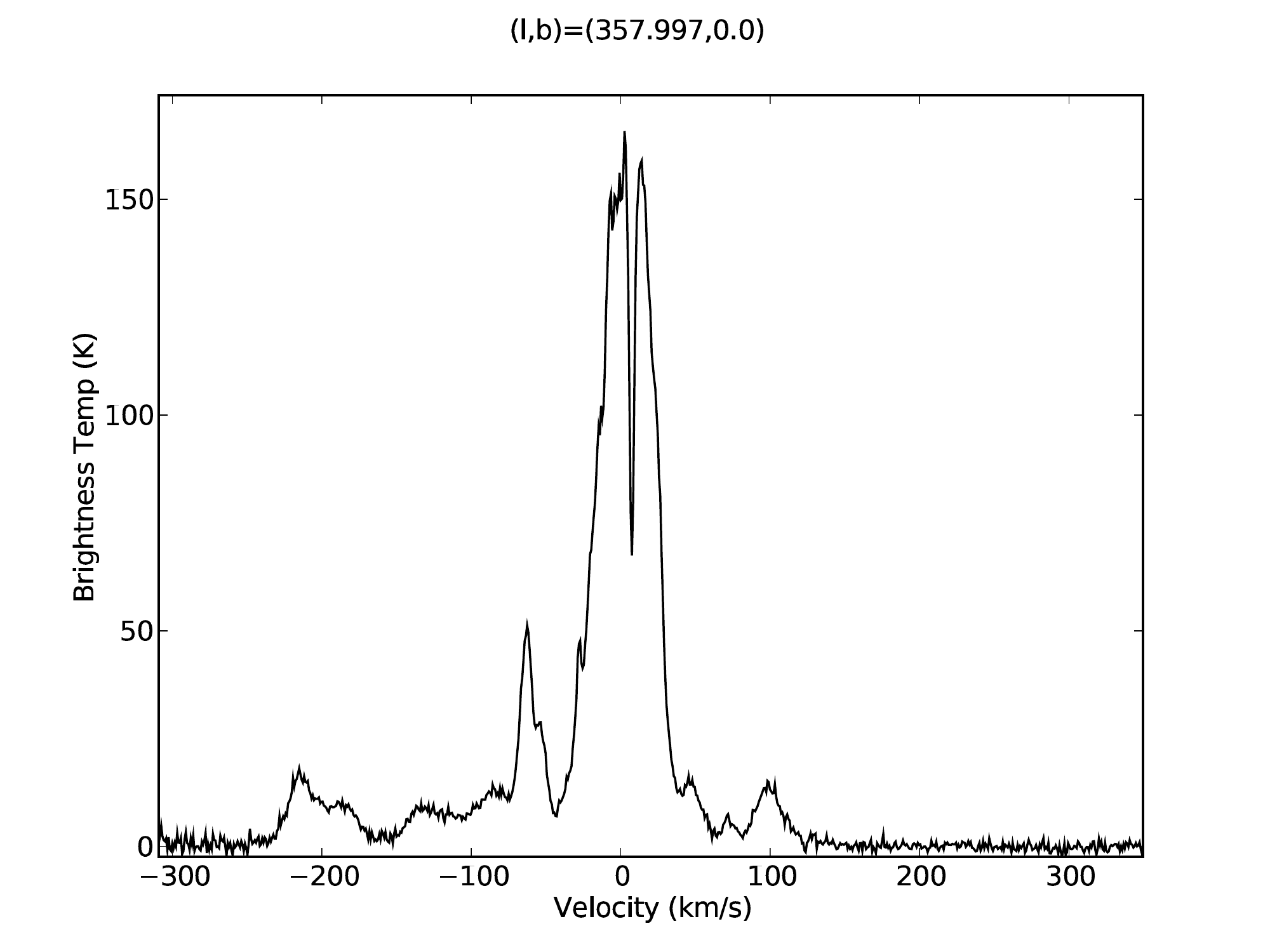}
\includegraphics[width=4in]{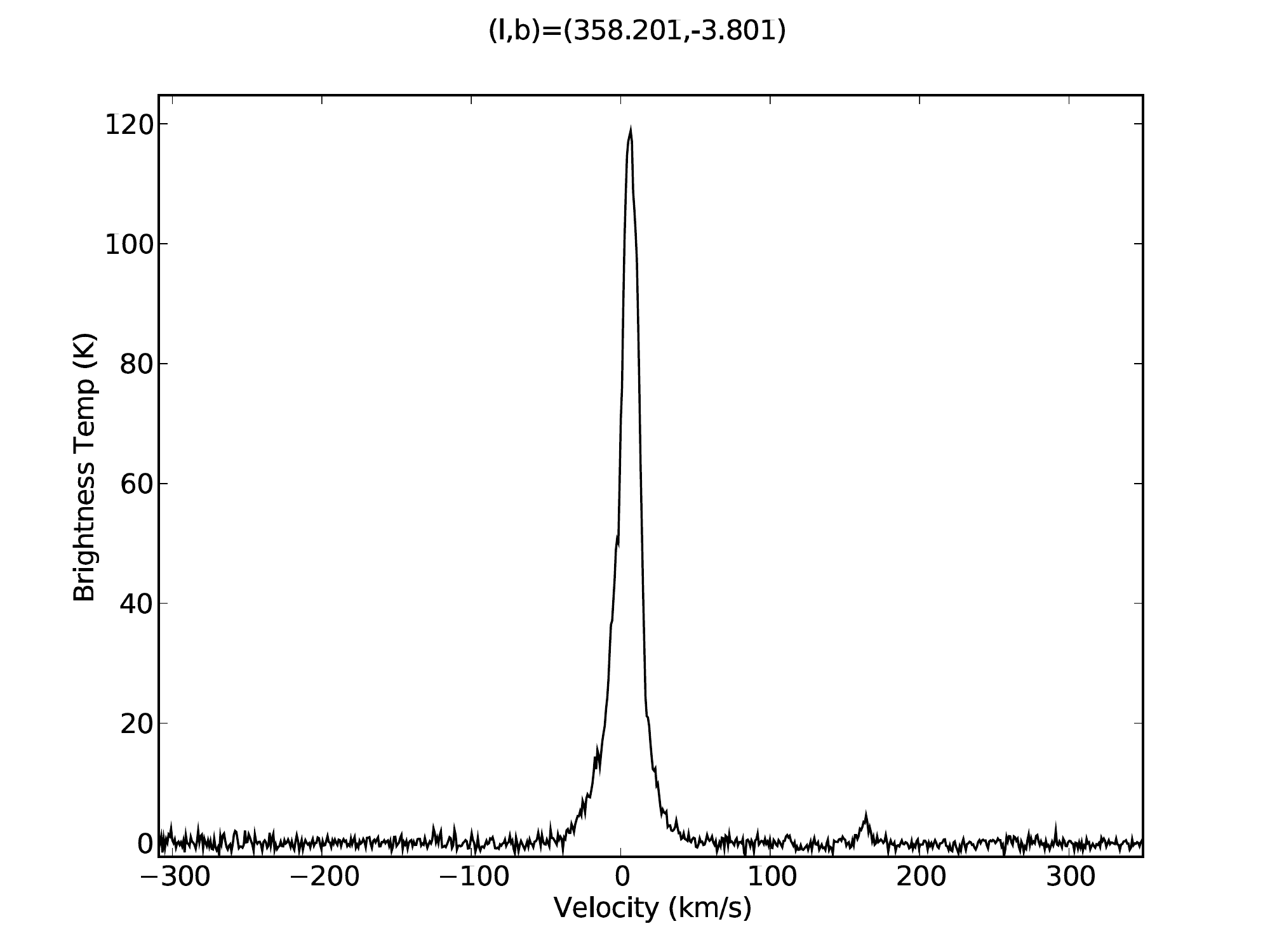}
\caption[]{Example spectra towards two positions, near ($l,b=
  -2.0\arcdeg, 0.0\arcdeg$) and far ($l,b=
  -1.8\arcdeg, -3.8\arcdeg$)
  from the Galactic plane.  These spectra give an indication of the
  noise in the spectral domain and the spectral baselines. 
\label{fig:spec}}
\end{figure}

Figure \ref{fig:spec} shows two example spectra at
$(l,b)=(358.0\arcdeg,0\arcdeg)$ and $(l,b)=(358.2\arcdeg,-3.8\arcdeg)$.
These demonstrate the spectral baseline quality and spectral noise
achieved in this survey.  The slight increase in rms at the extreme
velocities is apparent in the spectrum obtained at $(l,b)=(358.0\arcdeg,0\arcdeg)$.

The data cube contains some imaging artifacts.  These fall into two
main categories: those associated with strongly absorbed continuum
sources, and low-level artifacts due to limited {\em u-v} coverage that
lie near the noise level in individual channels.  The most obvious
imaging artifacts are those due to dynamic range limitations around
the strongly absorbed continuum sources near
$(l,b)=(0\arcdeg,0\arcdeg$) and $(l,b)=(-2.4\arcdeg,-0.1\arcdeg)$.
These regions are limited by incomplete cleaning, a problem that
cannot be solved with the deconvolution techniques used here, but may
be addressed in the future using Multi-scale clean
\citep{cornwell08}.  There are a few pixels at
$(l,b)\approx(0\arcdeg,0\arcdeg$) in the GASS data that were saturated
in the raw data at the time of observation. The limited dynamic range
of the area immediately surrounding Sgr A$^{*}$, plus the lack of an
accurate single dish image at this position, limits the usefulness of
this dataset for studying the \HI\ absorption towards Sgr A$^{*}$.
For studies of absorption in the inner degree, users are advised
to use the \citet{lang10} VLA survey.

The other category of imaging artifacts is apparent in images created
by integrating over large velocity ranges such as zeroth moment
images, which show residual striping and cross-hatching.  These
effects are almost certainly due to the limited {\em u-v}
coverage, although some of the more prominent regions of parallel
striping are clearly due to low-level RFI that has not been
completely flagged.  In individual velocity channels these effects
mostly lie below the noise.

Small spurious emission features appear at a velocity of $v=+48$ \kms\
at $(l,b)=(3.77\arcdeg, -2.24\arcdeg)$, $(l,b)=(4.13\arcdeg,
-2.57\arcdeg)$ and $(l,b)=(4.03\arcdeg, -1.73\arcdeg)$.  These bright,
$T_b \sim 25$ K, features are only three velocity channels wide and
are artifacts from the GASS data.  No counterparts to these features
are observed in the ATCA-only data.

\section{Images}

Images of every tenth velocity channel are shown in Figure
\ref{fig:chanmaps}.  The greyscale is different for each panel as
shown in the associated color wedges.  The images
highlight the wealth of small-scale features visible in these data.
In Figure \ref{fig:lv} we show longitude-velocity ({\em l-v})  images at latitude
intervals of $0.1\arcdeg$ over the latitude range $-1.4\arcdeg \leq b
\leq +1.5\arcdeg$.  These are especially useful for studying the
large-scale structure of the region and can be compared usefully with
existing CO data \citep[e.g.][]{bally87,oka98}.
\label{subsec:features}

\subsection{Noteworthy features}
Full analysis of these data will focus on the \HI\ structure and
dynamics of the Galactic Center and an analysis of individual
high-velocity features.  These will be the topics of forthcoming
papers.  Here we restrict ourselves to a brief discussion of some of
the most noteworthy features.

\subsubsection{Large-scale \HI\ Distribution}
Figure \ref{fig:chanmaps} shows the \HI\ sky distribution sampled
every 8 \kms.  The high velocities ($|v| > 100$ \kms) show the
well known \citep[e.g.][]{kerr67,vanderkruit70,burton78a} \HI\ emission
structure of positive velocity gas at $b<0\arcdeg$ and
$l>0\arcdeg$ and similarly, negative velocity gas at $b>0\arcdeg$
and $l<0\arcdeg$.  \citet{burton78a} and \citet{liszt80} have attributed this to a
tilted \HI\ inner disk which is oriented at 24\arcdeg\ with respect to
the Galactic plane. 

Other well-known features are seen in the longitude-velocity ({\em
  l-v}) images shown in Figure~\ref{fig:lv}, which are sampled at
intervals of $0.1^{\circ}$ in latitude. In these we can see the
prominent ``connecting arm'' at extreme positive longitudes and
positive velocities ($l>3\arcdeg$, $200~{\rm km~s^{-1}} \leq v \leq
300~{\rm km~s^{-1}}$) and the nearly symmetric ``looping'' ridge at
the extreme negative longitudes and velocities ($l<-1\arcdeg$, $v \sim
-210~{\rm km~s^{-1}}$) .  \citet{fux99} modeled these features as
dust lanes on the near and far side of the bar, respectively.  The
near-side dust lane is particularly prominent at negative longitudes,
whereas the far-side lane becomes quite clear at $b\approx 0^{\circ}$.

\subsubsection{The 3-kpc arms}
While the near 3-kpc arm has been well studied in \HI\ at low
resolution, both the near 3-kpc arm and the recently discovered far
3-kpc arm are unexplored in \HI\ at this resolution. Following the
linear {\em l-v} fits of \citet{dame08} we constructed zeroth order
moment maps of the \HI\ emission centered on the velocity mid-point of
the near and far arms at every longitude within the Galactic centre
region, using a width in velocity of 26 \kms\ as used by Dame \&
Thaddeus (2008). These are shown in Figure \ref{fig:3kpcarms}. For
most of the longitude range covered by our Galactic Center survey the
\HI\ emission from the far arm is blended in longitude and latitude
with both local gas and gas far outside the solar circle.  While the
near arm is clearly visible in our data, the far arm is obviously
confused by the Galactic centre gas ($l=\pm$1$^{\circ}$) and also gas
near $l\approx 3^{\circ}$, which is associated with Bania's Clump 2
(see \S \ref{subsubsec:bania} below).  In the intermediate longitudes,
particularly near $l= 2\arcdeg$ and $l= -2\arcdeg$, there is a narrow
strip of emission at $b\approx 0\arcdeg$ which may be attributed to
the far 3 kpc arm.  We note, however, that the best evidence for the
far 3-kpc arm presented by \citet{dame08} lies outside the longitude
range of our survey.  Figure \ref{fig:3kpc_lv} is a longitude-velocity
({\em l-v}) diagram at $b=0^{\circ}$ with the linear fits of
\citet{dame08} overlaid.  The near 3-kpc arm is clearly visible and
hints of the feature identified as the far 3-kpc arm by \citet{dame08}
are also visible. The latter shows evidence of a possible
split/bifurcation near $l\approx3\arcdeg$, as was discussed by
\citet{dame08}.

In Figure~\ref{fig:3kpc_lat}, we show the integrated latitude profiles
of the features shown in Figure~\ref{fig:3kpcarms}. These are summed
from the non-confused portions of each feature, and as such the far
3-kpc arm latitude profile contains \HI\ emission from only about 2
degrees of longitude.  As \citet{dame08} found for the CO emission,
the latitude widths of the two arms in differ by a factor of $\sim$2,
with full-widths at half-max (FWHMs) in \HI\ of 1.6$^{\circ}$ and
0.7$^{\circ}$ for the near and far arms, respectively. Assuming a
distance of the 3-kpc arm structure from the Galactic centre of 3.5 to
4 kpc, and a solar distance of 8.4 kpc \citep{ghez08,reid09}, these
correspond to a FWHM in $z$ of $126-141$ pc and $149-155$ pc for
the near and far arms, respectively. Our values differ slightly from
the estimates of \citet{dame08}, which were calculated over the
longitude range $5\arcdeg \leq l \leq 9\arcdeg$ and using slightly
different estimates for the solar distance and Galactocentric radius
of the 3-kpc arms.  If we use the same distance values as
\citet{dame08} (5.2 kpc and 11.8 kpc) we find $z$ FWHM values of 149
and 146, respectively.  \citet{dame08} used the comparable
$z$ heights of the two features as evidence for the far 3kpc arm.
They also postulated that the small FWHM is due to a lack of star formation to
heat and inflate the gas.  However \citet{green09b} showed high-mass
star formation to be present in both arms through the association of
6.7-GHz methanol masers, known tracers of high-mass star formation.
This implies that both arms have undergone star formation activity and
any effect on $z$ height is the same for both.  

\subsubsection{Bania's Clump 2}
\label{subsubsec:bania}
There are a few positions in the Galactic Center region that are
known to have very wide linewidths over an extent $>0.5\arcdeg$
\citep{bania77}.  These have been observed extensively in $^{12}$CO
\citep{bania86,stark86,liszt06}, OH \citep{boyce94} and other molecular line
tracers \citep{lee96,huettemeister98,liszt06}, as well as in \HI\
\citep[eg.][]{riffert97}.  The area known as Bania's Clump 2 at $l\sim
3.2\arcdeg$, $b\sim +0.3\arcdeg$ was mapped in \HI\ by
\citet{riffert97} using the Effelsberg 100m telescope, showing an \HI\
linewidth of $\sim 100$ \kms\ over the range $0~{\rm km~s^{-1}} \leq v
\leq 200$ \kms\ and implying an \HI\ mass of $1.5 \times 10^{5}~{\rm
  M_{\odot}}$.  High resolution CO maps by \citet{stark86} suggest the
region has a mass of $10^6~{\rm M_{\odot}}$ made up of approximately
16 individual clumps.

\begin{figure}
\centering
\epsscale{0.85}
\plottwo{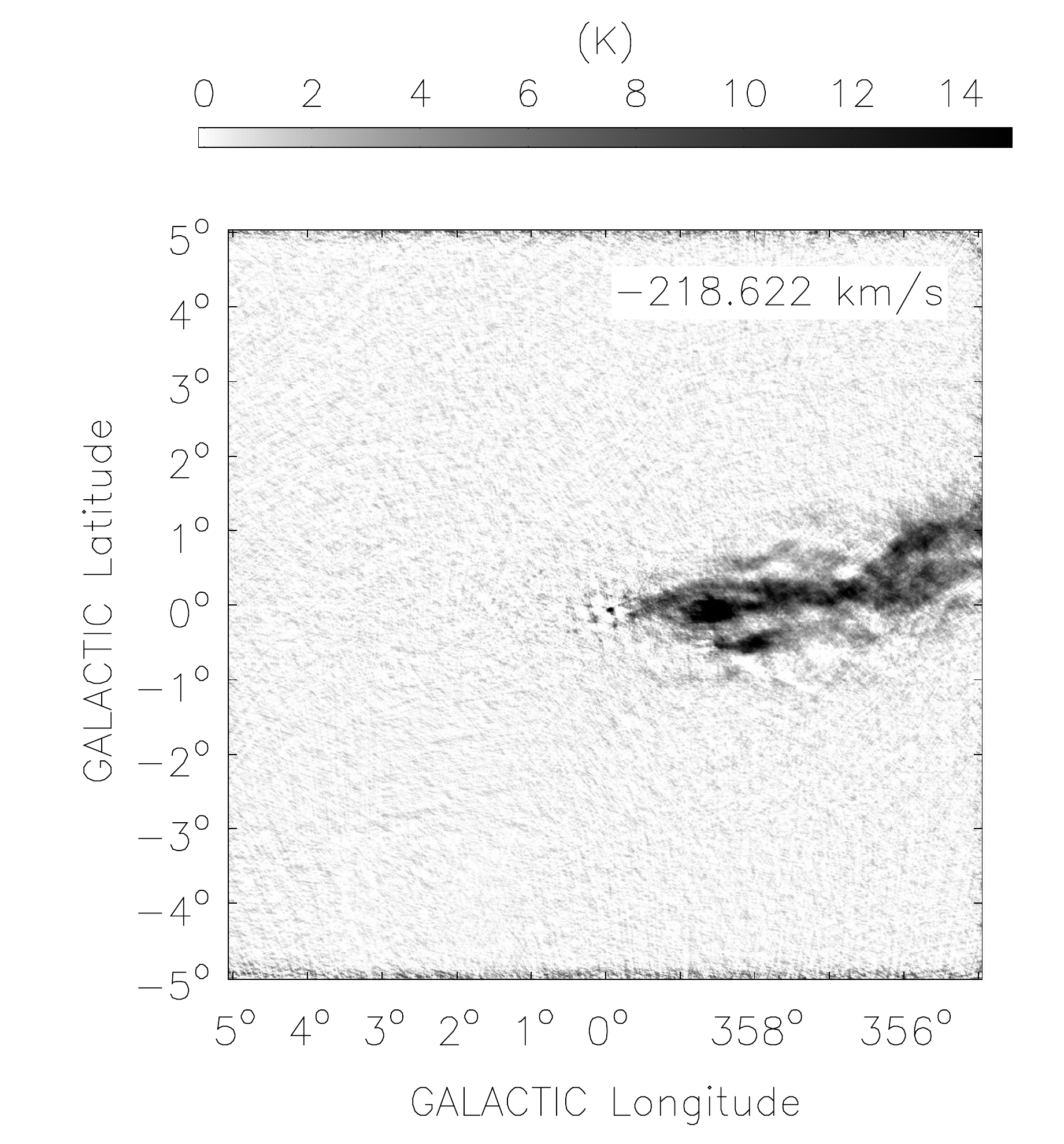}{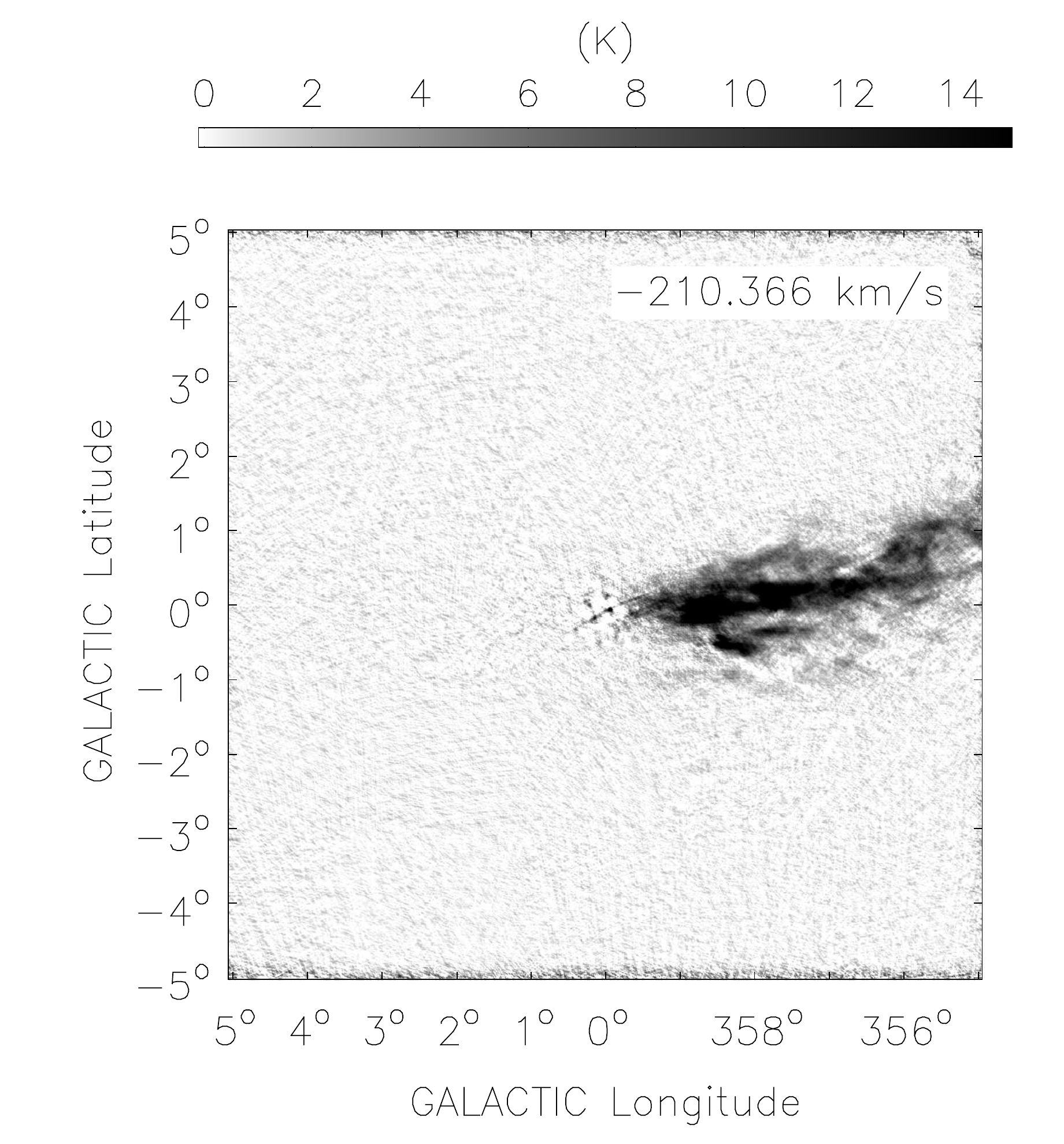}\\
\plottwo{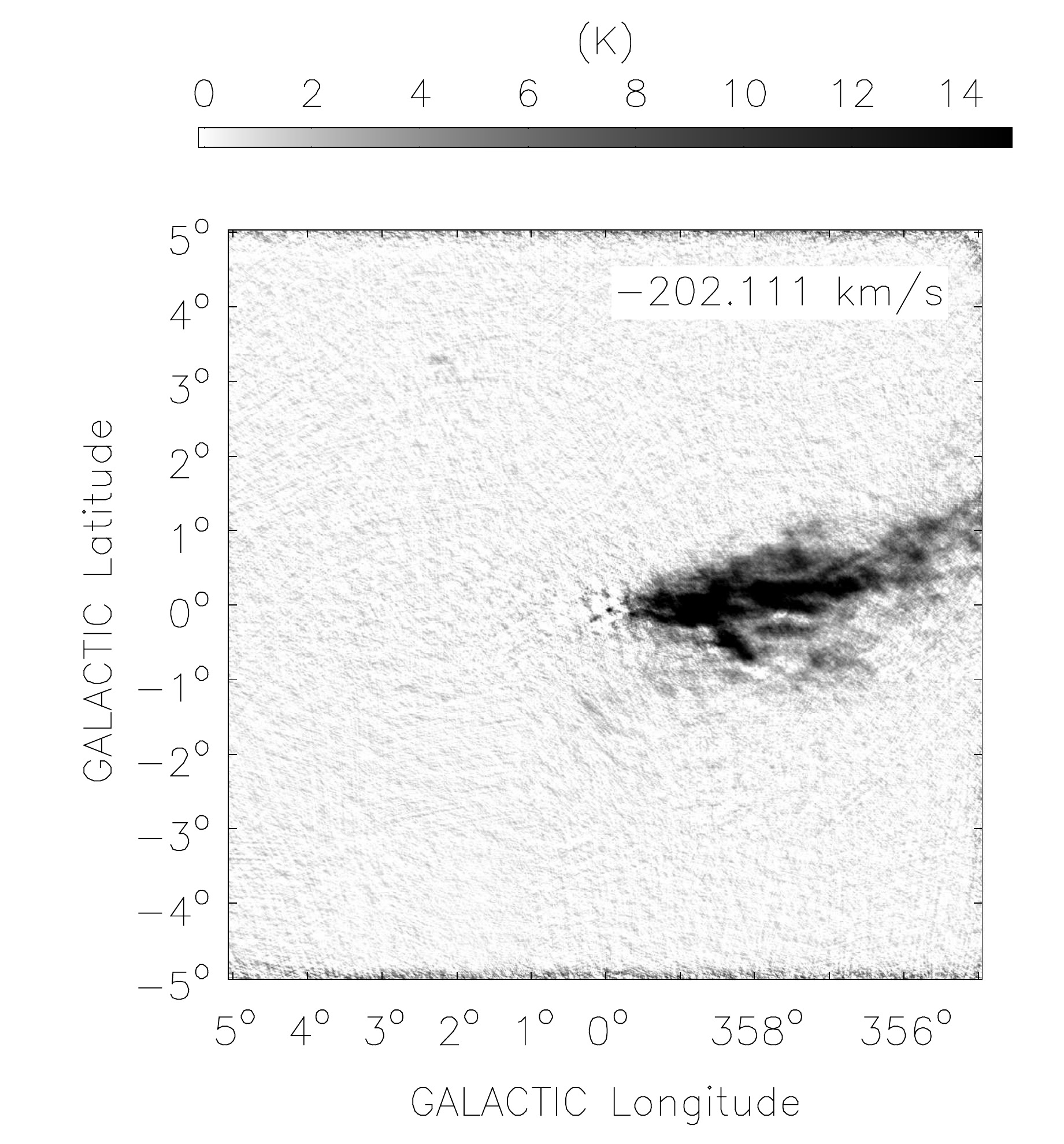}{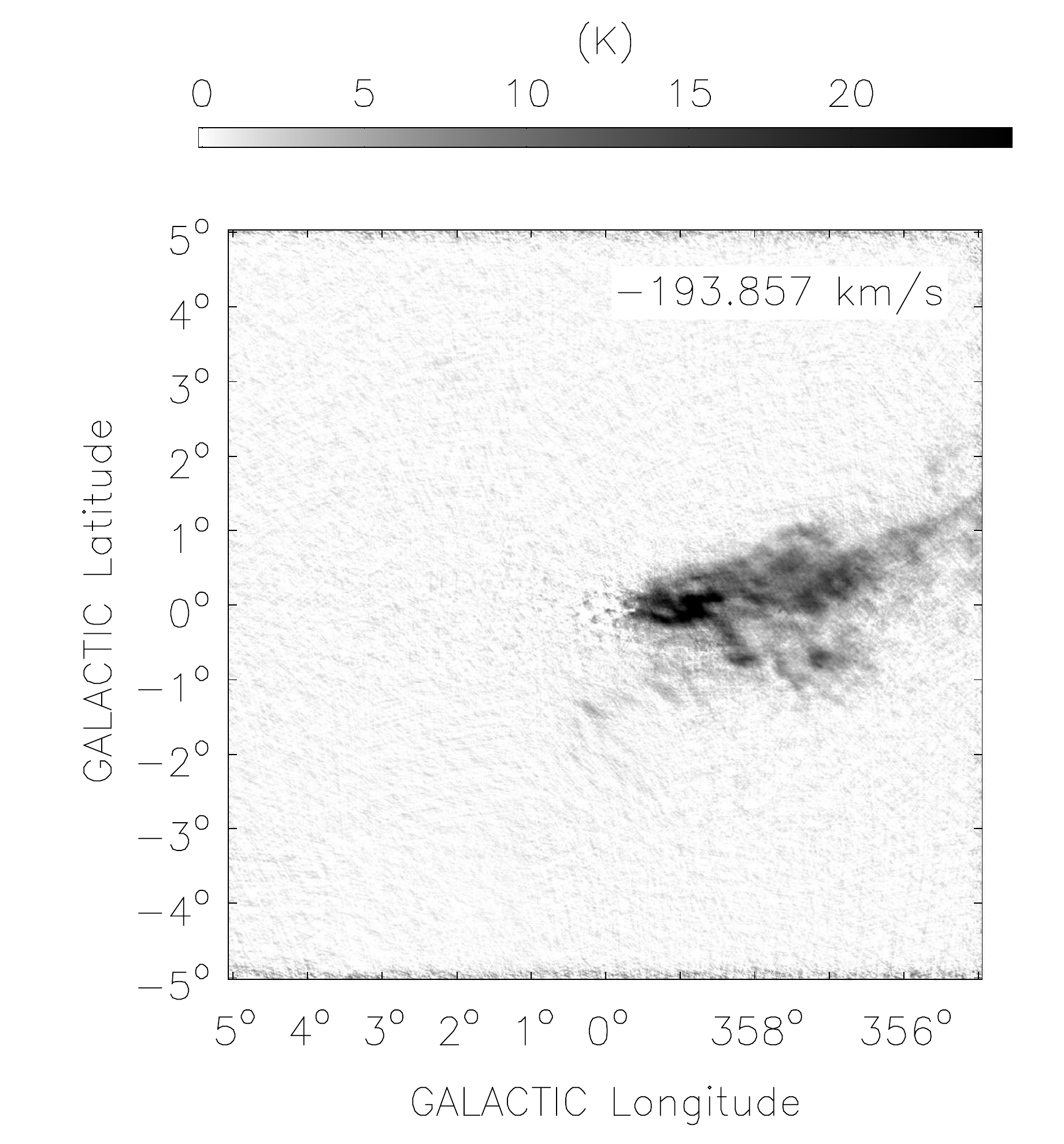}\\
\plottwo{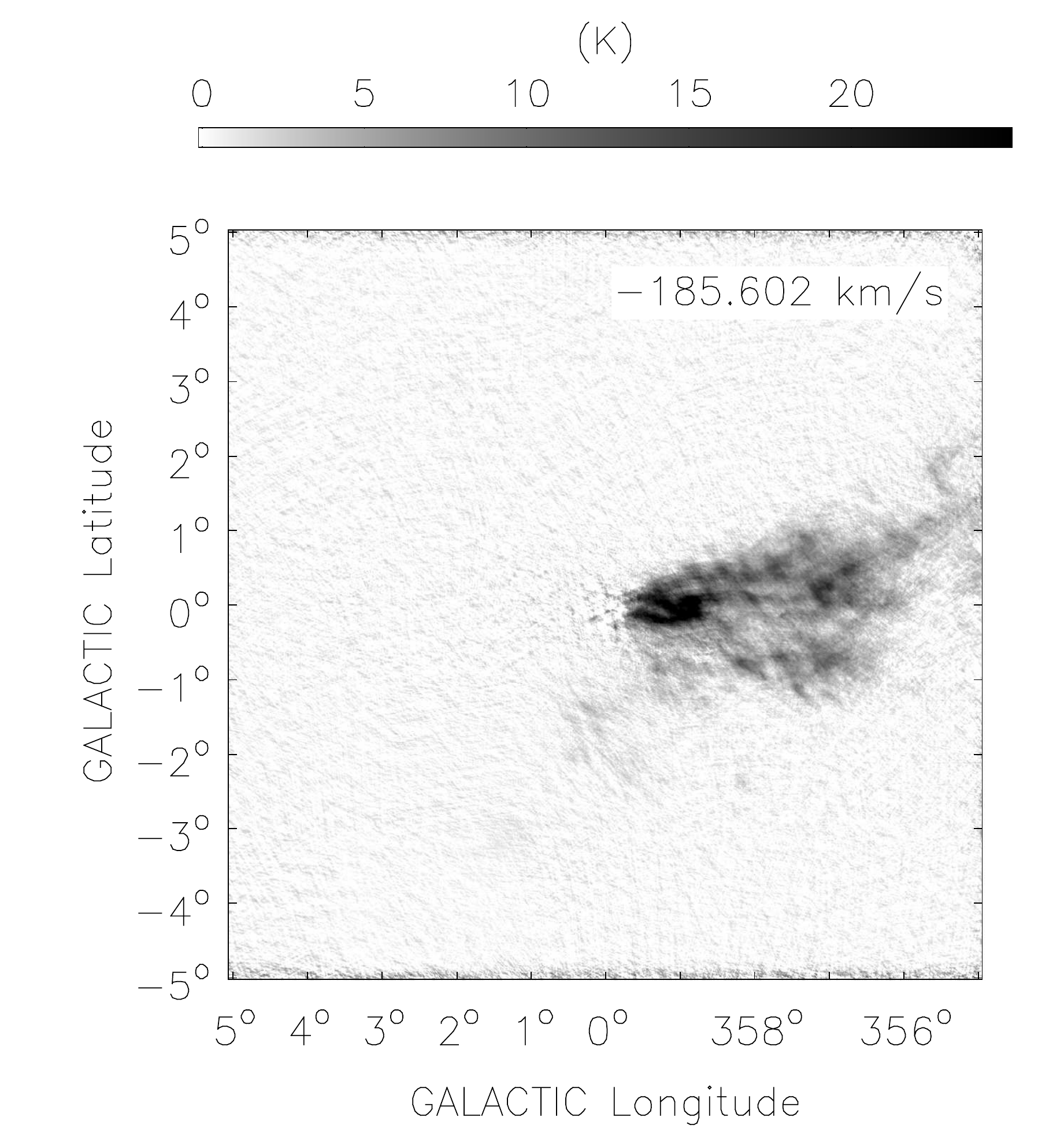}{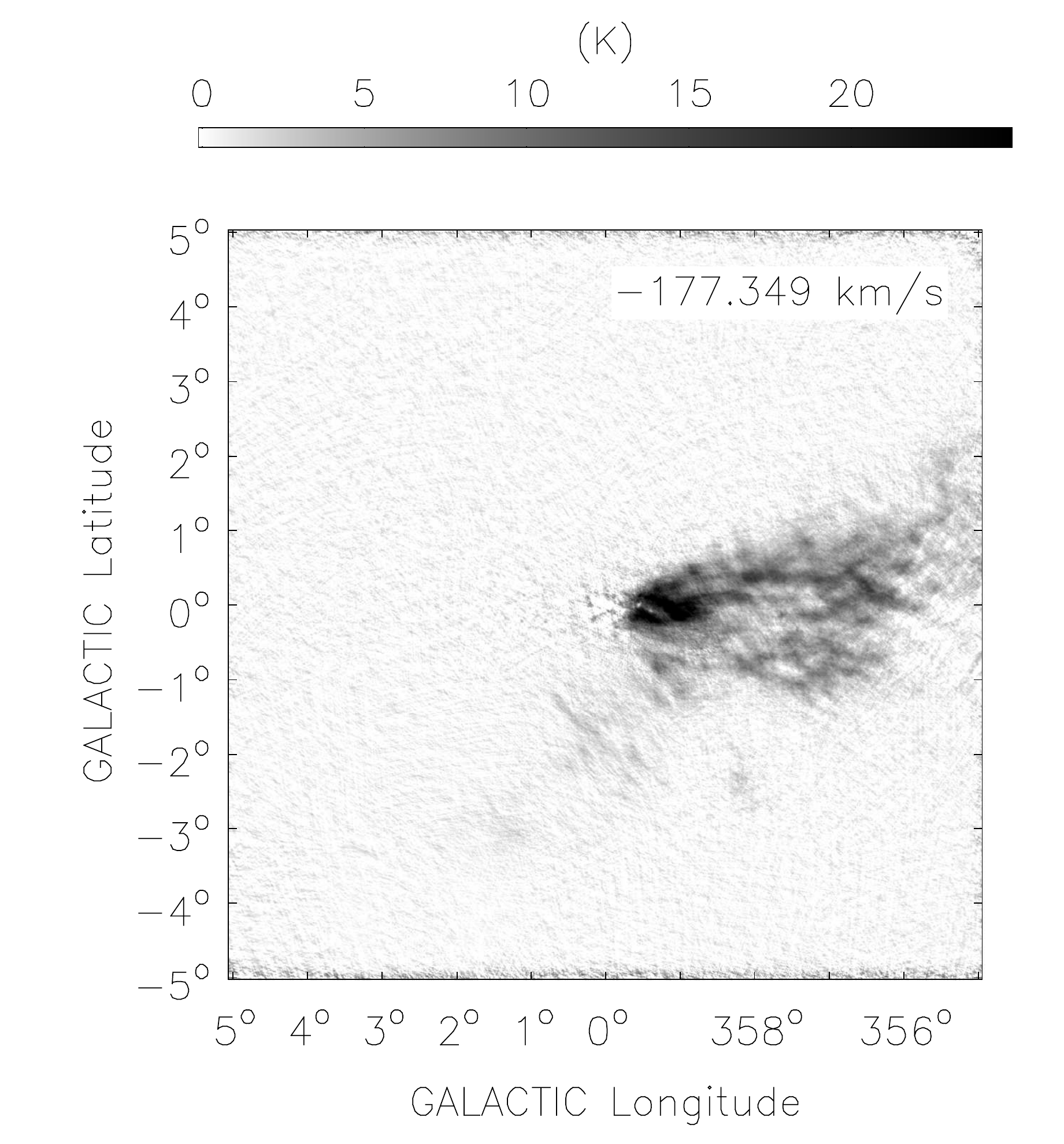}
\caption[]{Velocity channel images of the full survey area.  Images of
  every tenth channel are shown, such that the velocity spacing
  between images is $\sim 8$ \kms.  The greyscale is different for
  each panel, as shown in the wedges above, where black is bright and
  white is zero or negative. These images use a scaling
  power of -0.6. 
\label{fig:chanmaps}}
\end{figure}

The physical interpretation of the wide-linewidth molecular clouds has
been the subject of several studies.  \citet{stark86} interpreted
Clump 2 as a dust lane.  Liszt (2006, 2008)\nocite{liszt06,liszt08},
however, suggested that the linewidths and extended vertical structure of
the clouds can be explained by molecular cloud shredding due to orbits
associated with the bar.  In particular, Bania's Clump 2 is thought to
be a gas cloud that is about to enter a dust lane shock.  This is a
similar model to that of \citet{fux99}, where the dust lane shock is the
Connecting Arm, but in the \citet{liszt08} model the dust lane is at a
lower velocity, $v\sim 150$ \kms, than the Connecting Arm.  The
\citet{liszt08} model attributes the latitude extent of Clump 2 to a
transition from gas near the Galactic midplane to gas associated with
the dust lane, which is located below $b=0\arcdeg$.  This is supported
by the fact that the emission from Clump 2 does not extend below the
tilted disk, as observed at $v\sim 136$ \kms\ in
Figure~\ref{fig:chanmaps}g.

Bania's Clump 2 is clearly seen in the \HI\ channel maps shown in
Figs~\ref{fig:chanmaps}f \& \ref{fig:chanmaps}g over the velocity
range $79~{\rm km~s^{-1}} \leq v \leq 144$ \kms, as well as in the
{\em l-v} images shown in Figs~\ref{fig:lv}c \& \ref{fig:lv}d over the latitude
range $+0.2\arcdeg< b < +0.9\arcdeg$.  The cloud forms a very clear arc
and appears complemented by another arc opposite at $l\sim
1.5\arcdeg$.  The feature at $l\sim 1.5\arcdeg$ is another well-known
broad line molecular cloud \citep[e.g.][]{liszt08,tanaka07}.  The
looping structure of Clump 2 is also visible in CO \citep[see Fig 3
of][]{liszt08}.  No model has been suggested to explain the apparent
morphological connection between the $l\sim 1.5\arcdeg$ cloud and
Clump 2.

Slices taken in latitude and longitude across Clump 2, as shown in
Figure \ref{fig:clump_slices}, show very sharp edges towards the
``interior'', with a more gradual decline to the exterior.  Similar
structure was observed in the Galactic supershell GSH 277+00+36, where
the sharp edges were attributed to compression on the interior edge of
a supershell \citep{mcgriff03}.  In the present case, it may also be
possible that the complementary looping structures are walls of a
supershell.  In fact, the cloud at $l\sim 1.5\arcdeg$ itself was
described as a nascent superbubble by \citet{tanaka07}, who presented
observations of several small expanding objects within the larger
cloud complex.  However, there is no clear evidence for massive star
formation to power this feature \citep{rodriguez-fernandez06}.
Furthermore, there is no evidence of expansion over the wide range of
velocities observed.  The walls appear largely stationary with
velocity.  It therefore seems unlikely that these wide-line clouds are
part of an expanding supershell, but more likely the \citet{liszt08}
interpretation that both of these clumps are molecular clouds entering
the dust lane shock is correct.  In this case the compression along the
low longitude edge of Clump 2 would suggest that the gas there is
already being shocked and that the shock emanates from the side
nearest to the Galactic Center.  Interestingly there is little
evidence for extremely sharp walls in the cloud at $l\sim 1.5\arcdeg$.

\citet{bally10} focused on the structure of Bania's Clump 2, particularly
as it pertains to 1.1-mm dust emission from the Bolocam Galactic Plane
Survey.  \citet{bally10} show that Bania's Clump 2 contains dozens of 1.1-mm
sub-clumps, which are differing in near and mid-infrared emission,
suggesting that they are either inefficient at forming stars or are
pre-stellar.  \citet{bally10} go on to suggest that the lack of star
formation is due to high pressures or to large non-thermal motions, such
as might be expected if the cloud is located either where the {\em x1}
orbits, i.e.\ those parallel to bar, become self-intersecting, leading
to cloud collisions, or if the cloud is located where a spur
encounters a dust lane at the leading edge of the bar.  The \HI\
emission structure, particularly the sharp walls, is consistent with
the shock picture.

\subsection{Galactic outflows and loops}
Figure~\ref{fig:small_clouds} is a velocity channel image at $v=163$
\kms\ showing several very small cloud-like features lying at high
Galactic latitudes.  Clouds like these are apparent throughout the
cube at high ($|v|>130$ \kms) velocities, as well as at lower
velocities where they are quite distinct from the emission related to
large-scale Galactic structure, such as a tilted disk
\citep[e.g.][]{liszt80}.  We find over 60 separate compact clouds in
the survey area at velocities $|v|>80$ \kms.  The properties of the
detected clouds vary but typical angular sizes are $0.1 - 0.5$ deg,
with a few larger clouds.  Most of the detected clouds are faint, with
brightness temperatures of $\sim 3 - 10$ K.  Several clouds have very
narrow velocity linewidths ($\sim 2$ \kms), but most have linewidths in
the range $7 - 20$ \kms.  Rough mass estimates assuming optically thin
emission and a distance of 8.4 kpc \citep{reid09} are in the range $250
- 1500~{\rm M_{\odot}}$.  In terms of physical properties these clouds
are very similar to those found by \citet{ford10} towards inner Galaxy
tangent points.

In addition to the compact discrete clouds, there are cohesive structures like that shown in
Figure~\ref{fig:chanmaps}c (bottom two panels), which forms an extended loop at $l\sim
359\arcdeg$, $b\sim -3.5\arcdeg$ pointing back towards the Galactic
Center with narrow linewidth clumps along it.  These high
velocity features are discussed further in McClure-Griffiths et al
(2012, in prep) where we elaborate on the distribution and properties of the
clouds and consider the possibility that they are associated with gas
entrained in a Galactic outflow \citep[e.g.][]{fujita09}.

A well studied large-scale feature of the Galactic Center is the
so-called Galactic Center lobe \citep{sofue84,law10}, which is traced
in radio continuum, H$\alpha$, and infrared emission extending more
than a degree above the plane.  There is no obvious \HI\ signature of
the Galactic Center lobe. This is not surprising, however, given that
the hydrogen recombination lines are at very low LSR velocities of
$\sim -5$ \kms\ to $\sim 8$ \kms\ \citep{law09}, which are very
crowded with \HI\ emission.  The magnetic Loop 1 of molecular gas
proposed by \citet{fukui06}, however, is easily visible in the bottom
panels of Figure~\ref{fig:chanmaps}b at $v\sim-130$ \kms.  The other loop proposed by
same authors is less obvious in \HI.

\section{Summary}
The ATCA \HI\ survey of the Galactic Center covers the inner
$10^{\circ} \times 10^{\circ}$ of the  Milky Way, providing a new resource for
high-angular resolution surveys of the structure and dynamics of
neutral gas in the central 3 kiloparsecs of the Milky Way.  We present
the final \HI\ data cube with an angular resolution of $145\arcsec$ and
a velocity resolution of 1 \kms.  The mean rms $T_b$ across the full area is
$\sim 0.7$ K, varying from $0.5$ K to $2.0$ K, with the highest noise
towards $(l,b)=0\arcdeg,0\arcdeg$.  

This survey completes the International Galactic Plane Survey
\citep{mcgriff05,stil06,taylor03} \HI\ survey of the inner Milky Way,
providing the link between the first and fourth Galactic quadrants.
The data will be valuable for comparison with existing high angular
resolution surveys of Galactic Center at all wavelengths.  Comparison
with CO \citep{dame01,oka98} and infrared data \citep{molinari11,arendt08}
will help improve models of the inner Galaxy structure, while
comparison with radio continuum and high energy emission may reveal
the nature of the outflowing gas.  Here we have discussed briefly some
of the more noteworthy features including the near and far 3-kpc arms,
a wide linewidth molecular cloud (Bania's Clump 2) and small high
velocity clumps.  

\acknowledgements We gratefully acknowledge the extended support of
ATCA staff for the Southern Galactic Plane Survey projects,
without whom these surveys would not have been possible, in
particular, Mark Wieringa and Robin Wark.  We
are grateful to Katherine Newton-McGee for her assistance with
data editing and initial calibration.  NMMc-G and JAG thank Thomas Dame
for his comments on the \HI\ in the far 3-kpc arm.

\bibliographystyle{apj} 

\begin{thebibliography}{74}
\expandafter\ifx\csname natexlab\endcsname\relax\def\natexlab#1{#1}\fi

\bibitem[{{Arendt} {et~al.}(2008){Arendt}, {Stolovy}, {Ram{\'{\i}}rez},
  {Sellgren}, {Cotera}, {Law}, {Yusef-Zadeh}, {Smith}, \& {Gezari}}]{arendt08}
{Arendt}, R.~G., {Stolovy}, S.~R., {Ram{\'{\i}}rez}, S.~V., {et~al.} 2008,
  \apj, 682, 384

\bibitem[{{Babusiaux} \& {Gilmore}(2005)}]{babusiaux05}
{Babusiaux}, C., \& {Gilmore}, G. 2005, \mnras, 358, 1309

\bibitem[{Bally {et~al.}(2010)Bally, Aguirre, Battersby, \& et~al}]{bally10}
Bally, J., Aguirre, J., Battersby, C., \& et~al. 2010, \apj, 721, 137

\bibitem[{{Bally} {et~al.}(1987){Bally}, {Stark}, {Wilson}, \&
  {Henkel}}]{bally87}
{Bally}, J., {Stark}, A.~A., {Wilson}, R.~W., \& {Henkel}, C. 1987, \apjs, 65,
  13

\bibitem[{Bania(1977)}]{bania77}
Bania, T.~M. 1977, \apj, 216, 381

\bibitem[{Bania {et~al.}(1986)Bania, Stark, \& Heiligman}]{bania86}
Bania, T.~M., Stark, A.~A., \& Heiligman, G.~M. 1986, \apj, 307, 350

\bibitem[{Benjamin {et~al.}(2005)Benjamin, Churchwell, Babler, Indebetouw,
  Meade, Whitney, Watson, Wolfire, Wolff, Ignace, Bania, Bracker, Clemens,
  Chomiuk, Cohen, Dickey, Jackson, Kobulnicky, Mercer, Mathis, Stolovy, \&
  Uzpen}]{benjamin05}
Benjamin, R.~A., Churchwell, E., Babler, B.~L., {et~al.} 2005, \apj, 630, L149

\bibitem[{{Binney} {et~al.}(1991){Binney}, {Gerhard}, {Stark}, {Bally}, \&
  {Uchida}}]{binney91}
{Binney}, J., {Gerhard}, O.~E., {Stark}, A.~A., {Bally}, J., \& {Uchida}, K.~I.
  1991, \mnras, 252, 210

\bibitem[{{Bitran} {et~al.}(1997){Bitran}, {Alvarez}, {Bronfman}, {May}, \&
  {Thaddeus}}]{bitran97}
{Bitran}, M., {Alvarez}, H., {Bronfman}, L., {May}, J., \& {Thaddeus}, P. 1997,
  \aaps, 125, 99

\bibitem[{Bland-Hawthorn \& Cohen(2003)}]{bland-hawthorn03}
Bland-Hawthorn, J., \& Cohen, M. 2003, \apj, 582, 246

\bibitem[{{Blitz} {et~al.}(1993){Blitz}, {Binney}, {Lo}, {Bally}, \&
  {Ho}}]{blitz93}
{Blitz}, L., {Binney}, J., {Lo}, K.~Y., {Bally}, J., \& {Ho}, P.~T.~P. 1993,
  \nat, 361, 417

\bibitem[{{Blitz} \& {Spergel}(1991)}]{blitz91}
{Blitz}, L., \& {Spergel}, D.~N. 1991, \apj, 370, 205

\bibitem[{{Boyce} \& {Cohen}(1994)}]{boyce94}
{Boyce}, P.~J., \& {Cohen}, R.~J. 1994, \aaps, 107, 563

\bibitem[{{Braunfurth} \& {Rohlfs}(1981)}]{braunfurth81}
{Braunfurth}, E., \& {Rohlfs}, K. 1981, \aaps, 44, 437

\bibitem[{{Burton} \& {Liszt}(1978)}]{burton78a}
{Burton}, W.~B., \& {Liszt}, H.~S. 1978, \apj, 225, 815

\bibitem[{Burton \& Liszt(1983)}]{burton83}
Burton, W.~B., \& Liszt, H.~S. 1983, \aaps, 52, 63

\bibitem[{Burton \& Liszt(1992)}]{burton92}
---. 1992, \aaps, 95, 9

\bibitem[{{Cabrera-Lavers} {et~al.}(2008){Cabrera-Lavers},
  {Gonz{\'a}lez-Fern{\'a}ndez}, {Garz{\'o}n}, {Hammersley}, \&
  {L{\'o}pez-Corredoira}}]{cabrera-lavers08}
{Cabrera-Lavers}, A., {Gonz{\'a}lez-Fern{\'a}ndez}, C., {Garz{\'o}n}, F.,
  {Hammersley}, P.~L., \& {L{\'o}pez-Corredoira}, M. 2008, \aap, 491, 781

\bibitem[{Cohen(1975)}]{cohen75}
Cohen, R. 1975, \mnras, 171, 659

\bibitem[{{Cornwell}(2008)}]{cornwell08}
{Cornwell}, T.~J. 2008, ISTSP, 2, 793

\bibitem[{Dame {et~al.}(2001)Dame, Hartmann, \& Thaddeus}]{dame01}
Dame, T.~M., Hartmann, D., \& Thaddeus, P. 2001, \apj, 547, 792

\bibitem[{Dame \& Thaddeus(2008)}]{dame08}
Dame, T.~M., \& Thaddeus, P. 2008, \apj, 683, L143

\bibitem[{{Dwek} {et~al.}(1995){Dwek}, {Arendt}, {Hauser}, {Kelsall}, {Lisse},
  {Moseley}, {Silverberg}, {Sodroski}, \& {Weiland}}]{dwek95}
{Dwek}, E., {Arendt}, R.~G., {Hauser}, M.~G., {et~al.} 1995, \apj, 445, 716

\bibitem[{{Ferri\`{e}re} {et~al.}(2007){Ferri\`{e}re}, {Gillard}, \&
  {Jean}}]{ferriere07}
{Ferri\`{e}re}, K.~M., {Gillard}, W., \& {Jean}, P. 2007, \aap, 467, 611

\bibitem[{Ford {et~al.}(2010)Ford, Lockman, \& McClure-Griffiths}]{ford10}
Ford, H.~A., Lockman, F.~J., \& McClure-Griffiths, N.~M. 2010, \apj, 722, 367

\bibitem[{Fujita {et~al.}(2009)Fujita, Martin, Low, New, \& Weaver}]{fujita09}
Fujita, A., Martin, C.~L., Low, M.-M.~M., New, K. C.~B., \& Weaver, R. 2009,
  \apj, 698, 693

\bibitem[{{Fukui} {et~al.}(2006){Fukui}, {Yamamoto}, {Fujishita}, {Kudo},
  {Torii}, {Nozawa}, {Takahashi}, {Matsumoto}, {Machida}, {Kawamura},
  {Yonekura}, {Mizuno}, {Onishi}, \& {Mizuno}}]{fukui06}
{Fukui}, Y., {Yamamoto}, H., {Fujishita}, M., {et~al.} 2006, Science, 314, 106

\bibitem[{{Fux}(1999)}]{fux99}
{Fux}, R. 1999, \aap, 345, 787

\bibitem[{{Genzel} {et~al.}(1994){Genzel}, {Hollenbach}, \&
  {Townes}}]{genzel94}
{Genzel}, R., {Hollenbach}, D., \& {Townes}, C.~H. 1994, RPPh, 57, 417

\bibitem[{Ghez {et~al.}(2008)Ghez, Salim, Weinberg, Lu, Do, Dunn, Matthews,
  Morris, Yelda, Becklin, Kremenek, Milosavljevic, \& Naiman}]{ghez08}
Ghez, A.~M., Salim, S., Weinberg, N.~N., {et~al.} 2008, \apj, 0808, 1044

\bibitem[{Green {et~al.}(2009)Green, McClure-Griffiths, Caswell, Ellingsen,
  Fuller, Quinn, \& Voronkov}]{green09b}
Green, J.~A., McClure-Griffiths, N.~M., Caswell, J.~L., {et~al.} 2009, \apjl,
  696, L156

\bibitem[{{Hammersley} {et~al.}(2000){Hammersley}, {Garz{\'o}n}, {Mahoney},
  {L{\'o}pez-Corredoira}, \& {Torres}}]{hammersley00}
{Hammersley}, P.~L., {Garz{\'o}n}, F., {Mahoney}, T.~J.,
  {L{\'o}pez-Corredoira}, M., \& {Torres}, M.~A.~P. 2000, \mnras, 317, L45

\bibitem[{{H{\"o}gbom}(1974)}]{hogbom74}
{H{\"o}gbom}, J.~A. 1974, \aaps, 15, 417

\bibitem[{Huettemeister {et~al.}(1998)Huettemeister, Dahmen, Mauersberger,
  Henkel, Wilson, \& Martin-Pintado}]{huettemeister98}
Huettemeister, S., Dahmen, G., Mauersberger, R., {et~al.} 1998, \aap, 334, 646

\bibitem[{{Kalberla} {et~al.}(2005){Kalberla}, {Burton}, {Hartmann}, {Arnal},
  {Bajaja}, {Morras}, \& {Poppel}}]{kalberla05}
{Kalberla}, P. M.~W., {Burton}, W.~B., {Hartmann}, D., {et~al.} 2005, \aap,
  440, 775

\bibitem[{Kalberla {et~al.}(2010)Kalberla, McClure-Griffiths, Pisano,
  Calabretta, Ford, Lockman, Staveley-Smith, Kerp, Winkel, Murphy, \&
  Newton-McGee}]{kalberla10}
Kalberla, P. M.~W., McClure-Griffiths, N.~M., Pisano, D.~J., {et~al.} 2010,
  \aap, 521, 17

\bibitem[{Kerr(1967)}]{kerr67}
Kerr, F.~J. 1967, in IAU Symposium 31: Radio Astronomy and the Galactic System,
  ed. H.~van Woerden, 239

\bibitem[{Lang {et~al.}(2010)Lang, Goss, Cyganowski, \& Clubb}]{lang10}
Lang, C.~C., Goss, W.~M., Cyganowski, C., \& Clubb, K.~I. 2010, \apjs, 191, 275

\bibitem[{Law(2010)}]{law10}
Law, C.~J. 2010, The Astrophysical Journal, 708, 474

\bibitem[{{Law} {et~al.}(2009){Law}, {Backer}, {Yusef-Zadeh}, \&
  {Maddalena}}]{law09}
{Law}, C.~J., {Backer}, D., {Yusef-Zadeh}, F., \& {Maddalena}, R. 2009, \apj,
  695, 1070

\bibitem[{{Lee}(1996)}]{lee96}
{Lee}, C.~W. 1996, \apjs, 105, 129

\bibitem[{Liszt(2006)}]{liszt06}
Liszt, H.~S. 2006, \aap, 447, 533

\bibitem[{Liszt(2008)}]{liszt08}
---. 2008, \aap, 486, 467

\bibitem[{Liszt \& Burton(1978)}]{liszt78}
Liszt, H.~S., \& Burton, W.~B. 1978, \apj, 226, 790

\bibitem[{Liszt \& Burton(1980)}]{liszt80}
---. 1980, \apj, 236, 779

\bibitem[{{Martin} {et~al.}(2004){Martin}, {Walsh}, {Xiao}, {Lane}, {Walker},
  \& {Stark}}]{martin04}
{Martin}, C.~L., {Walsh}, W.~M., {Xiao}, K., {et~al.} 2004, \apjs, 150, 239

\bibitem[{{McClure-Griffiths} {et~al.}(2003){McClure-Griffiths}, {Dickey},
  {Gaensler}, \& {Green}}]{mcgriff03}
{McClure-Griffiths}, N.~M., {Dickey}, J.~M., {Gaensler}, B.~M., \& {Green},
  A.~J. 2003, \apj, 594, 833

\bibitem[{{McClure-Griffiths} {et~al.}(2005){McClure-Griffiths}, {Dickey},
  {Gaensler}, {Green}, Haverkorn, \& Strasser}]{mcgriff05}
{McClure-Griffiths}, N.~M., {Dickey}, J.~M., {Gaensler}, B.~M., {et~al.} 2005,
  \apjs, 158, 178

\bibitem[{McClure-Griffiths {et~al.}(2009)McClure-Griffiths, Pisano,
  Calabretta, Ford, Lockman, Staveley-Smith, Kalberla, Bailin, Dedes,
  Janowiecki, Gibson, Murphy, Nakanishi, \& Newton-McGee}]{mcgriff09}
McClure-Griffiths, N.~M., Pisano, D.~J., Calabretta, M.~R., {et~al.} 2009,
  \apjs, 181, 398

\bibitem[{Molinari {et~al.}(2011)Molinari, Bally, Noriega-Crespo, \&
  et~al.}]{molinari11}
Molinari, S., Bally, J., Noriega-Crespo, A., \& et~al. 2011, \apjl, 735, L33

\bibitem[{Molinari {et~al.}(2010)Molinari, Swinyard, Bally, \&
  et~al.}]{molinari10}
Molinari, S., Swinyard, B., Bally, J., \& et~al. 2010, \aap, 518, L100

\bibitem[{Morris \& Serabyn(1996)}]{morris96}
Morris, M., \& Serabyn, E. 1996, \araa, 34, 645

\bibitem[{Oka {et~al.}(1998)Oka, Hasegawa, Sato, Tsuboi, \& Miyazaki}]{oka98}
Oka, T., Hasegawa, T., Sato, F., Tsuboi, M., \& Miyazaki, A. 1998, \apjs, 118,
  455

\bibitem[{Oort {et~al.}(1958)Oort, Kerr, \& Westerhout}]{oort58}
Oort, J.~H., Kerr, F.~J., \& Westerhout, G. 1958, \mnras, 118, 379

\bibitem[{{Peters}(1975)}]{peters75}
{Peters}, III, W.~L. 1975, \apj, 195, 617

\bibitem[{{Reid} {et~al.}(2009){Reid}, {Menten}, {Zheng}, {Brunthaler},
  {Moscadelli}, {Xu}, {Zhang}, {Sato}, {Honma}, {Hirota}, {Hachisuka}, {Choi},
  {Moellenbrock}, \& {Bartkiewicz}}]{reid09}
{Reid}, M.~J., {Menten}, K.~M., {Zheng}, X.~W., {et~al.} 2009, \apj, 700, 137

\bibitem[{Reynolds(1994)}]{reynolds94}
Reynolds, J.~E. 1994, ATNF Technical Document Series, Tech. Rep. AT/39.3/0400,
  Australia Telescope National Facility, Sydney

\bibitem[{Riffert {et~al.}(1997)Riffert, Kumar, \& Huchtmeier}]{riffert97}
Riffert, H., Kumar, P., \& Huchtmeier, W.~K. 1997, Monthly Notices of the Royal
  Astronomical Society, 284, 749

\bibitem[{Rodriguez-Fernandez {et~al.}(2006)Rodriguez-Fernandez, Combes,
  Martin-Pintado, Wilson, \& Apponi}]{rodriguez-fernandez06}
Rodriguez-Fernandez, N.~J., Combes, F., Martin-Pintado, J., Wilson, T.~L., \&
  Apponi, A. 2006, \aap, 455, 963

\bibitem[{{Romero-G{\'o}mez} {et~al.}(2011){Romero-G{\'o}mez}, {Athanassoula},
  {Antoja}, \& {Figueras}}]{romero-gomez11}
{Romero-G{\'o}mez}, M., {Athanassoula}, E., {Antoja}, T., \& {Figueras}, F.
  2011, \mnras, 418, 1176

\bibitem[{{Sanders} \& {Wrixon}(1972)}]{sanders72}
{Sanders}, R.~H., \& {Wrixon}, G.~T. 1972, \aap, 18, 467

\bibitem[{{Sault}(1994)}]{sault94}
{Sault}, R.~J. 1994, \aaps, 107, 55

\bibitem[{{Sault} {et~al.}(1995){Sault}, {Teuben}, \& {Wright}}]{sault95}
{Sault}, R.~J., {Teuben}, P.~J., \& {Wright}, M.~C.~H. 1995, in in ASP. Conf.
  Ser. 77, Astronomical Data Analysis Software and Systems IV, ed.
  {R.~A.~Shaw}, {H.~E.~Payne}, \& {J.~J.~E.~Hayes} (San Francisco, CA: ASP),
  433

\bibitem[{{Sawada} {et~al.}(2004){Sawada}, {Hasegawa}, {Handa}, \&
  {Cohen}}]{sawada04}
{Sawada}, T., {Hasegawa}, T., {Handa}, T., \& {Cohen}, R.~J. 2004, \mnras, 349,
  1167

\bibitem[{{Sofue} \& {Handa}(1984)}]{sofue84}
{Sofue}, Y., \& {Handa}, T. 1984, \nat, 310, 568

\bibitem[{{Stanimirovi\'c}(2002)}]{stanimirovic02}
{Stanimirovi\'c}, S. 2002, in ASP Conf. Ser. 278: Single-Dish Radio Astronomy:
  Techniques and Applications, ed. {S.~Stanimirovic}, {D.~Altschuler},
  {P.~Goldsmith}, \& {C.~Salter} (San Francisco, CA: ASP), 375

\bibitem[{Stark \& Bania(1986)}]{stark86}
Stark, A.~A., \& Bania, T.~M. 1986, \apj, 306, L17

\bibitem[{{Stil} {et~al.}(2006){Stil}, {Taylor}, {Dickey}, {Kavars}, {Martin},
  {Rothwell}, {Boothroyd}, {Lockman}, \& {McClure-Griffiths}}]{stil06}
{Stil}, J.~M., {Taylor}, A.~R., {Dickey}, J.~M., {et~al.} 2006, \aj, 132, 1158

\bibitem[{Tanaka {et~al.}(2007)Tanaka, Kamegai, Nagai, \& Oka}]{tanaka07}
Tanaka, K., Kamegai, K., Nagai, M., \& Oka, T. 2007, \pasj, 59, 323

\bibitem[{{Taylor} {et~al.}(2003){Taylor}, {Gibson}, {Peracaula}, {Martin},
  {Landecker}, {Brunt}, {Dewdney}, {Dougherty}, {Gray}, {Higgs}, {Kerton},
  {Knee}, {Kothes}, {Purton}, {Uyaniker}, {Wallace}, {Willis}, \&
  {Durand}}]{taylor03}
{Taylor}, A.~R., {Gibson}, S.~J., {Peracaula}, M., {et~al.} 2003, \aj, 125,
  3145

\bibitem[{van~der Kruit(1970)}]{vanderkruit70}
van~der Kruit, P.~C. 1970, \aap, 4, 462

\bibitem[{van~der Kruit(1971)}]{vanderkruit71}
---. 1971, \aap, 13, 405

\bibitem[{{van Woerden} {et~al.}(1957){van Woerden}, {Rougoor}, \&
  {Oort}}]{vanwoerden57}
{van Woerden}, H., {Rougoor}, G.~W., \& {Oort}, J.~H. 1957, CRAS, 244, 1961

\bibitem[{Wilson {et~al.}(2011)Wilson, Ferris, Axtens, \& et~al.}]{wilson11}
Wilson, W.~E., Ferris, R., Axtens, P., \& et~al. 2011, \mnras, 416, 832

\end{thebibliography}

\normalsize


\clearpage

\clearpage

\begin{figure}
\figurenum{4b}
\centering
\plottwo{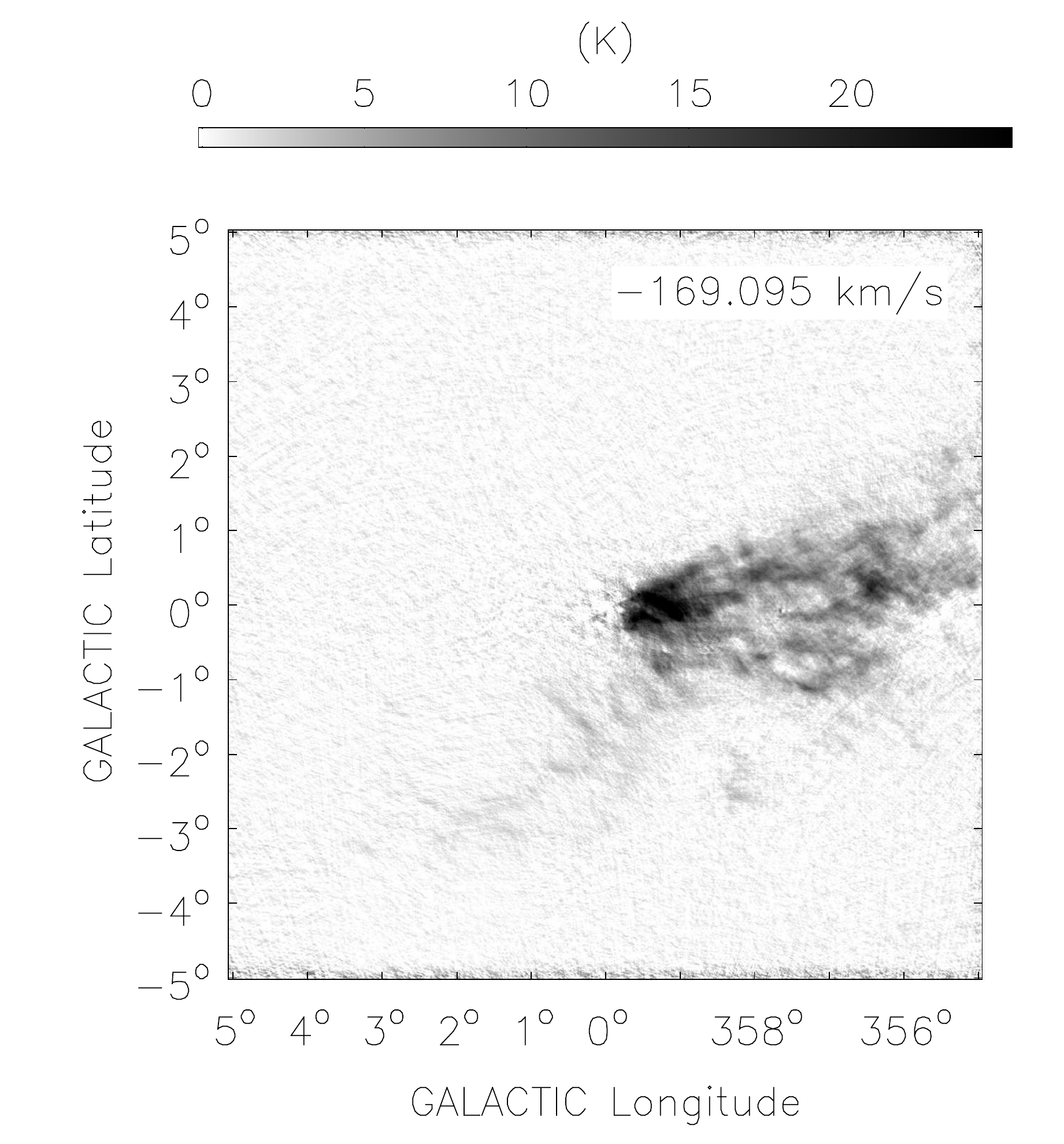}{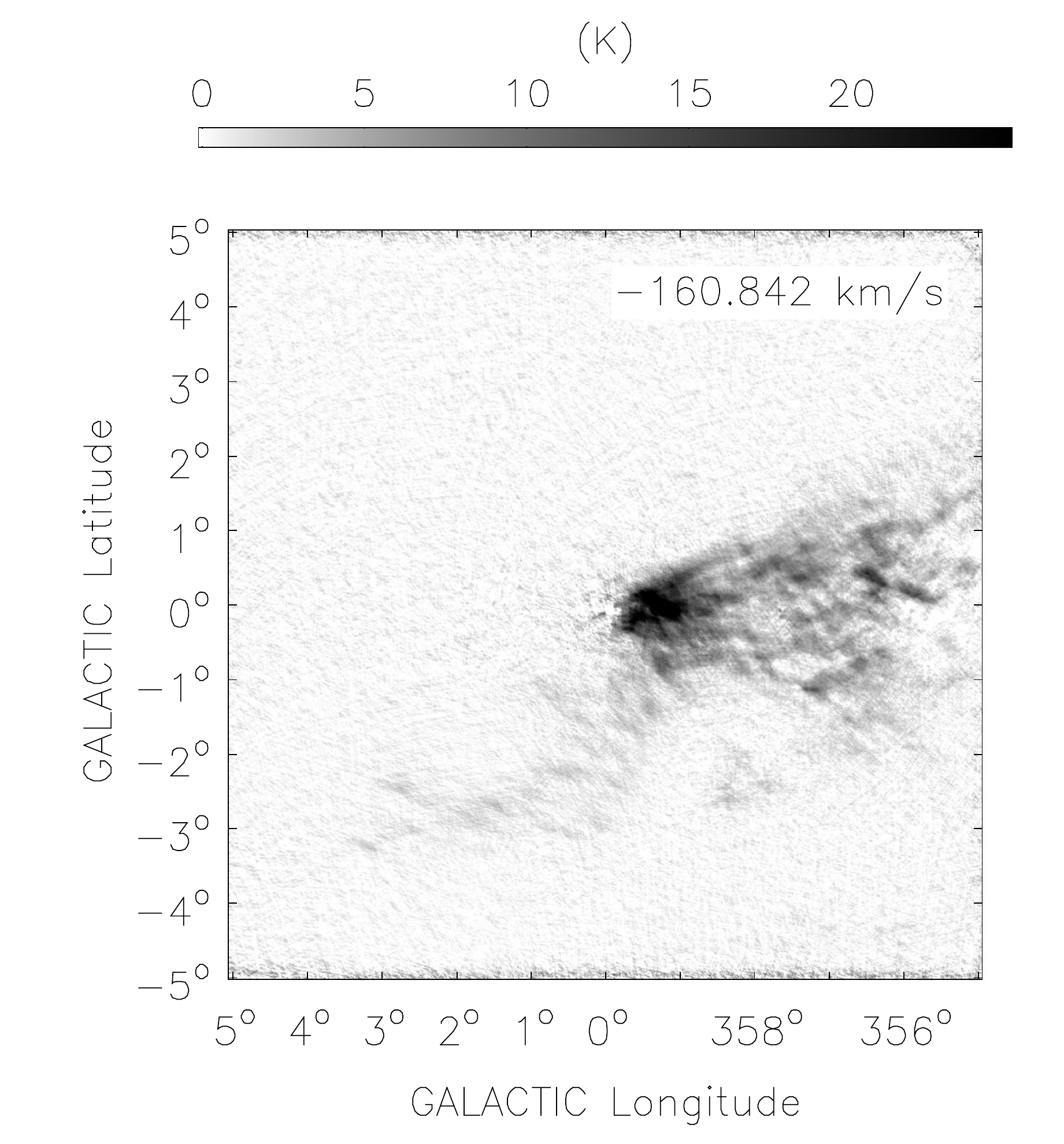}\\
\plottwo{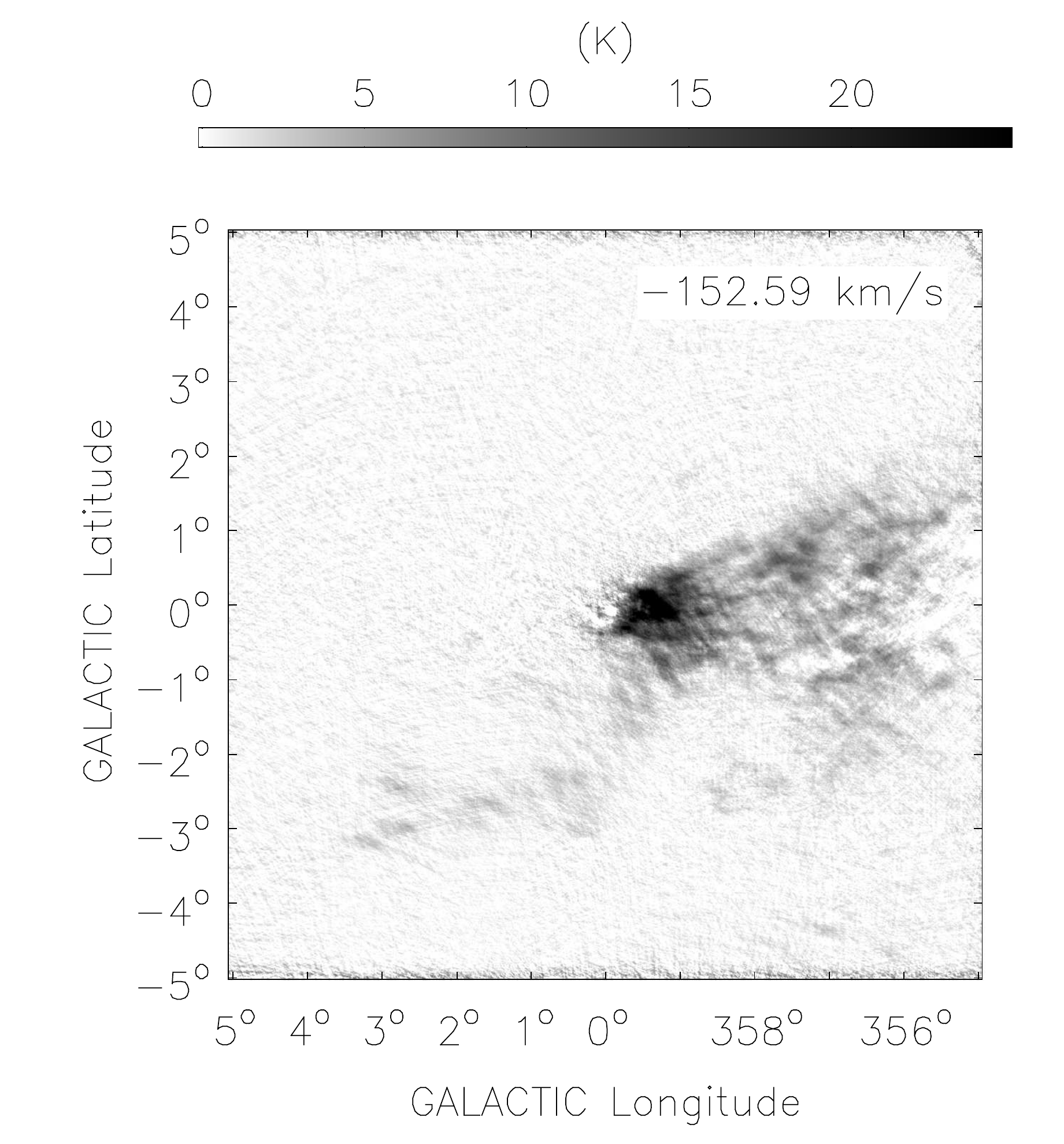}{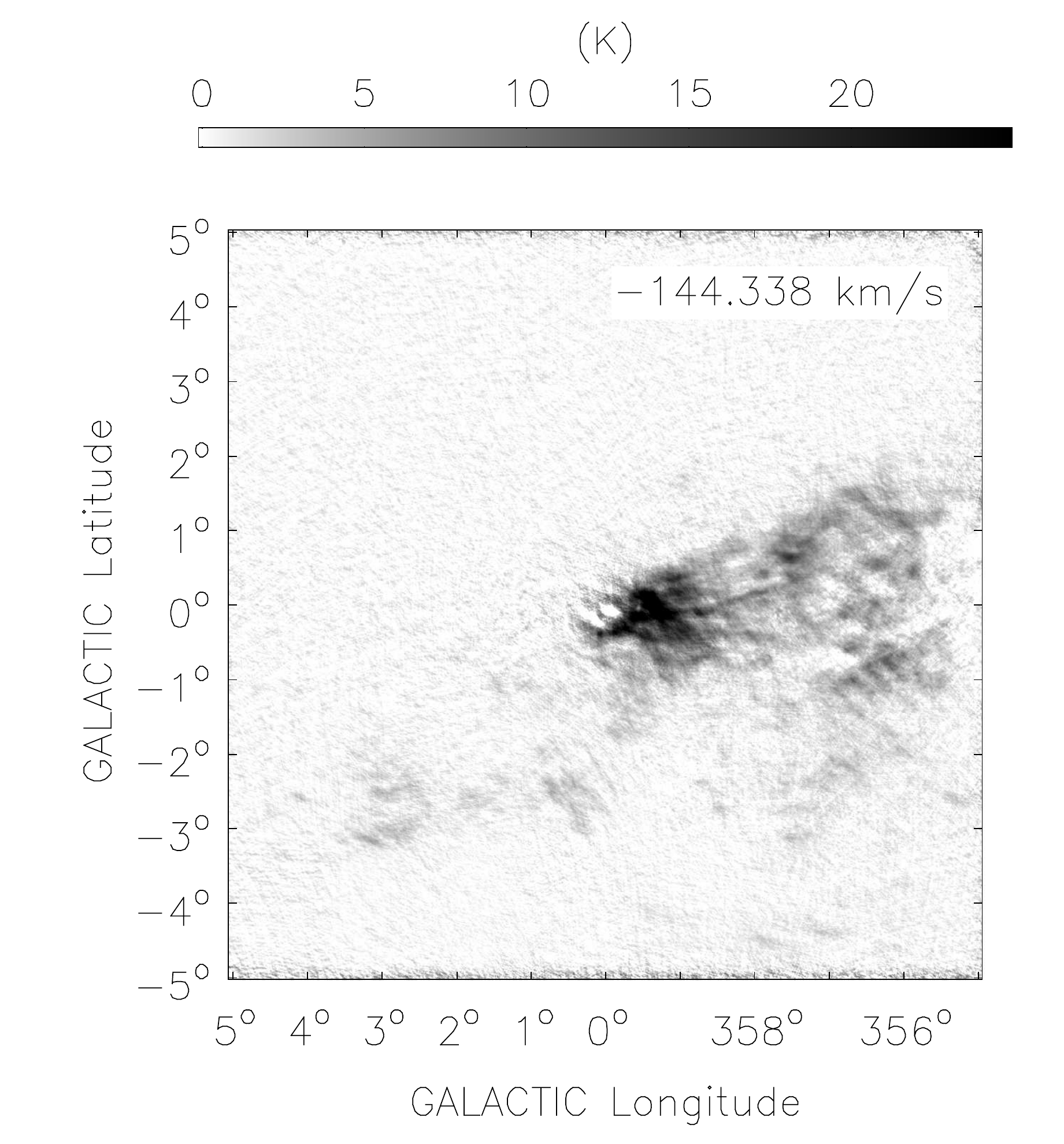}\\
\plottwo{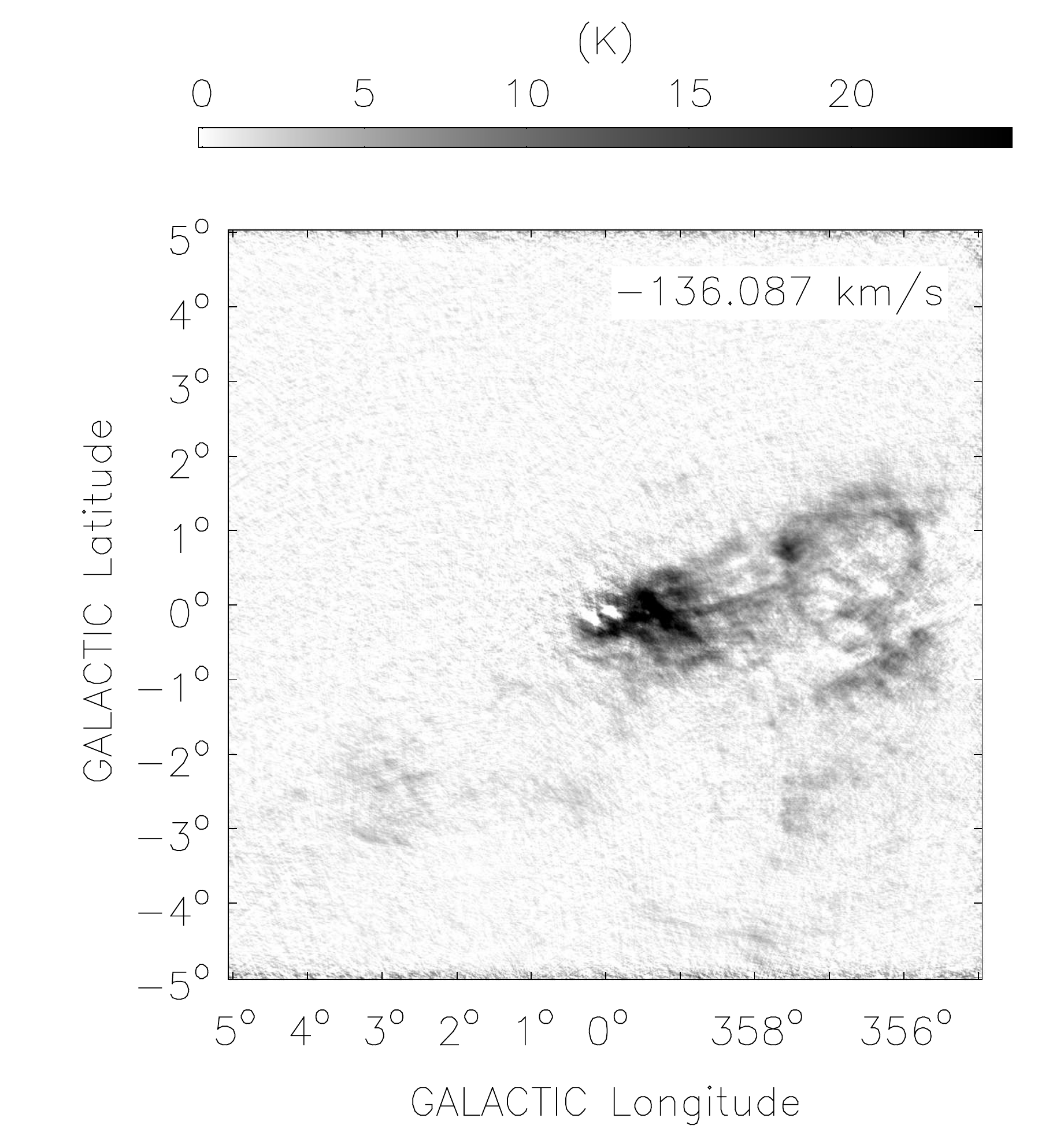}{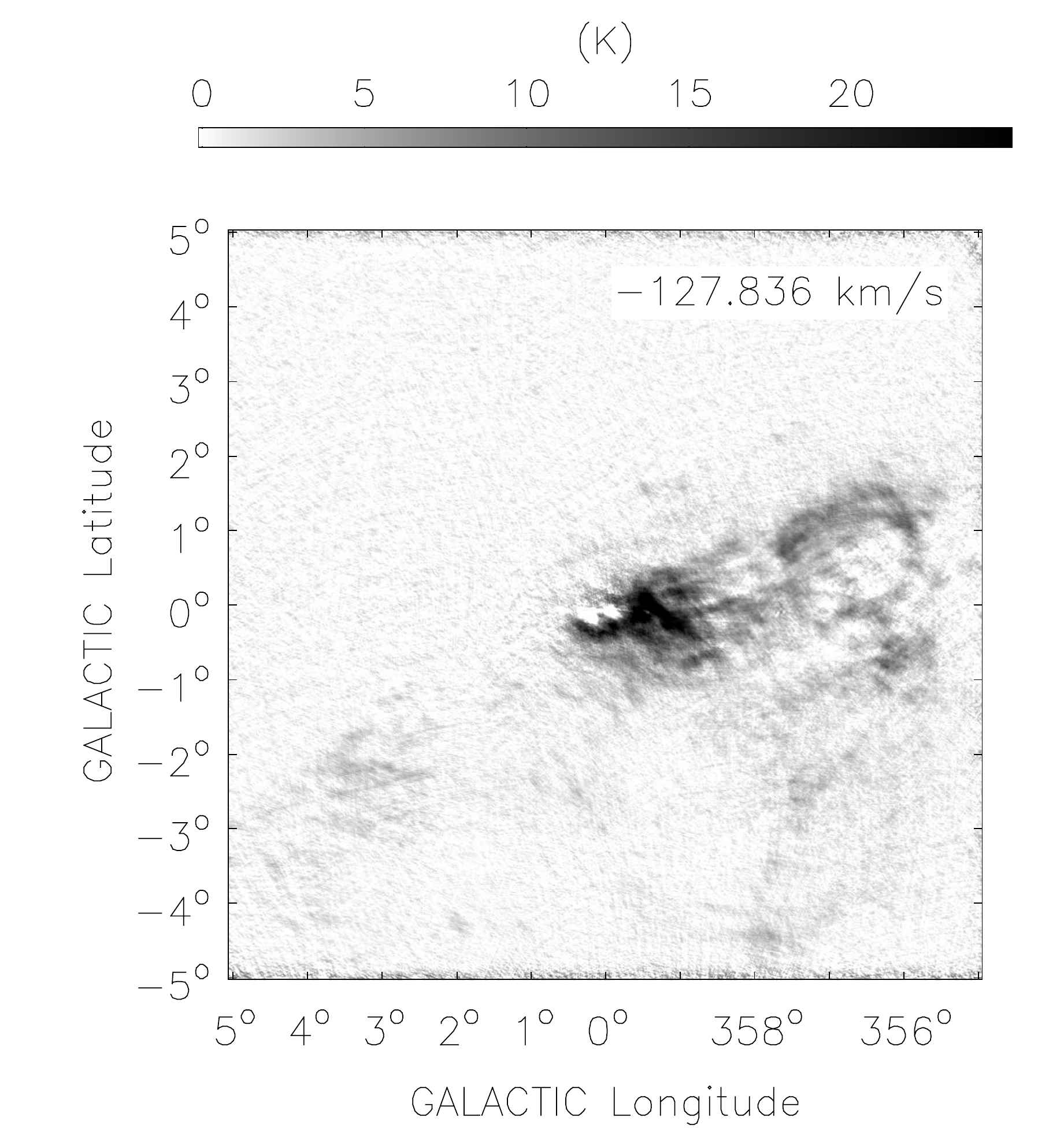}
\caption[]{}
\end{figure}
\clearpage
\begin{figure}
\figurenum{4c}
\centering
\plottwo{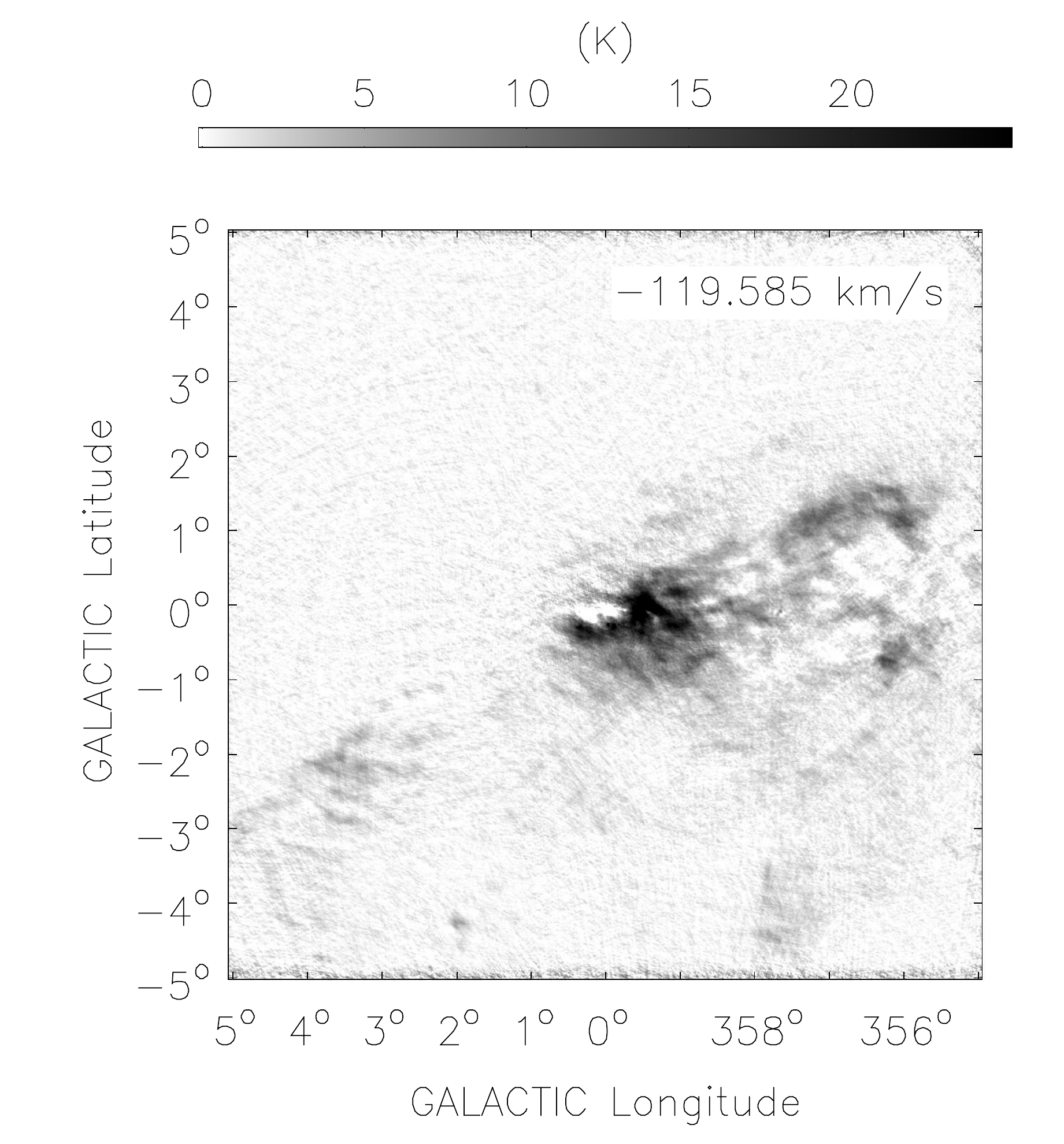}{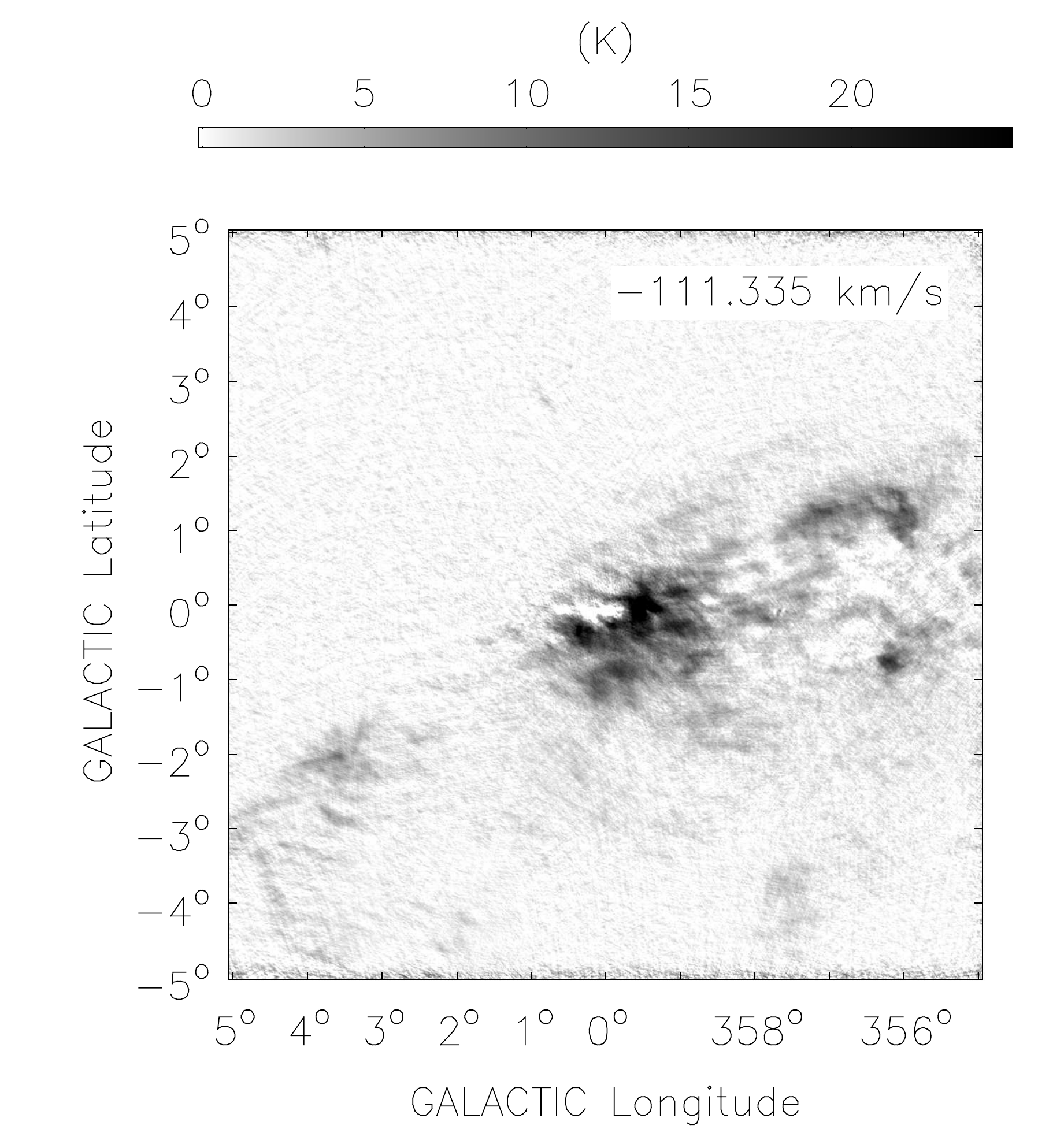}
\plottwo{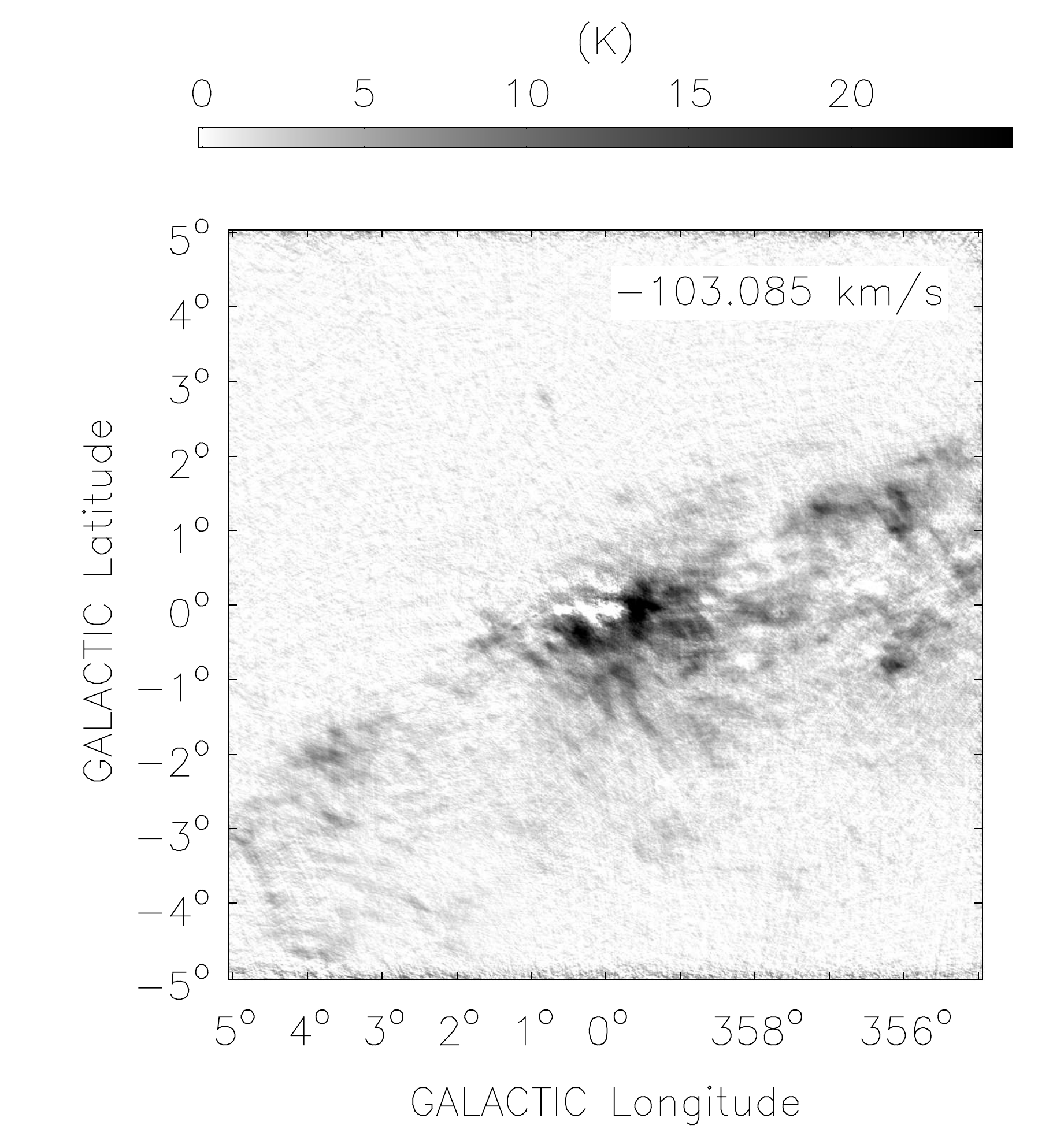}{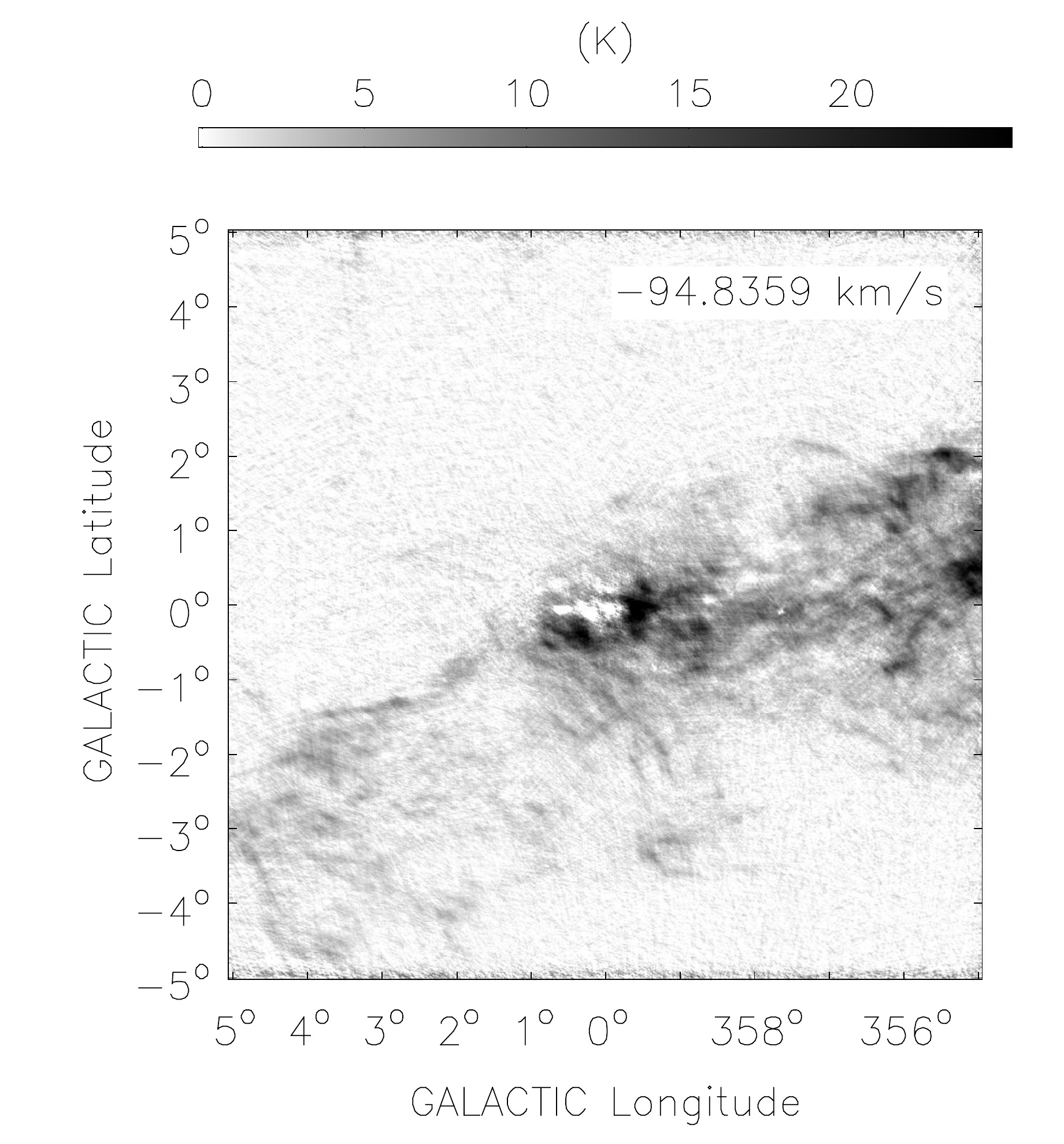}
\plottwo{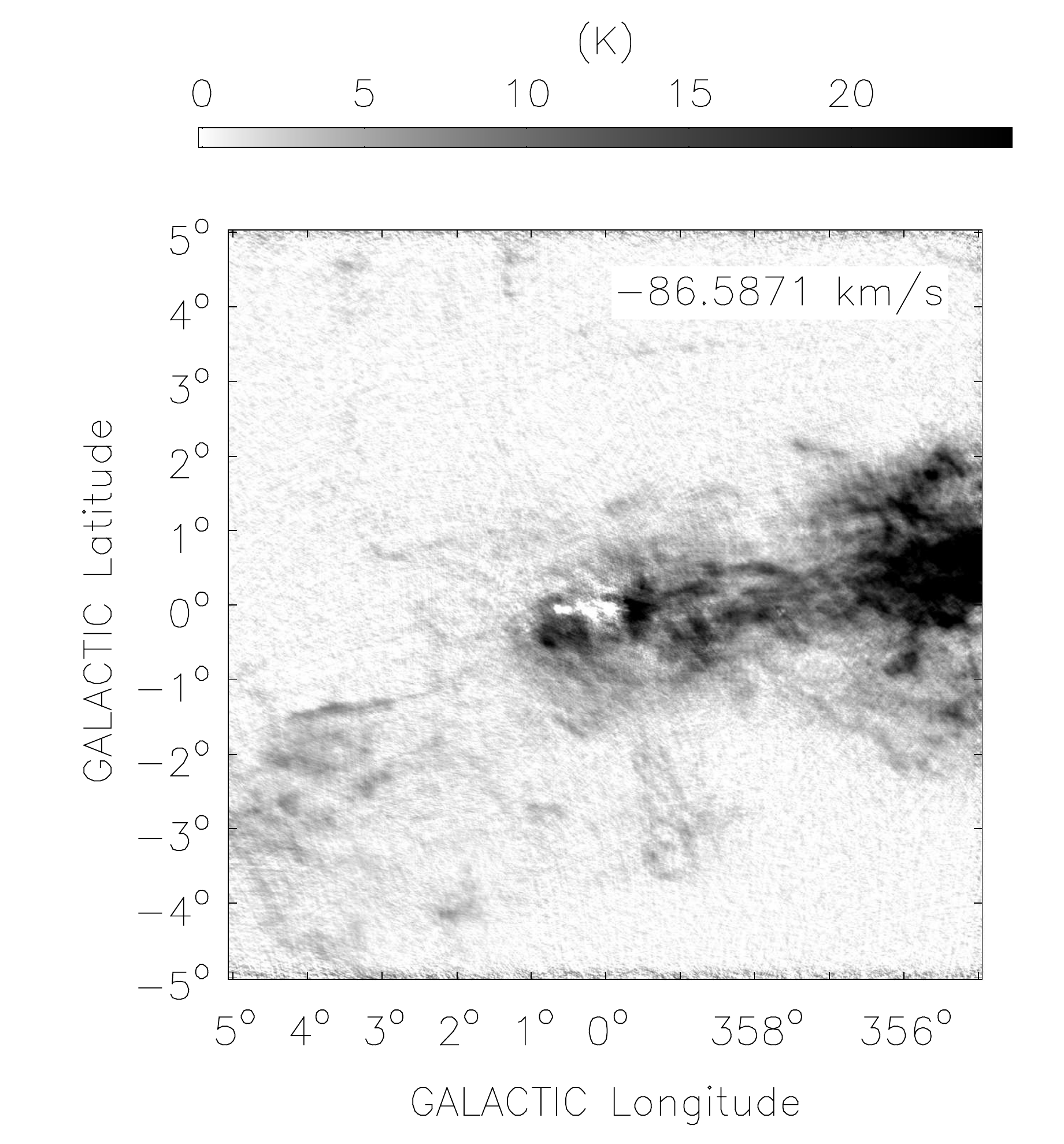}{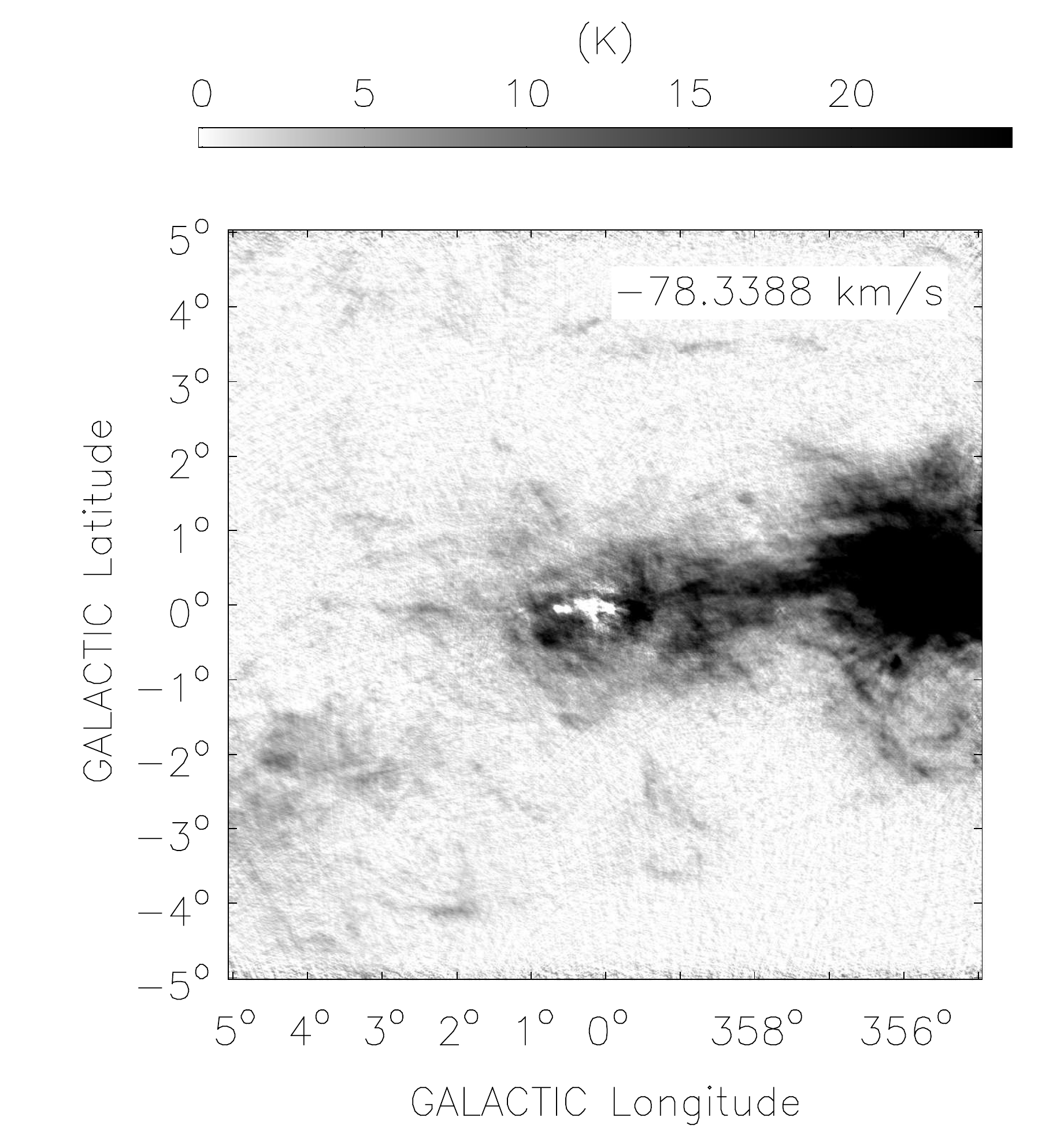}
\caption[]{}
\end{figure}
\clearpage
\begin{figure}
\figurenum{4d}
\centering
\plottwo{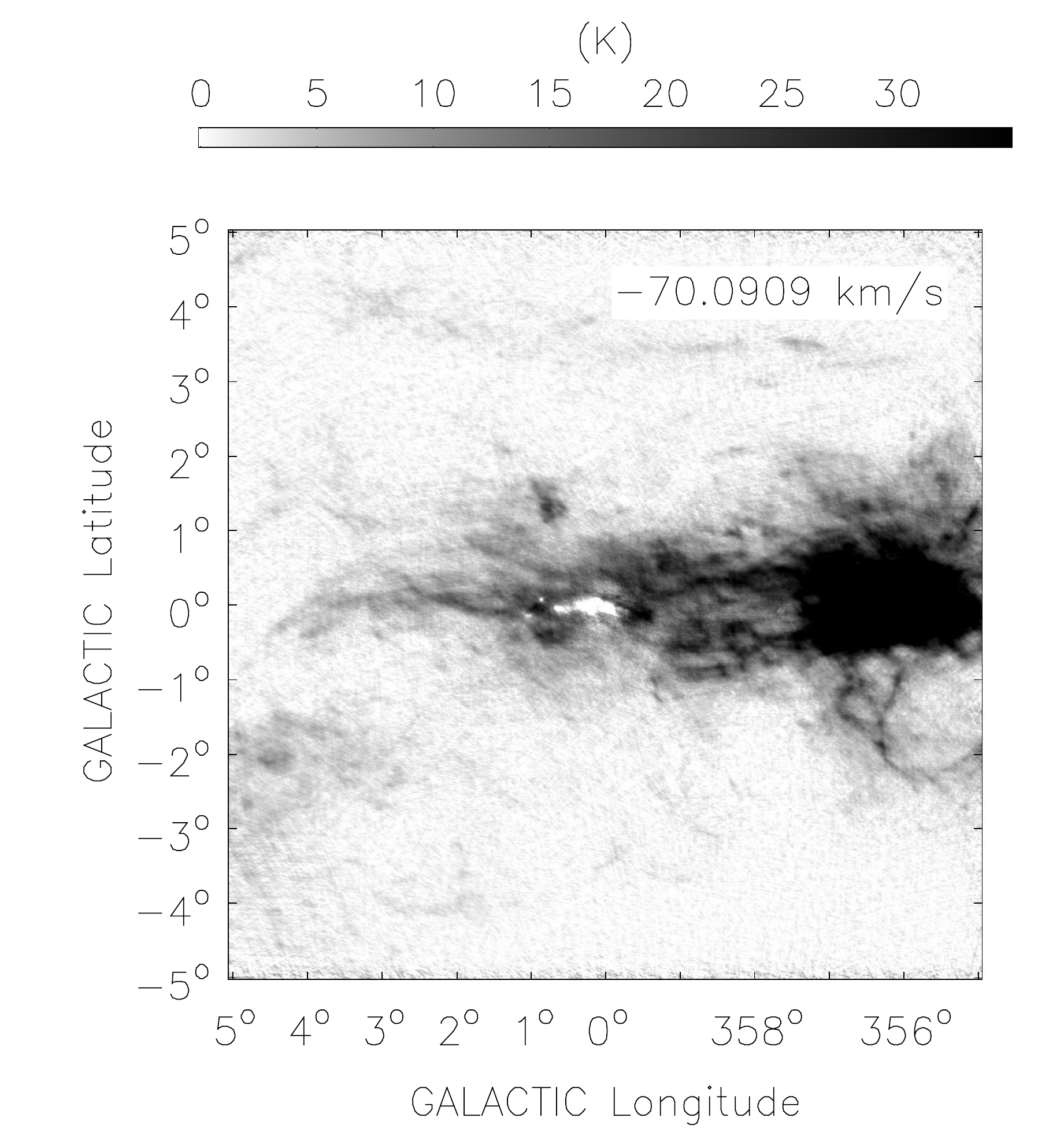}{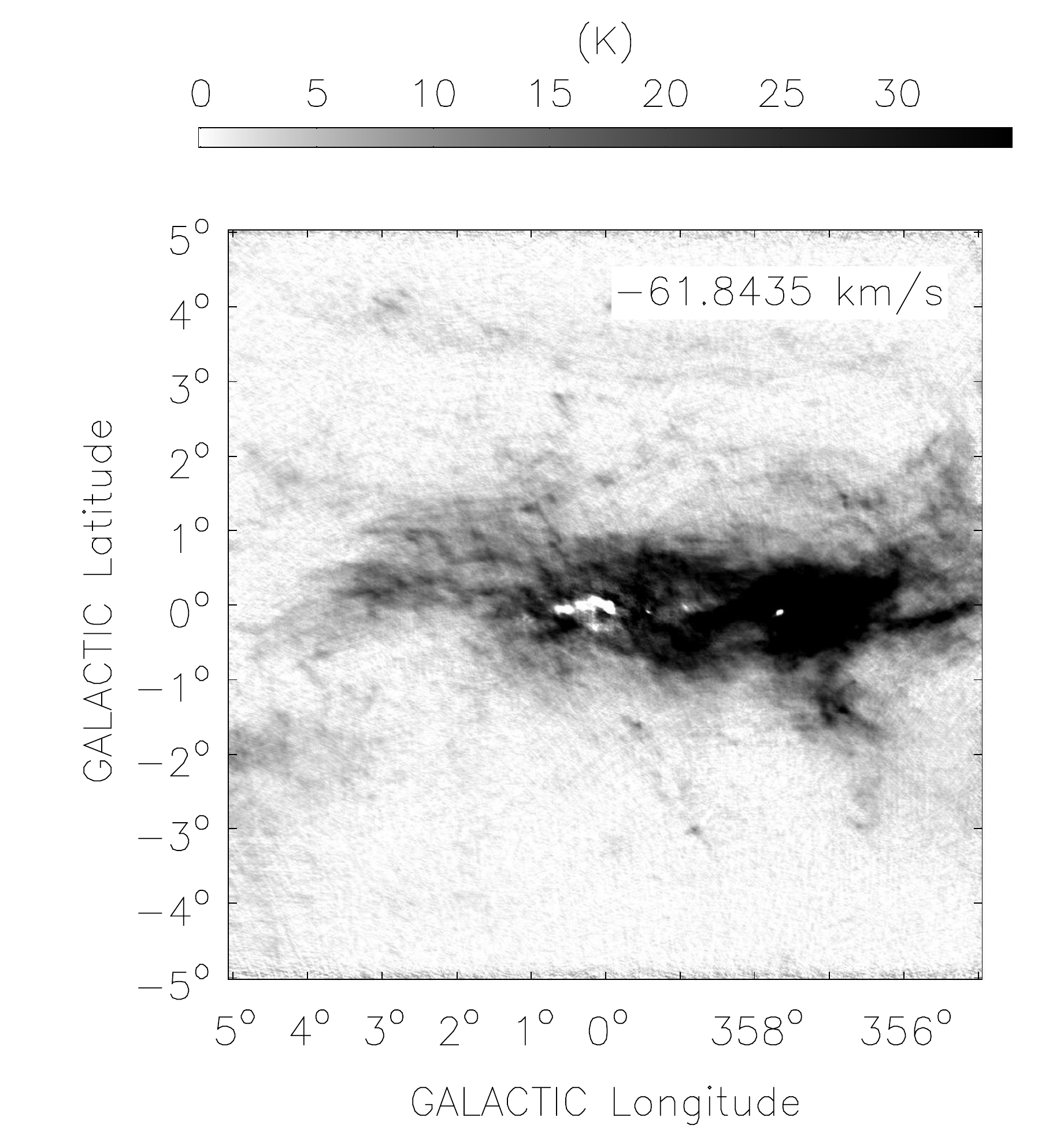}
\plottwo{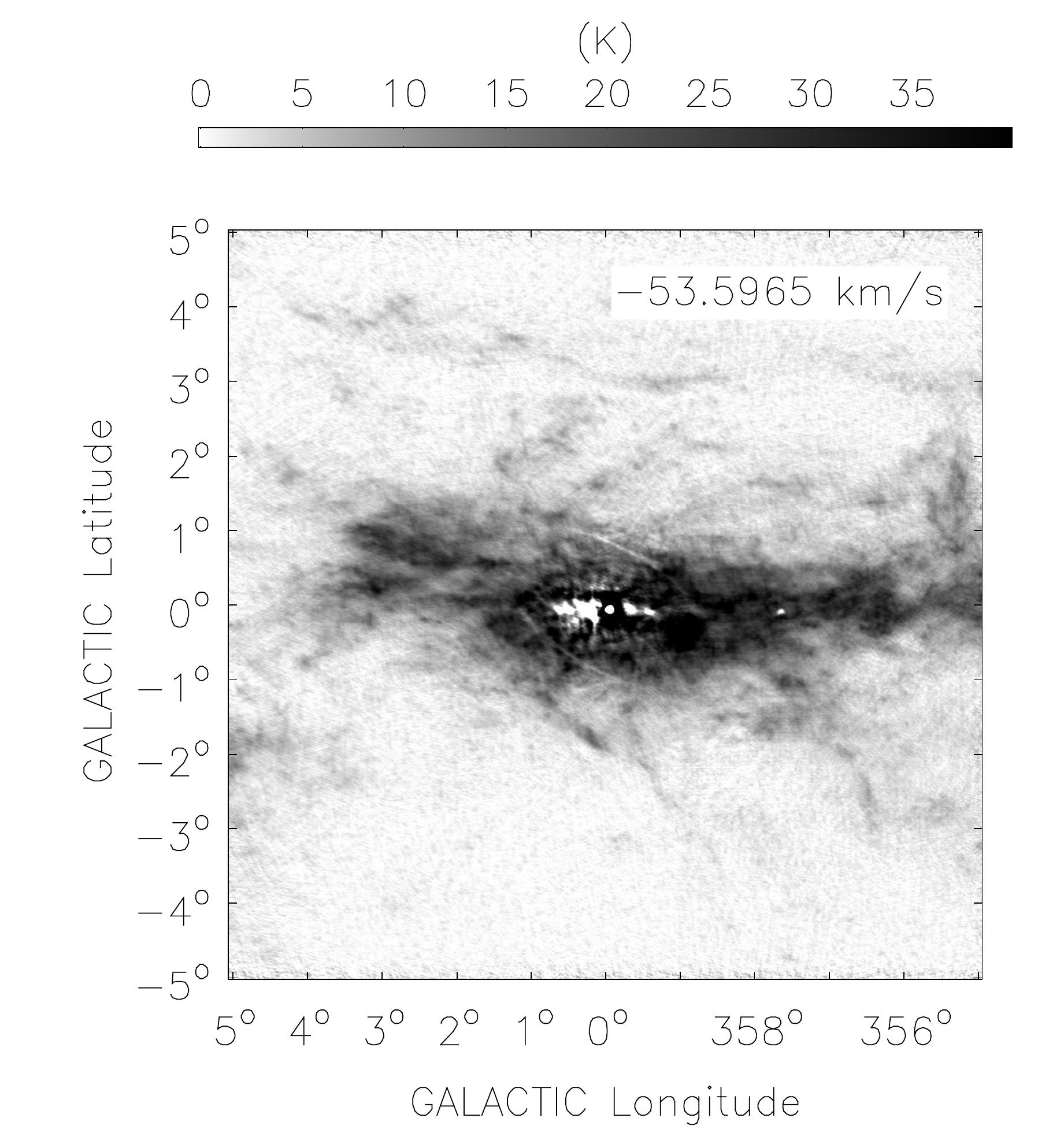}{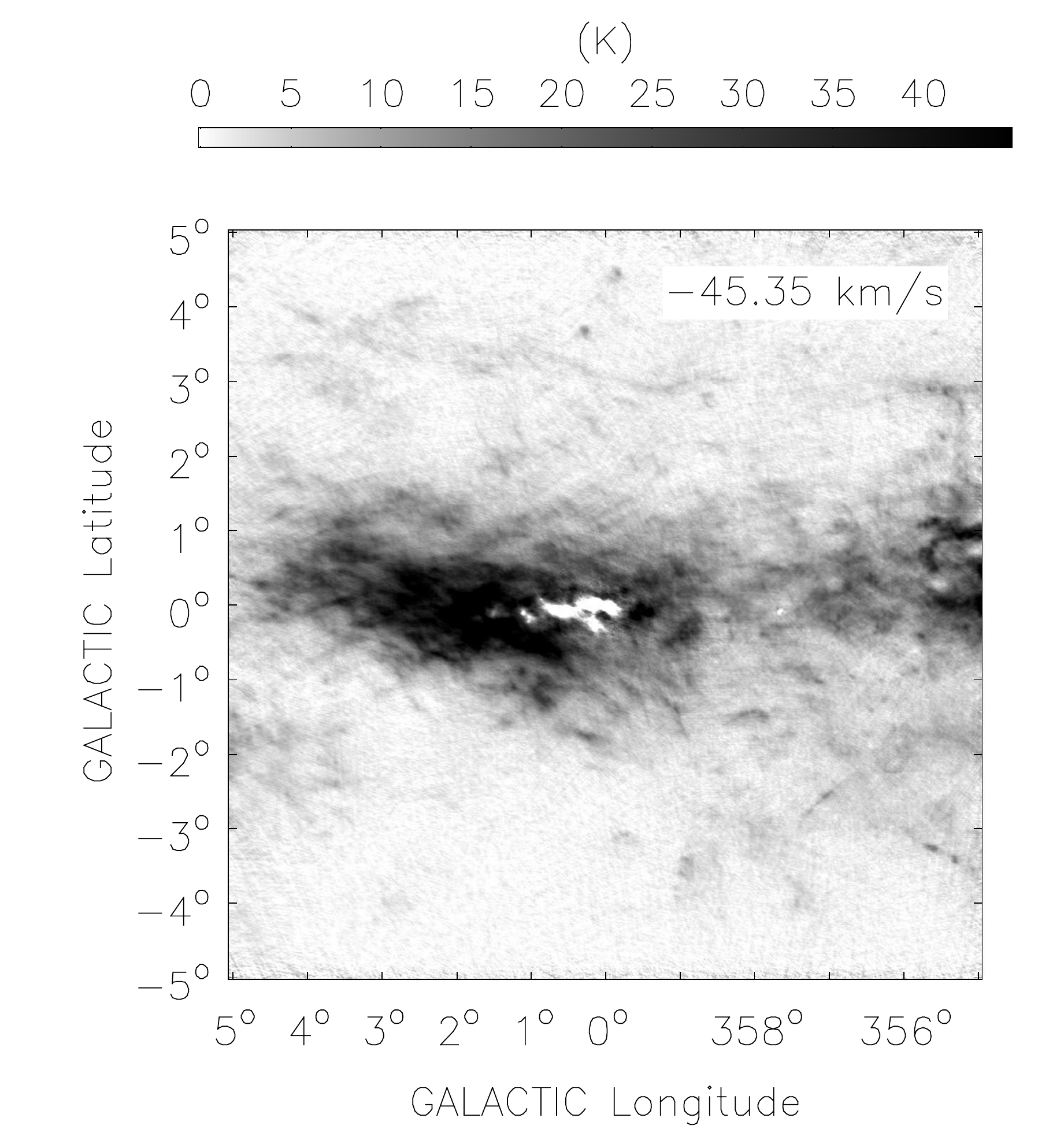}
\plottwo{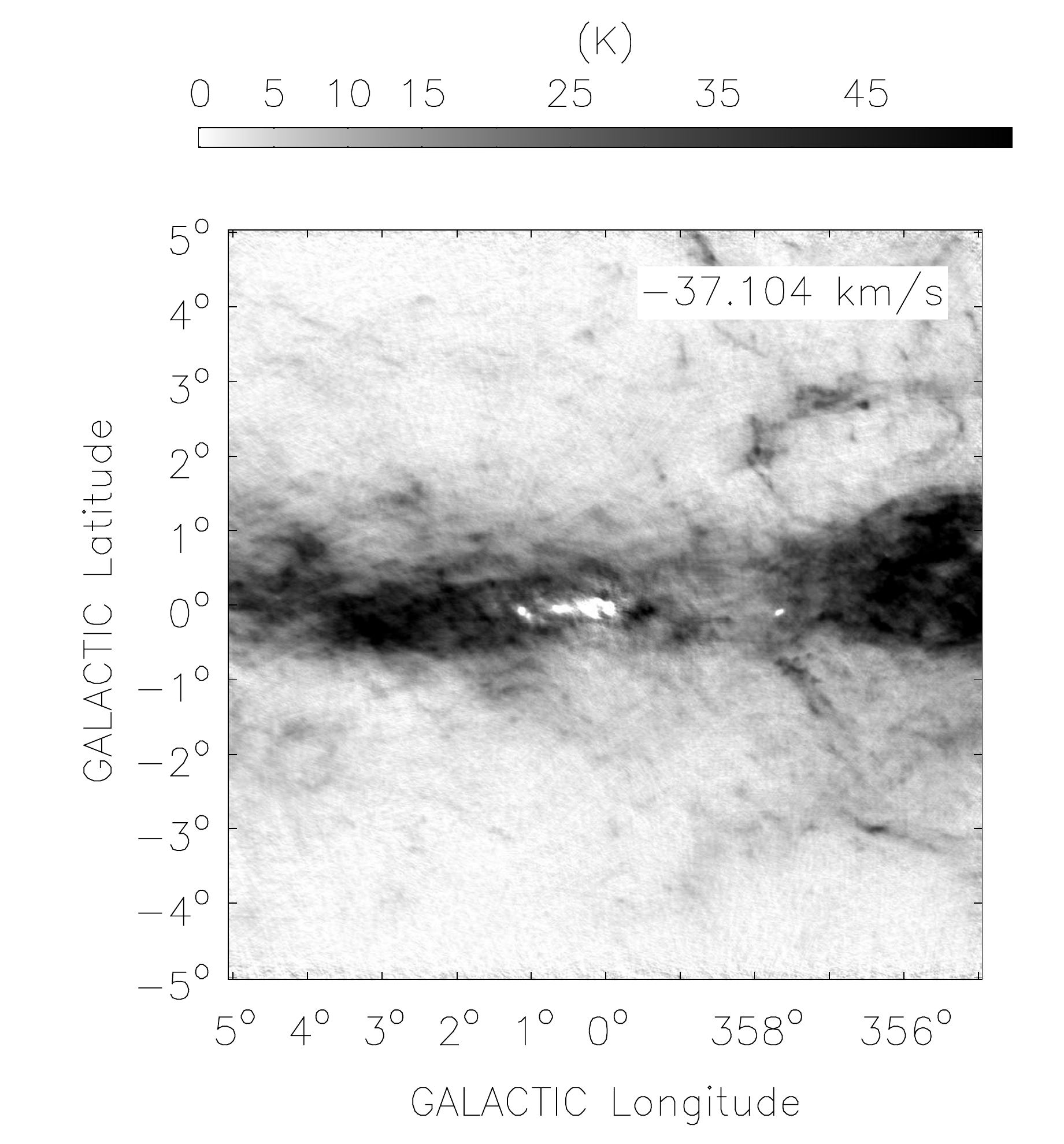}{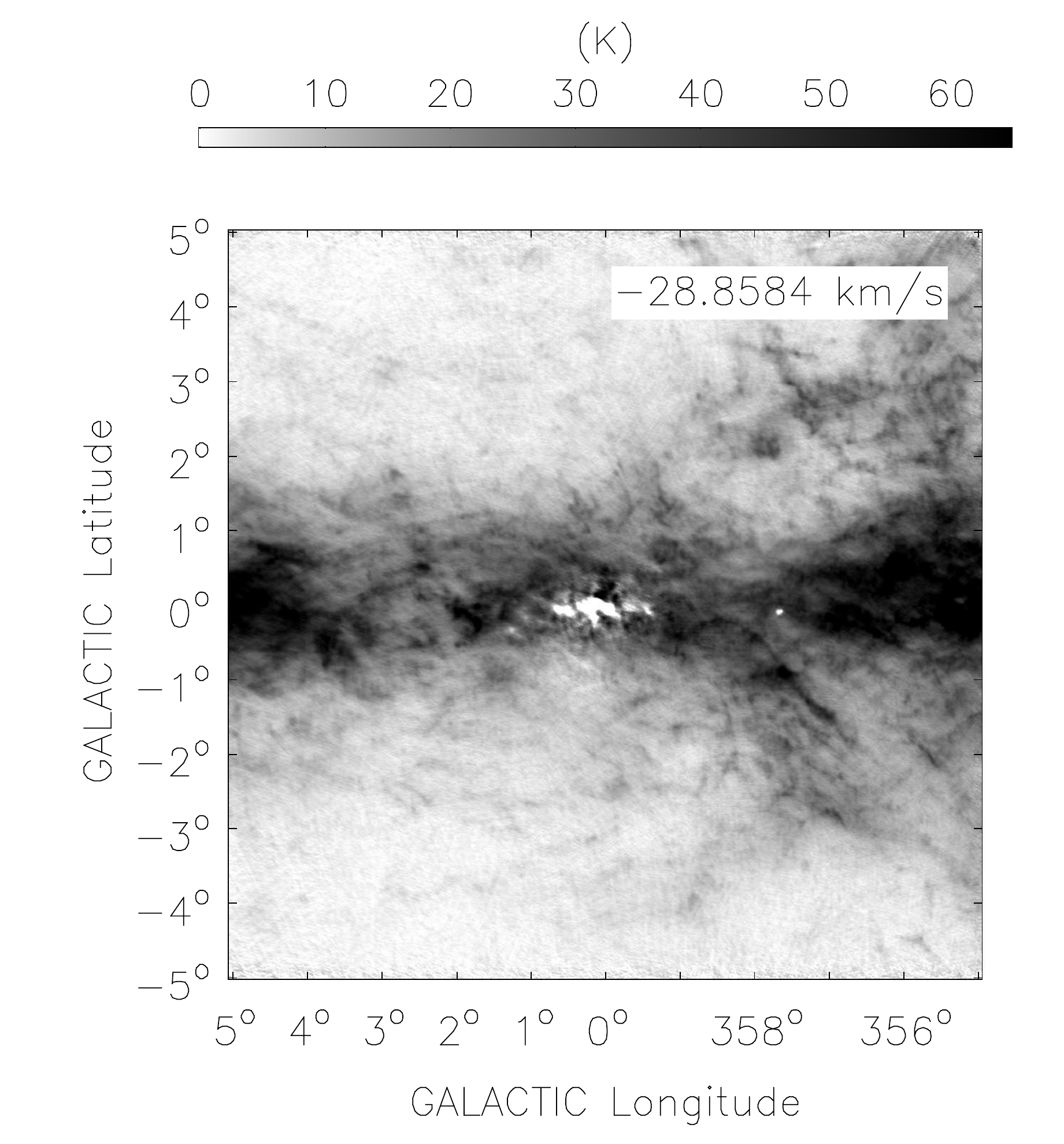}
\caption[]{}
\end{figure}
\clearpage
\begin{figure}
\figurenum{4e}
\centering
\plottwo{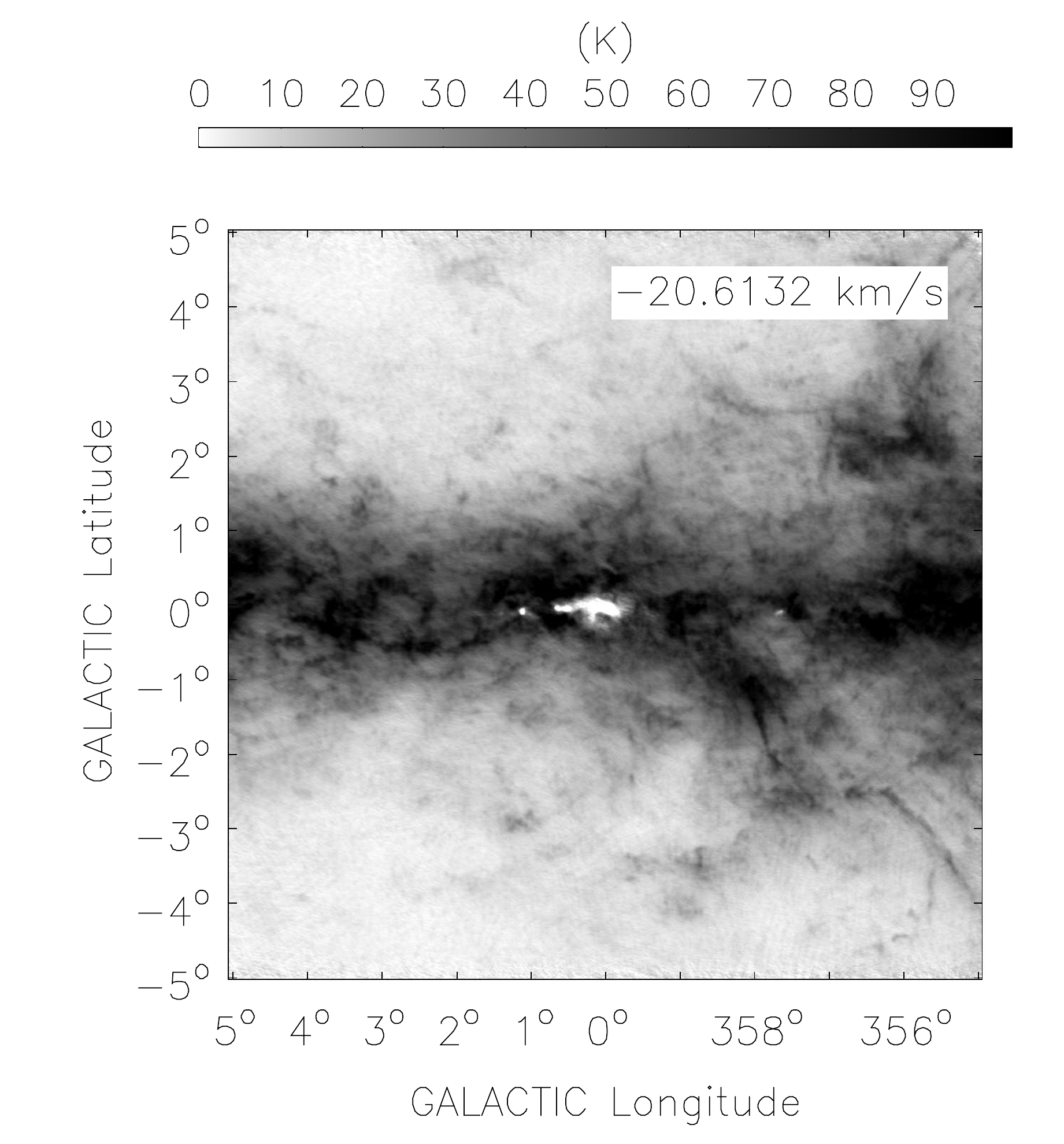}{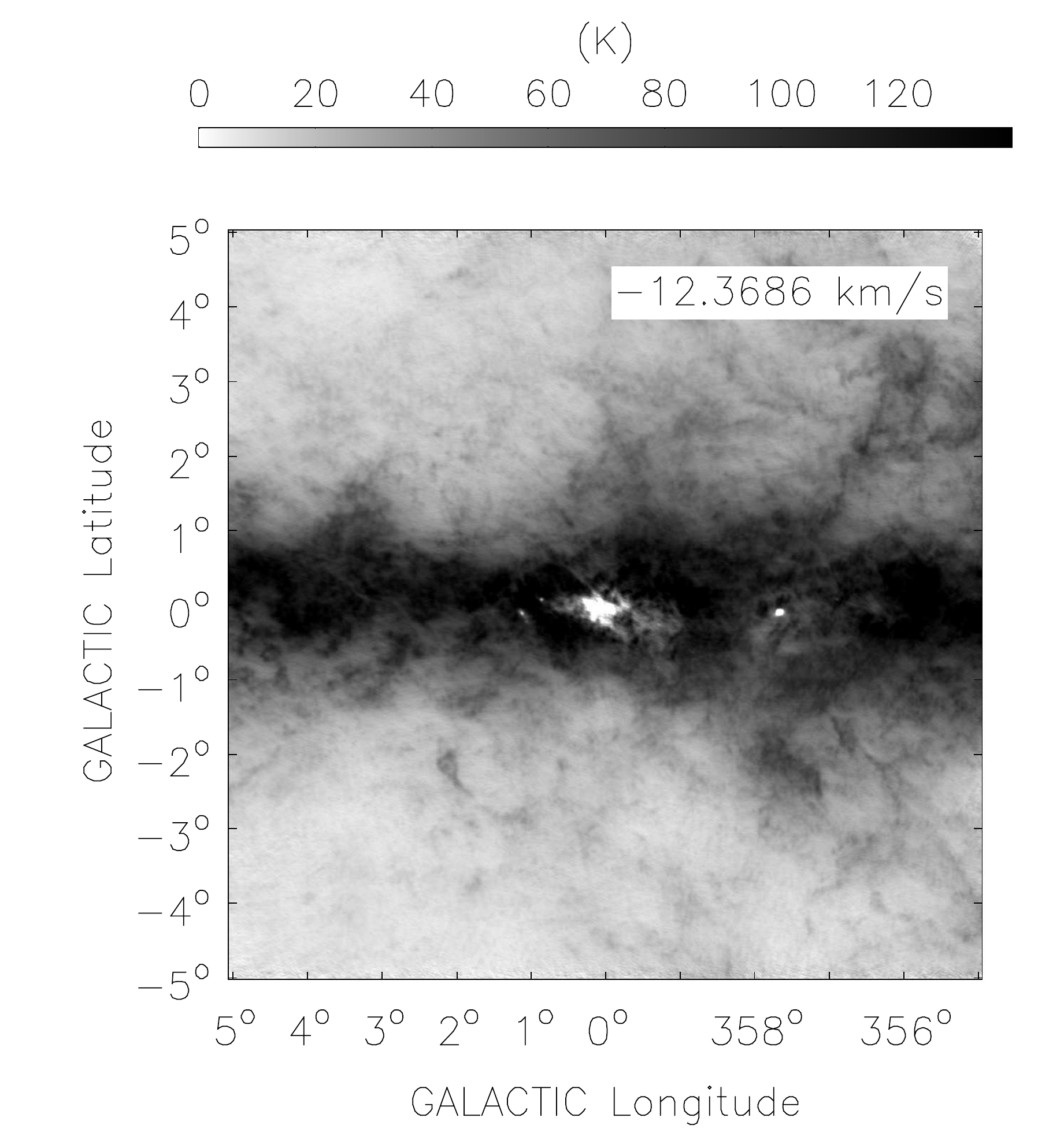}
\plottwo{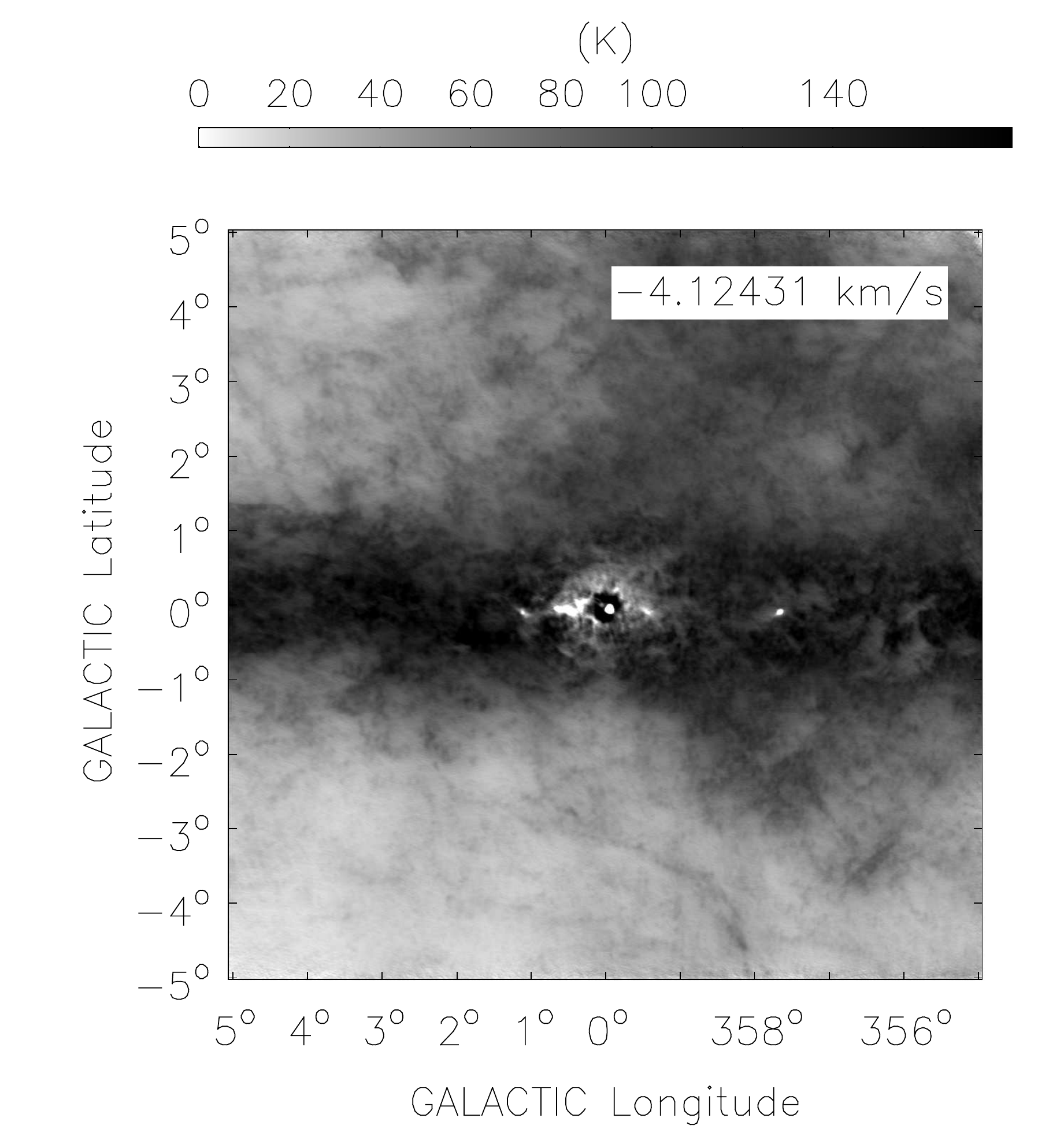}{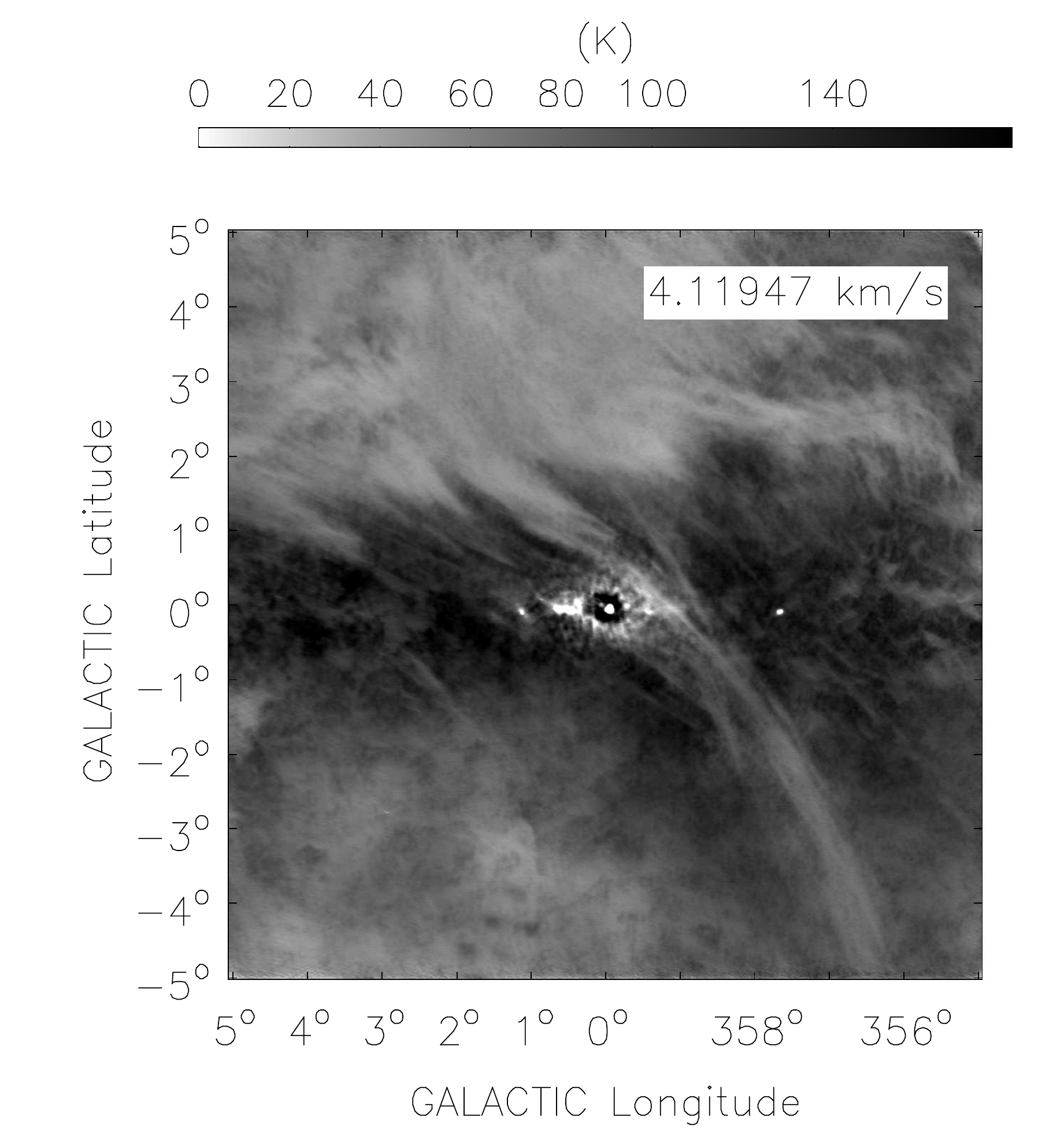}
\plottwo{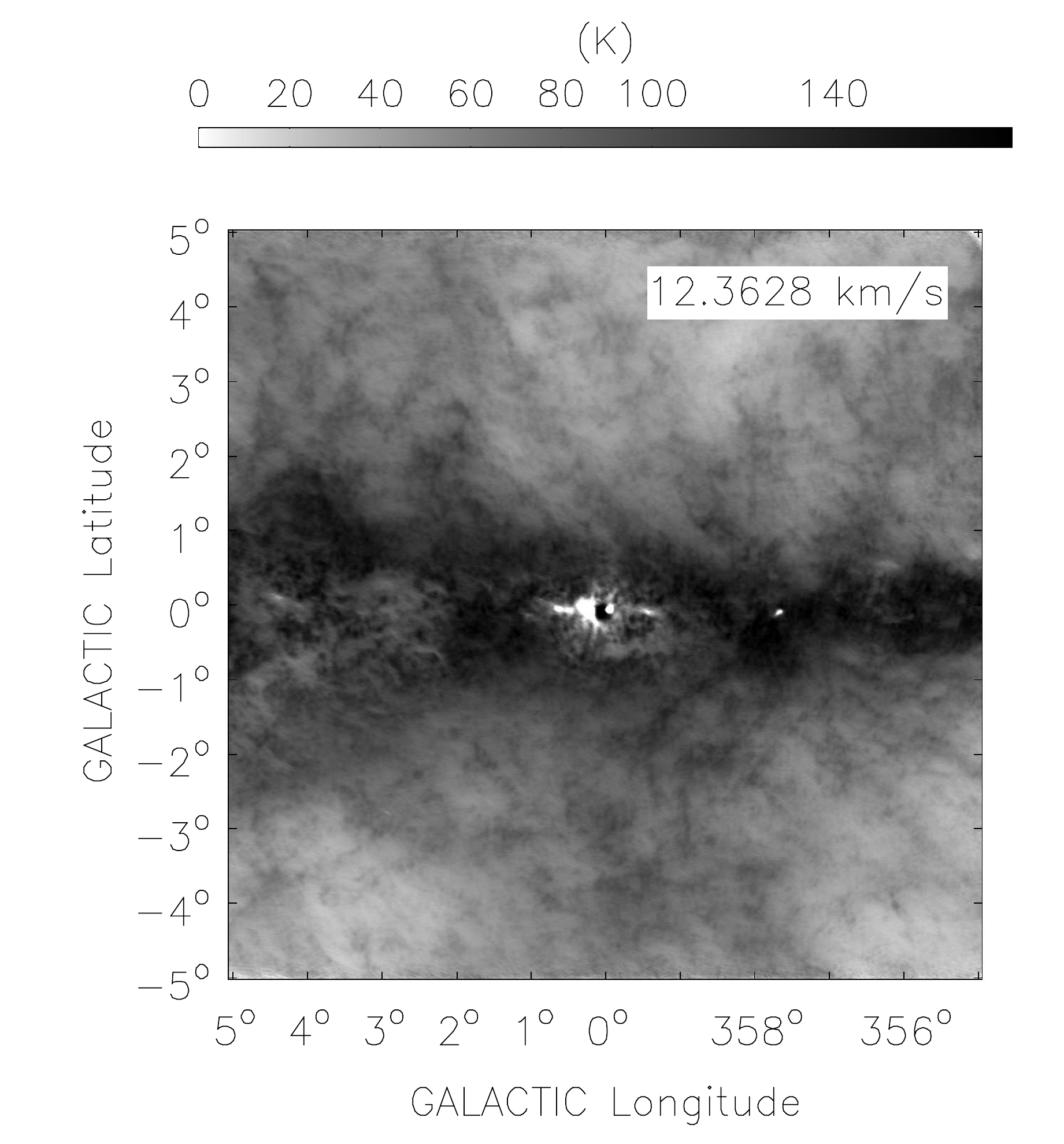}{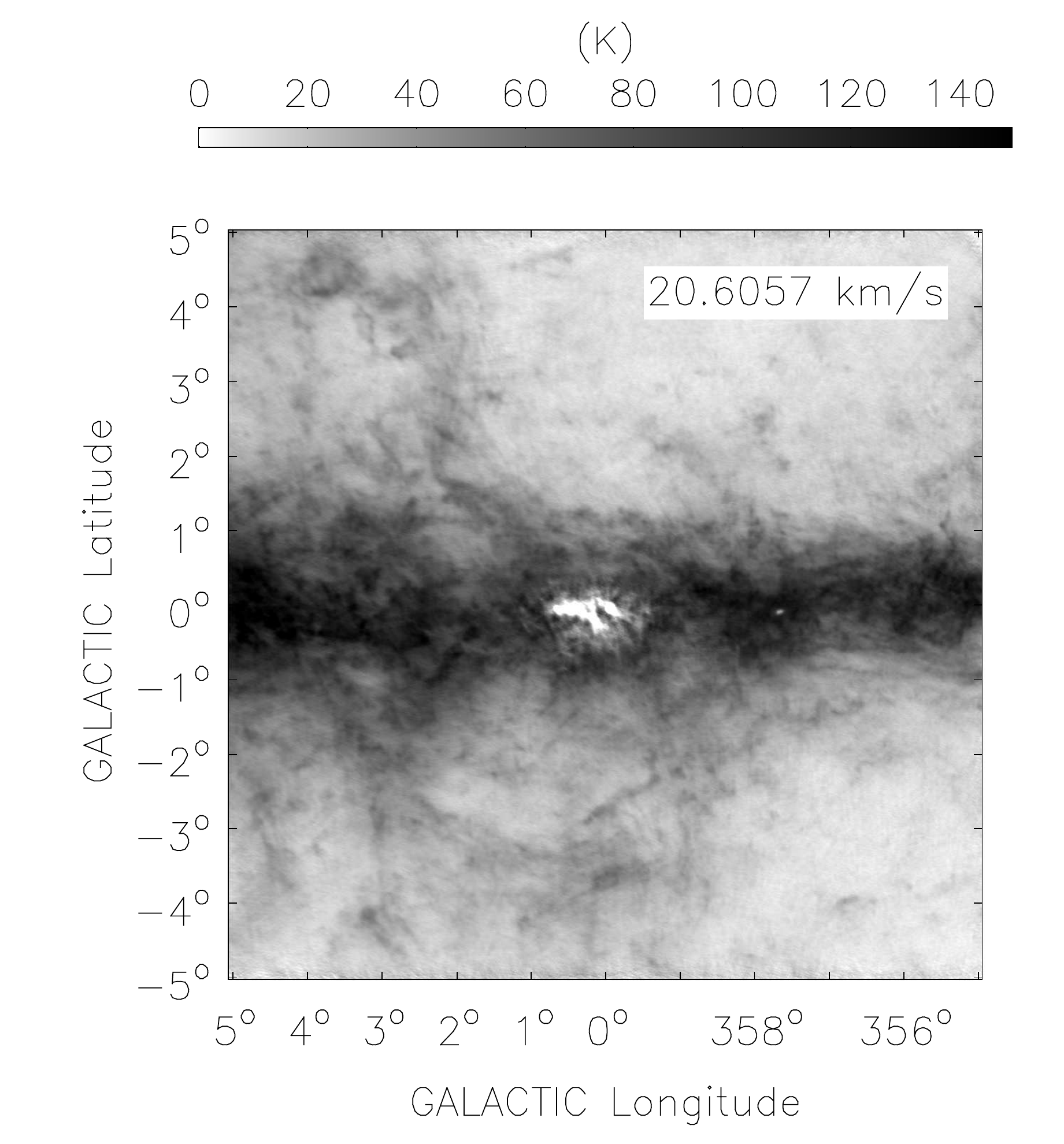}
\caption[]{}
\end{figure}
\begin{figure}
\figurenum{4f}
\centering
\plottwo{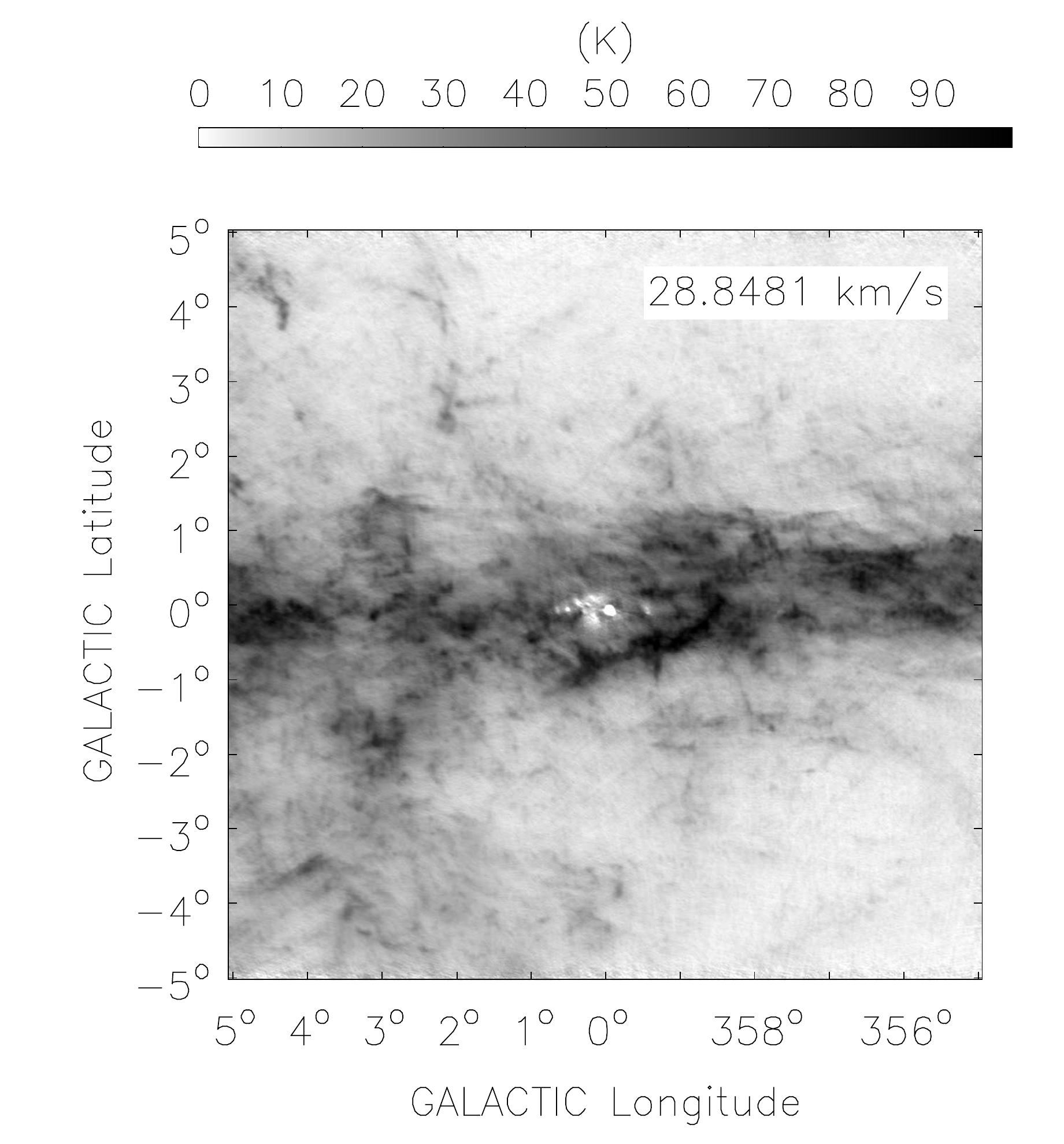}{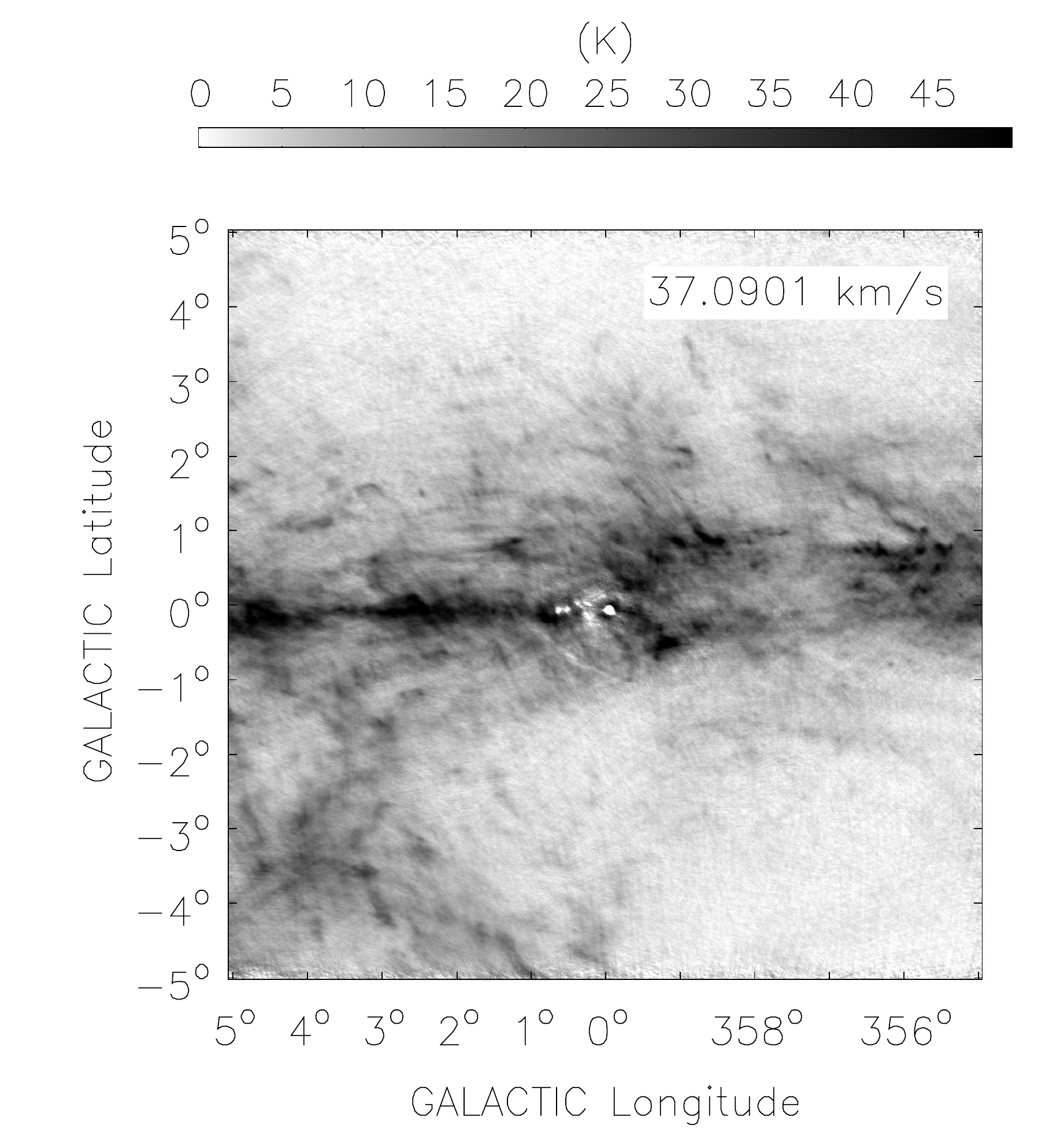}
\plottwo{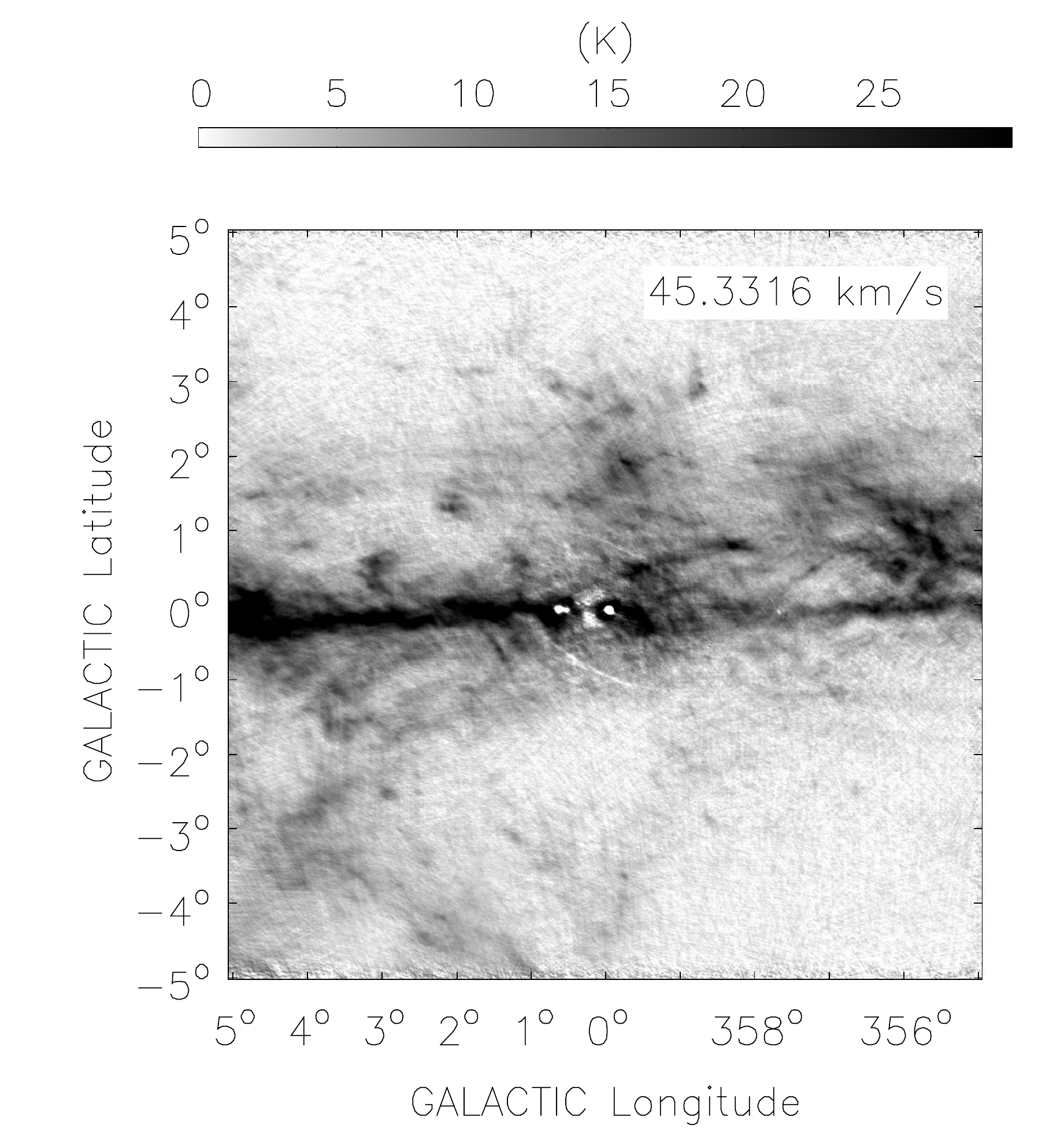}{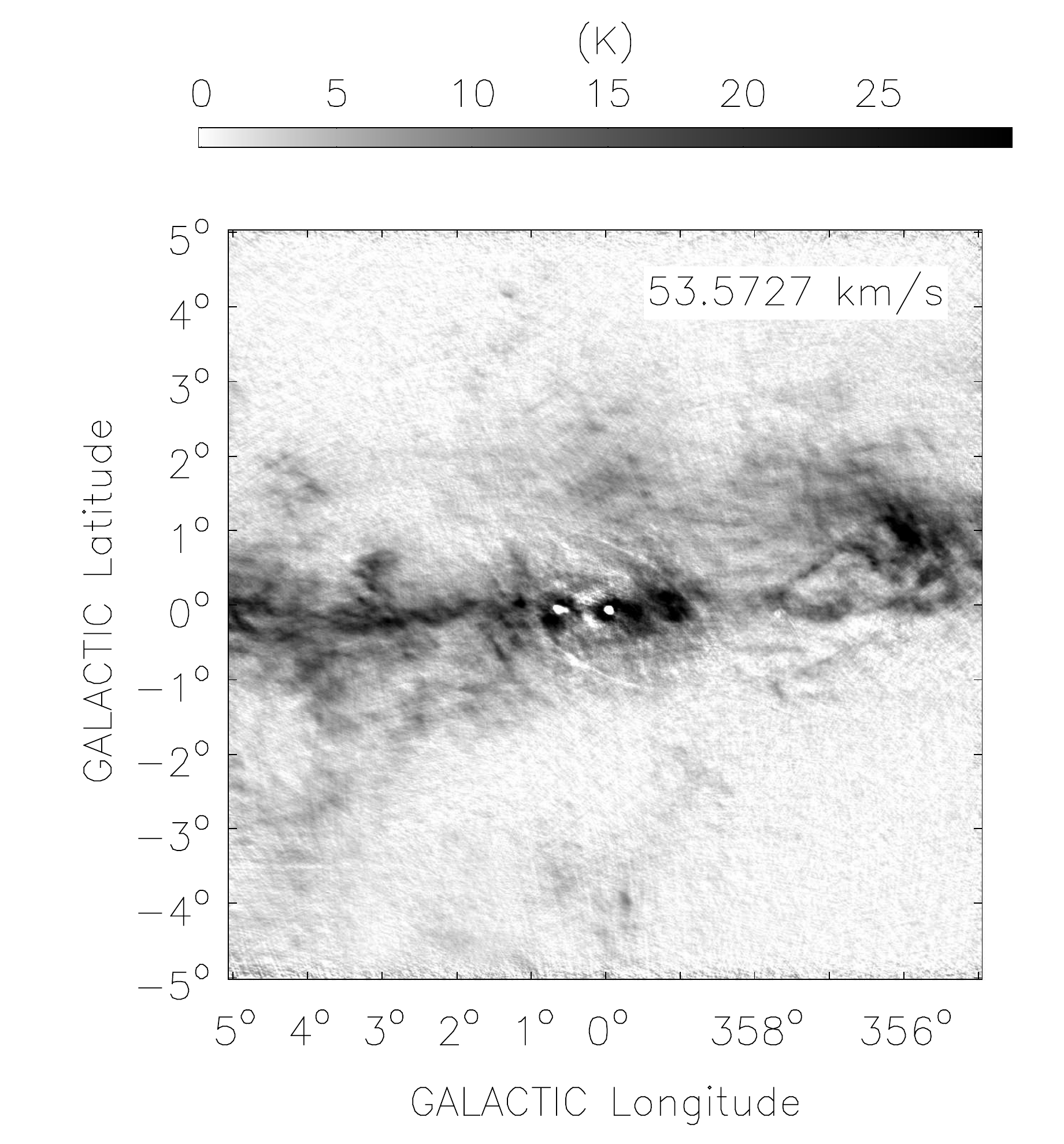}
\plottwo{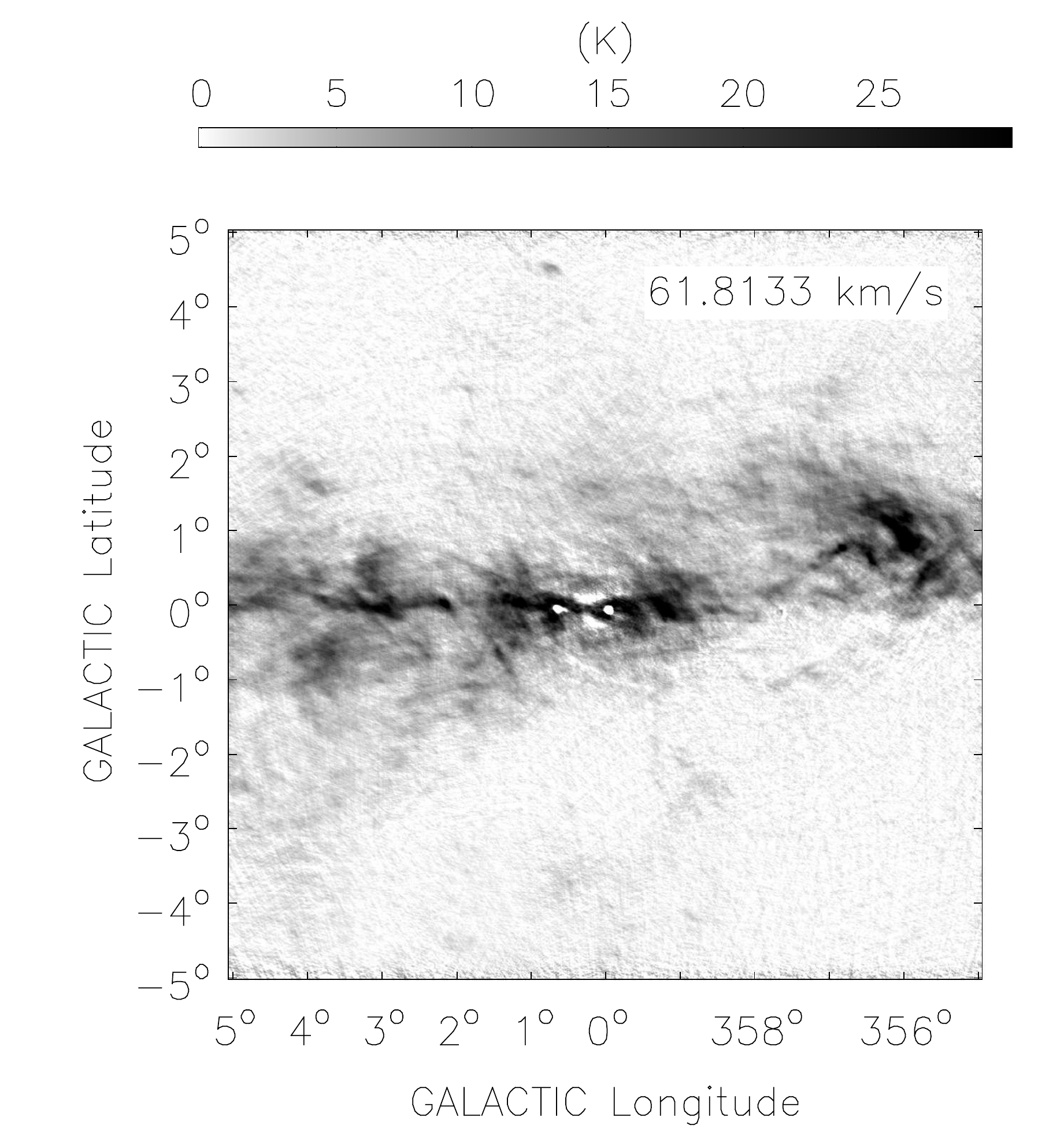}{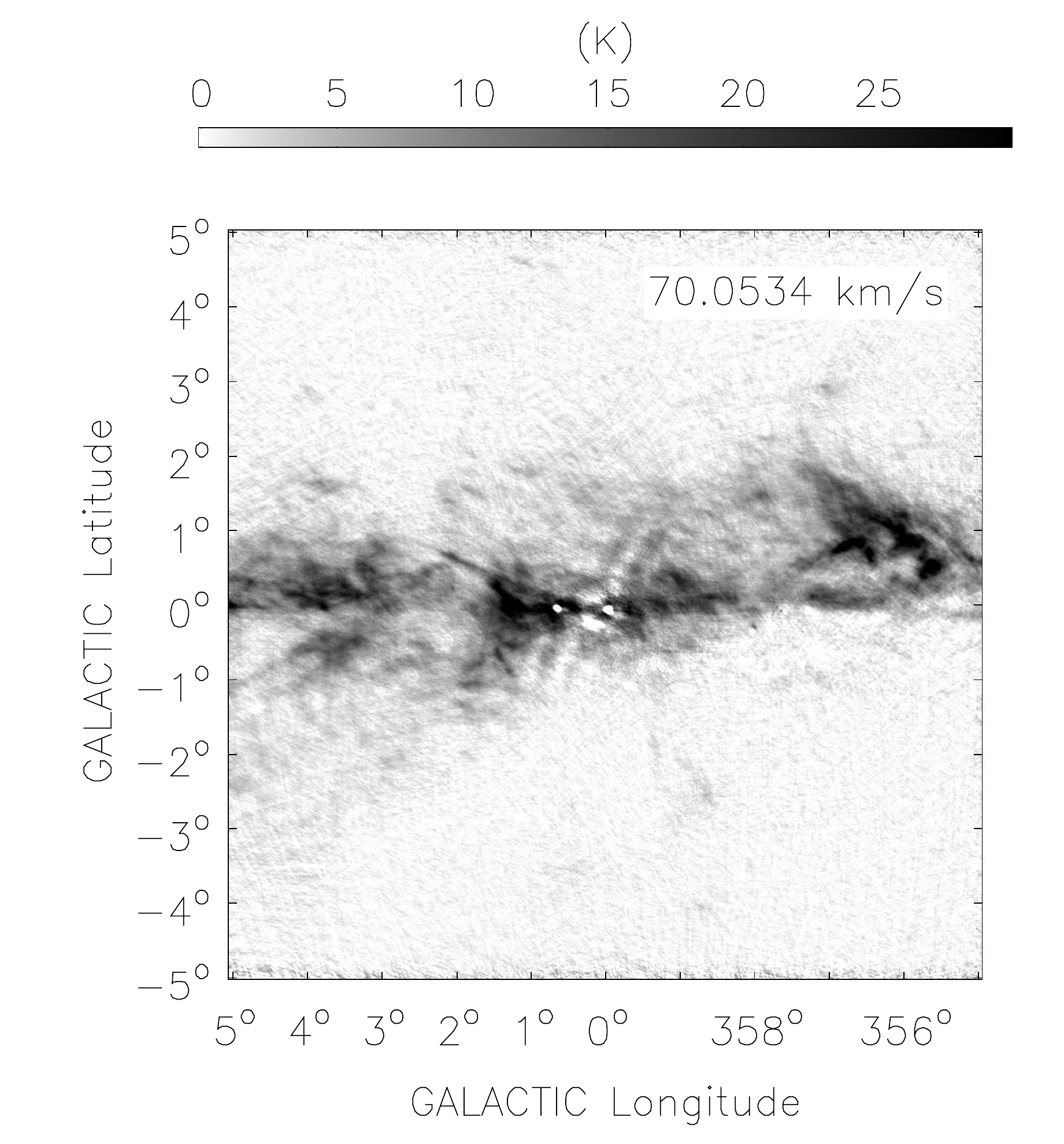}
\caption[]{}
\end{figure}
\begin{figure}
\figurenum{4g}
\centering
\plottwo{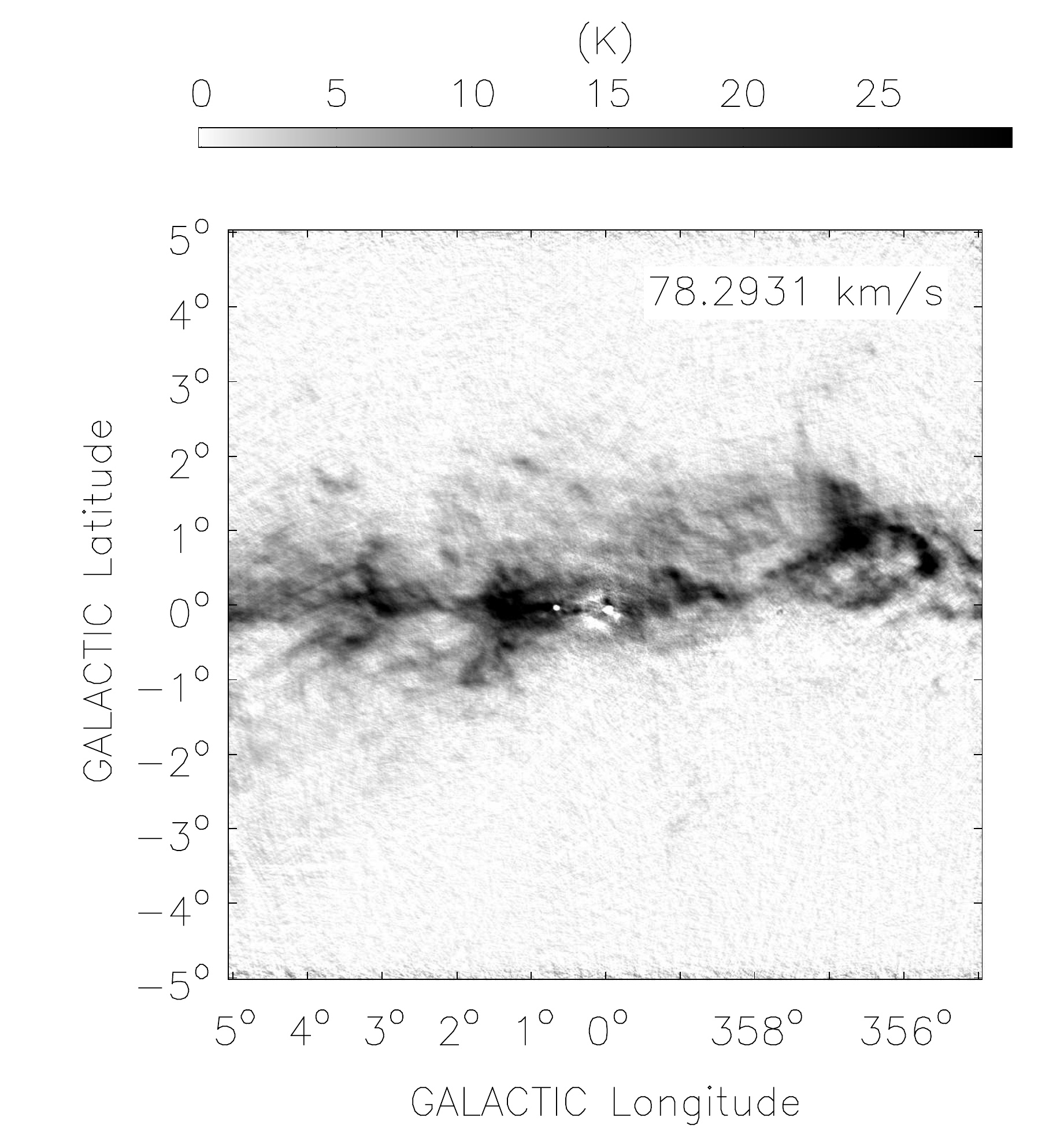}{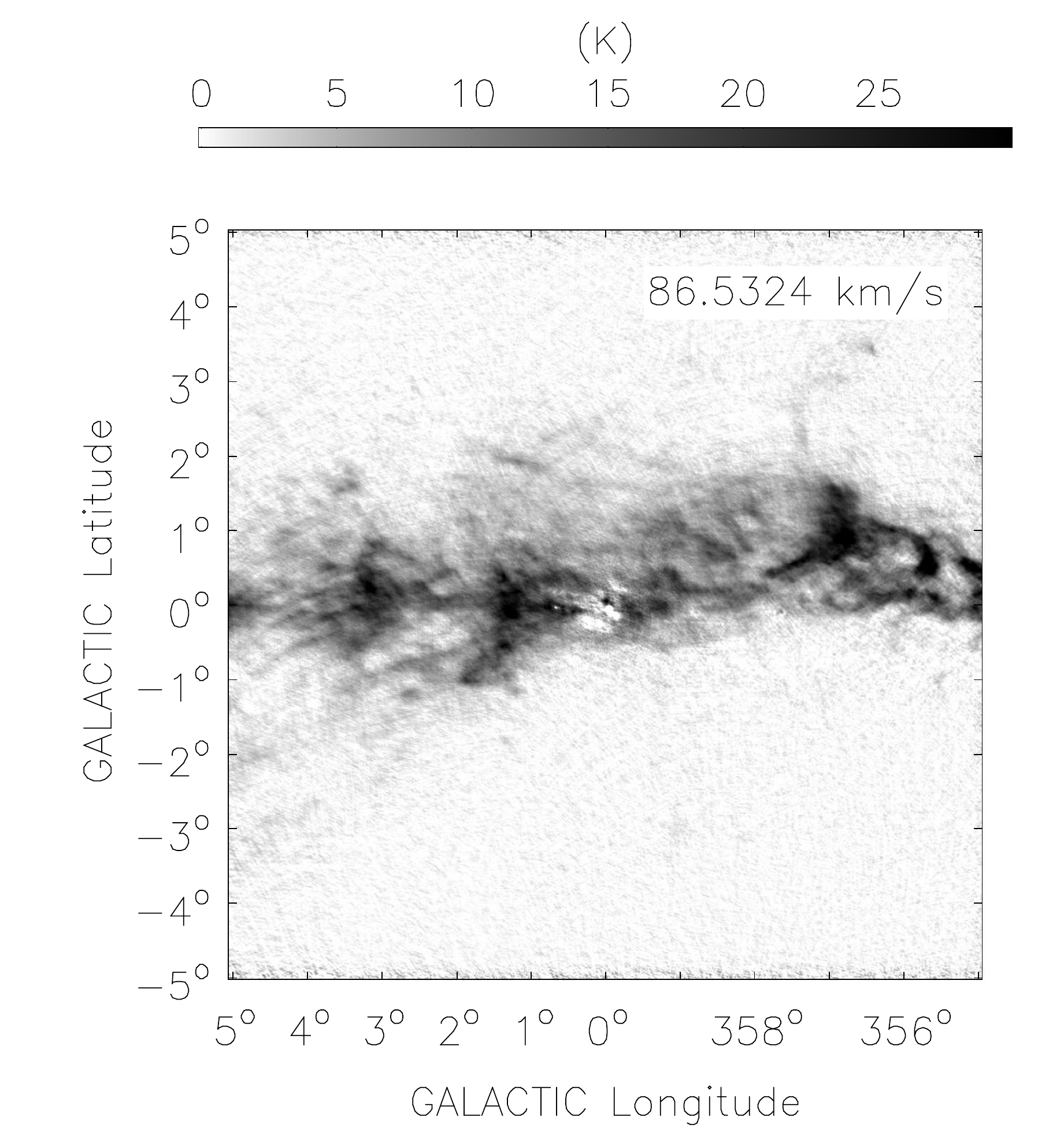}
\plottwo{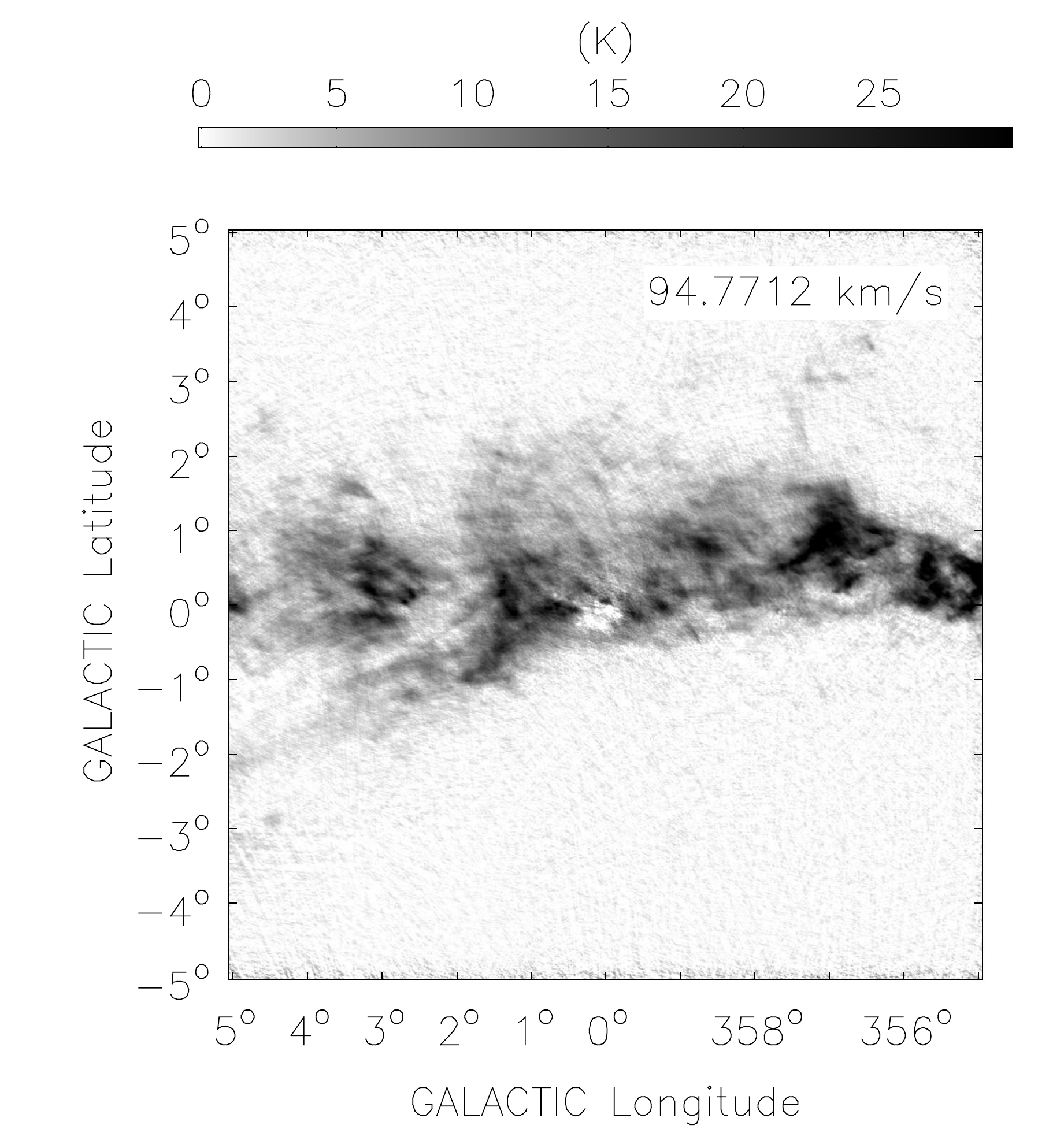}{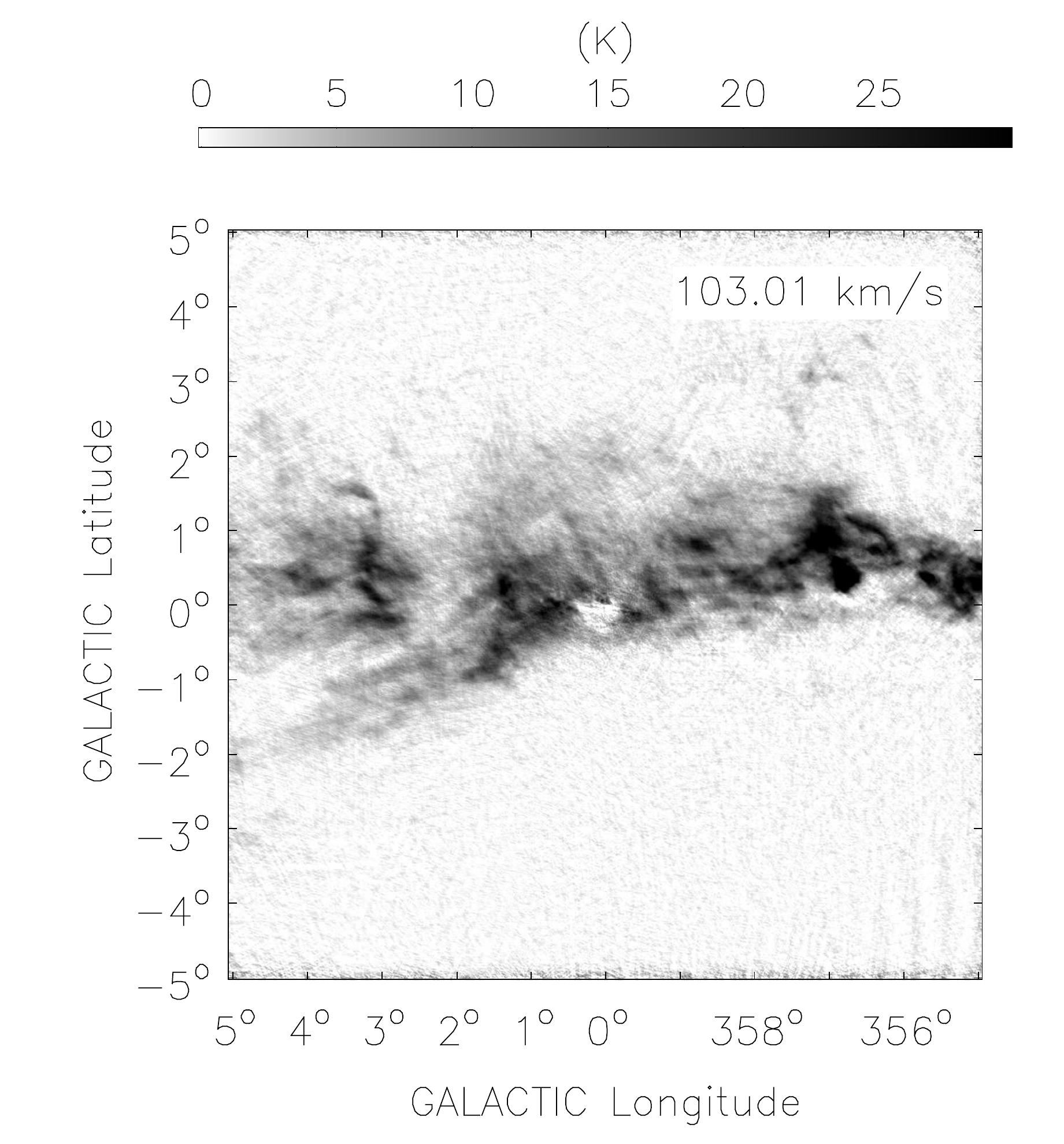}
\plottwo{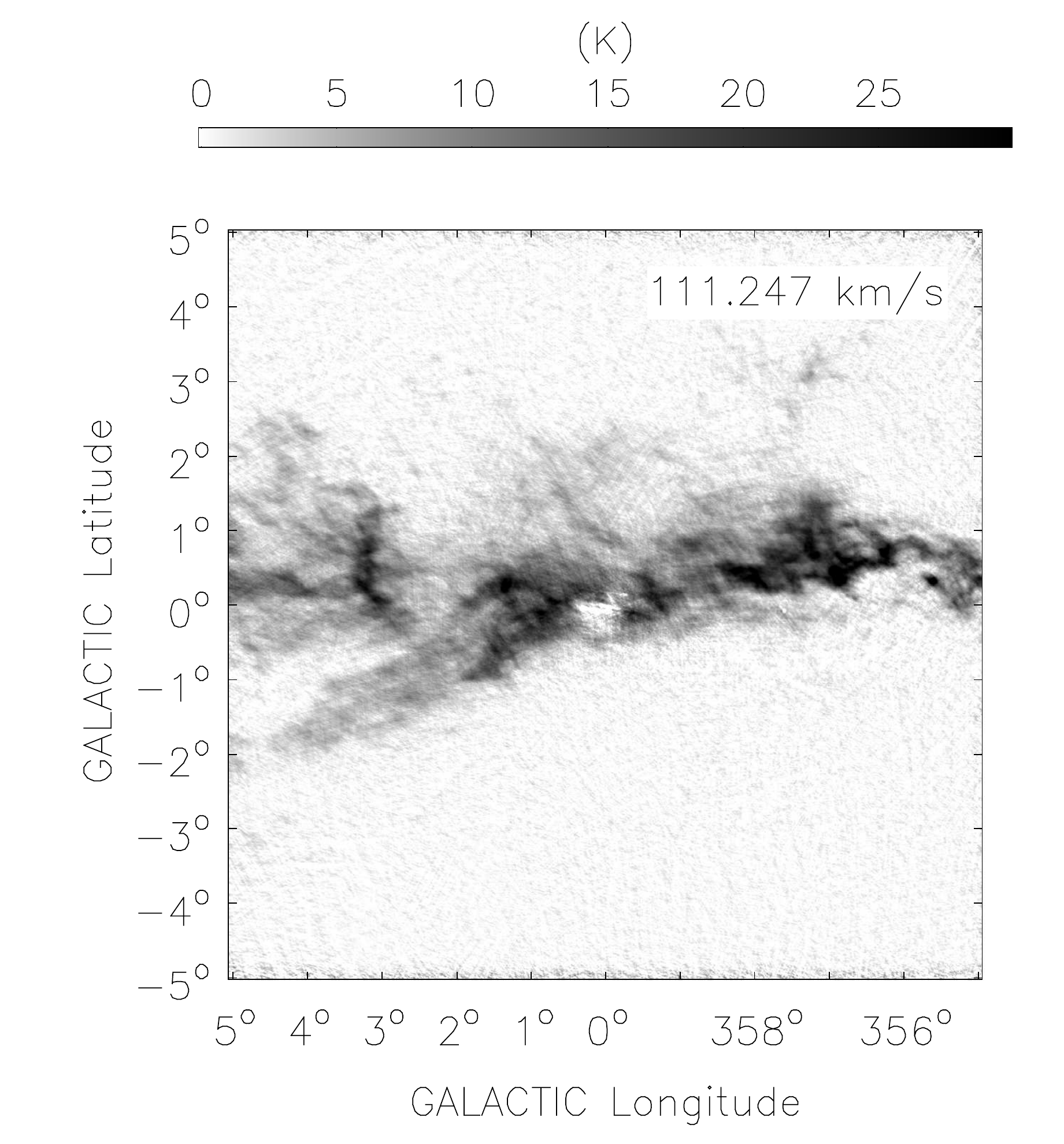}{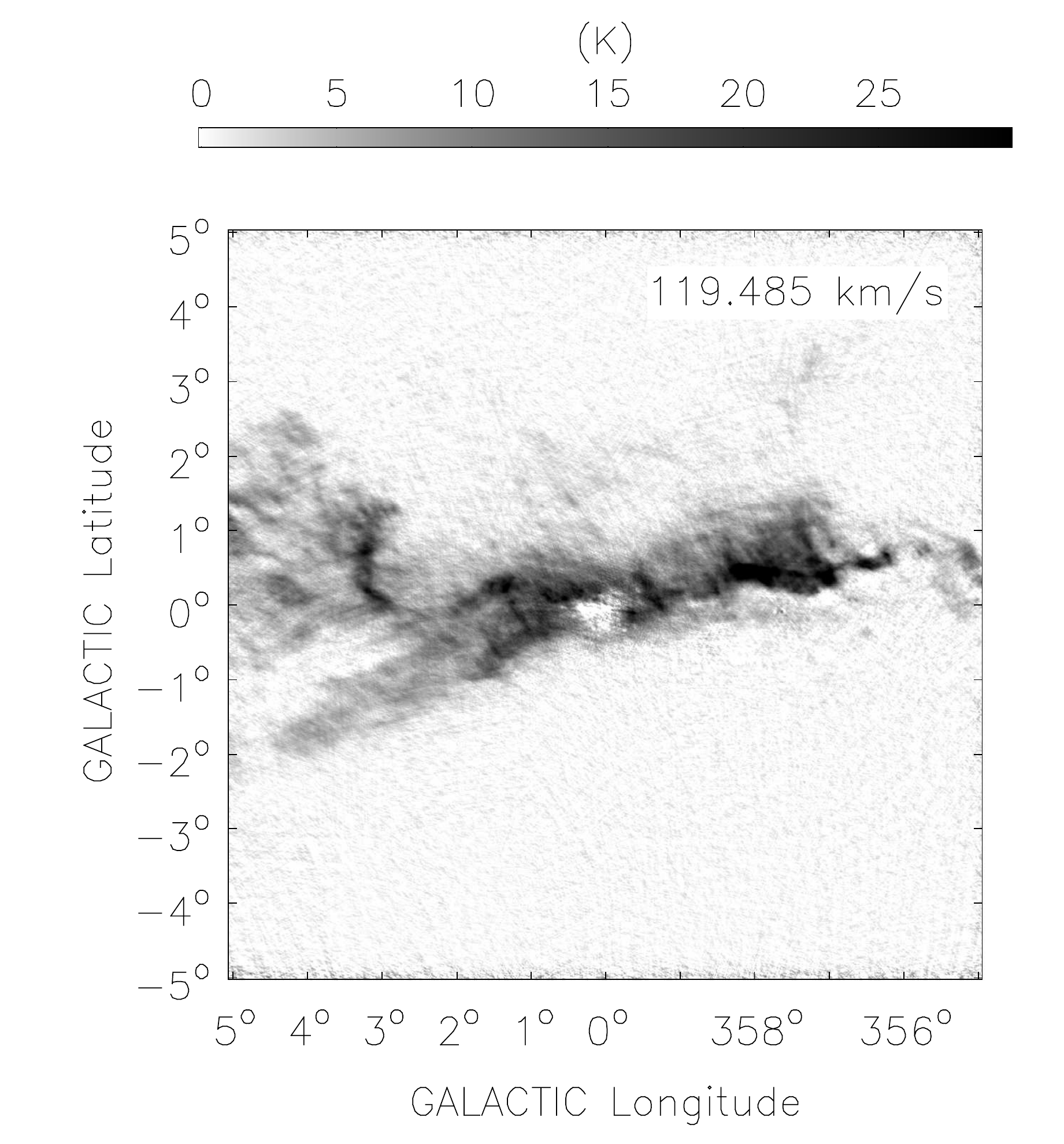}
\caption[]{}
\end{figure}
\begin{figure}
\figurenum{4h}
\centering
\plottwo{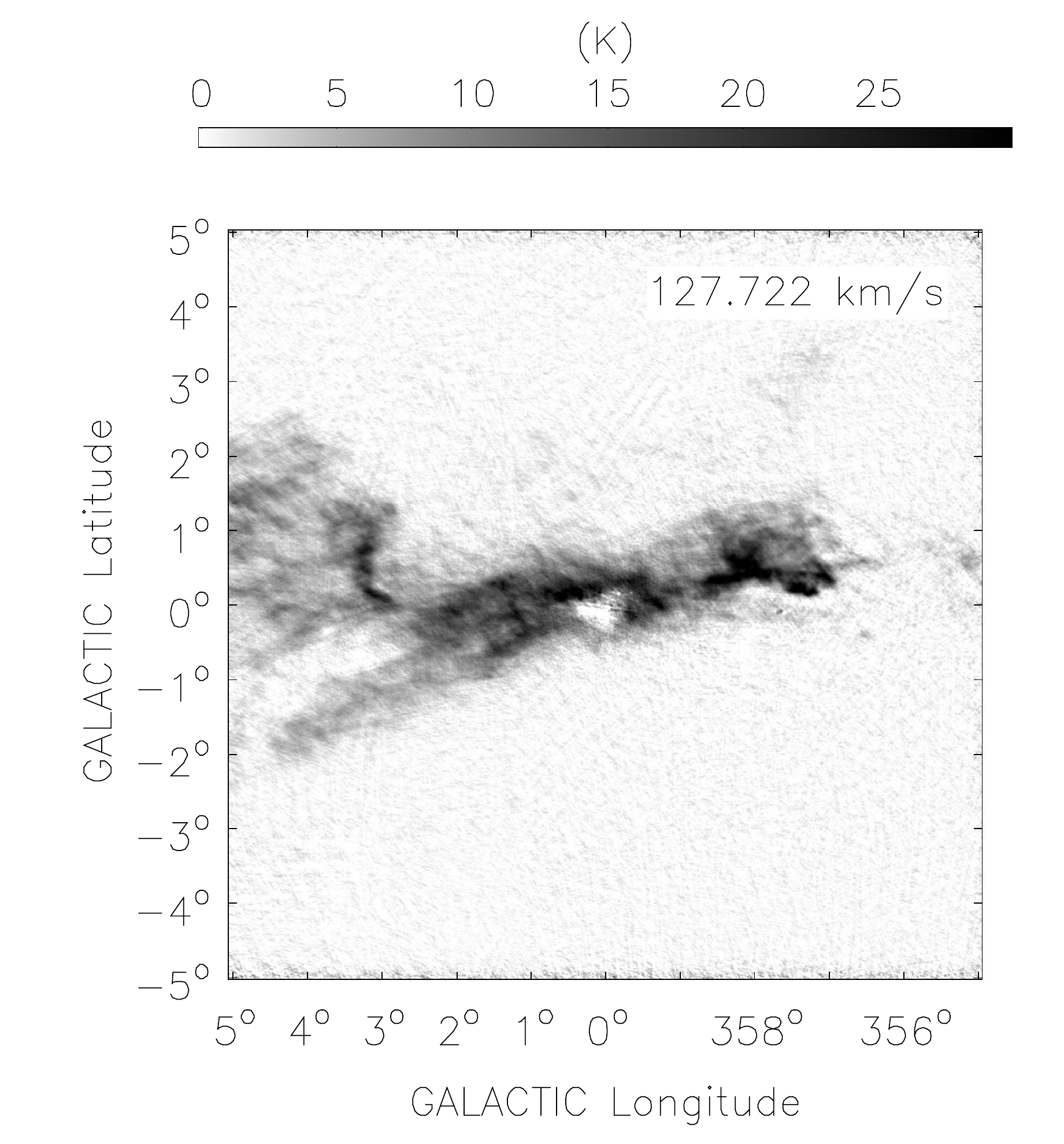}{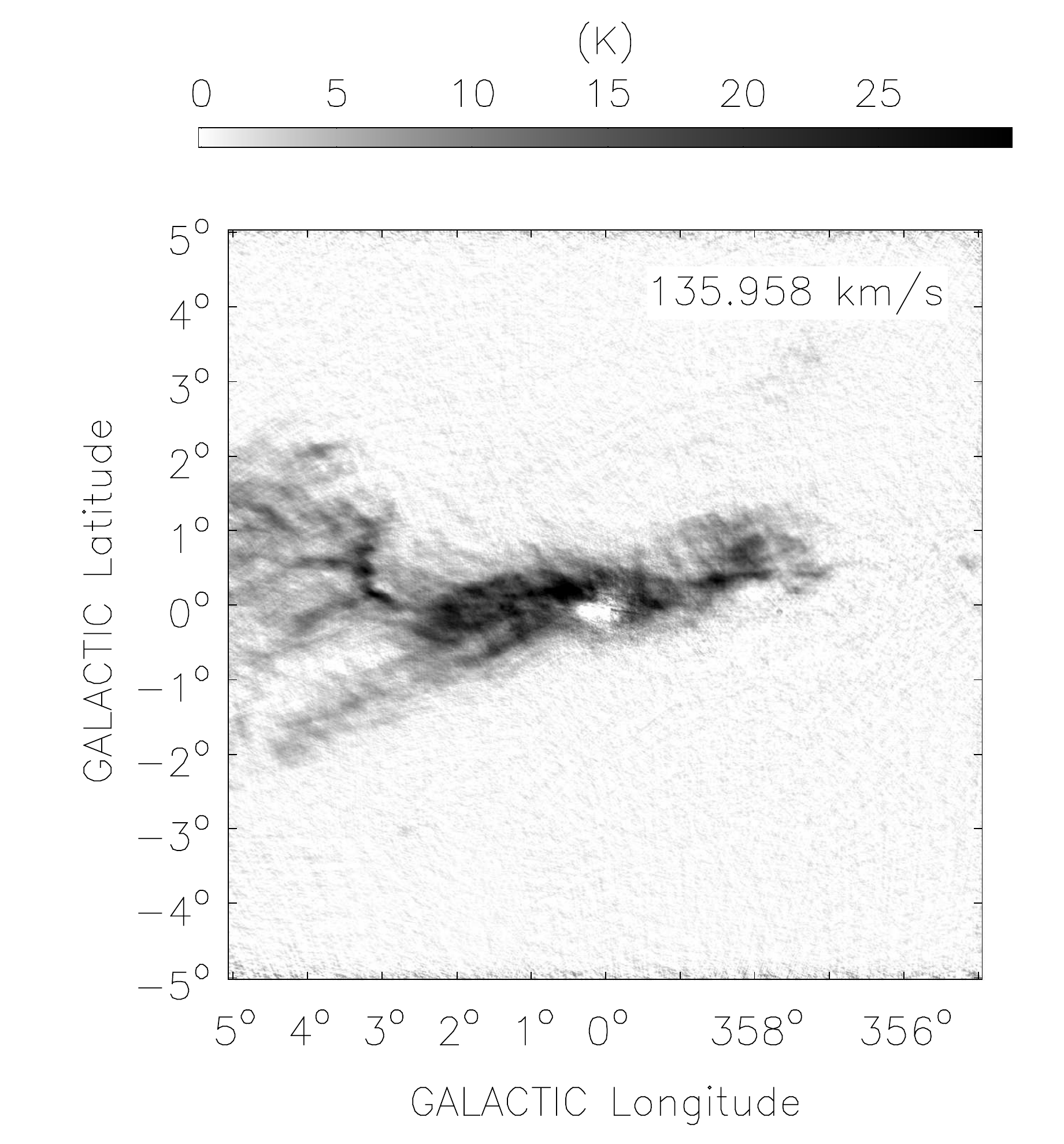}
\plottwo{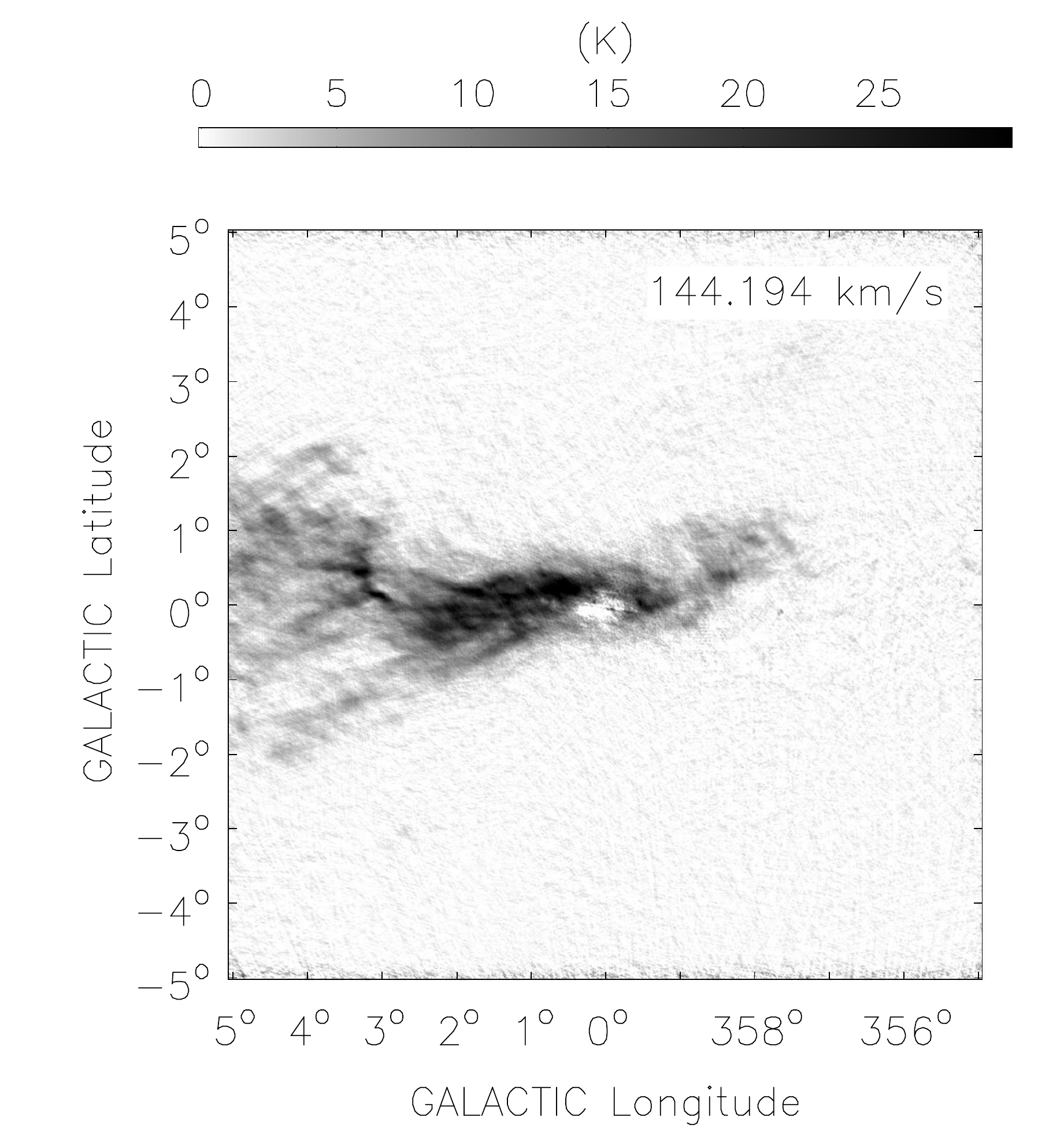}{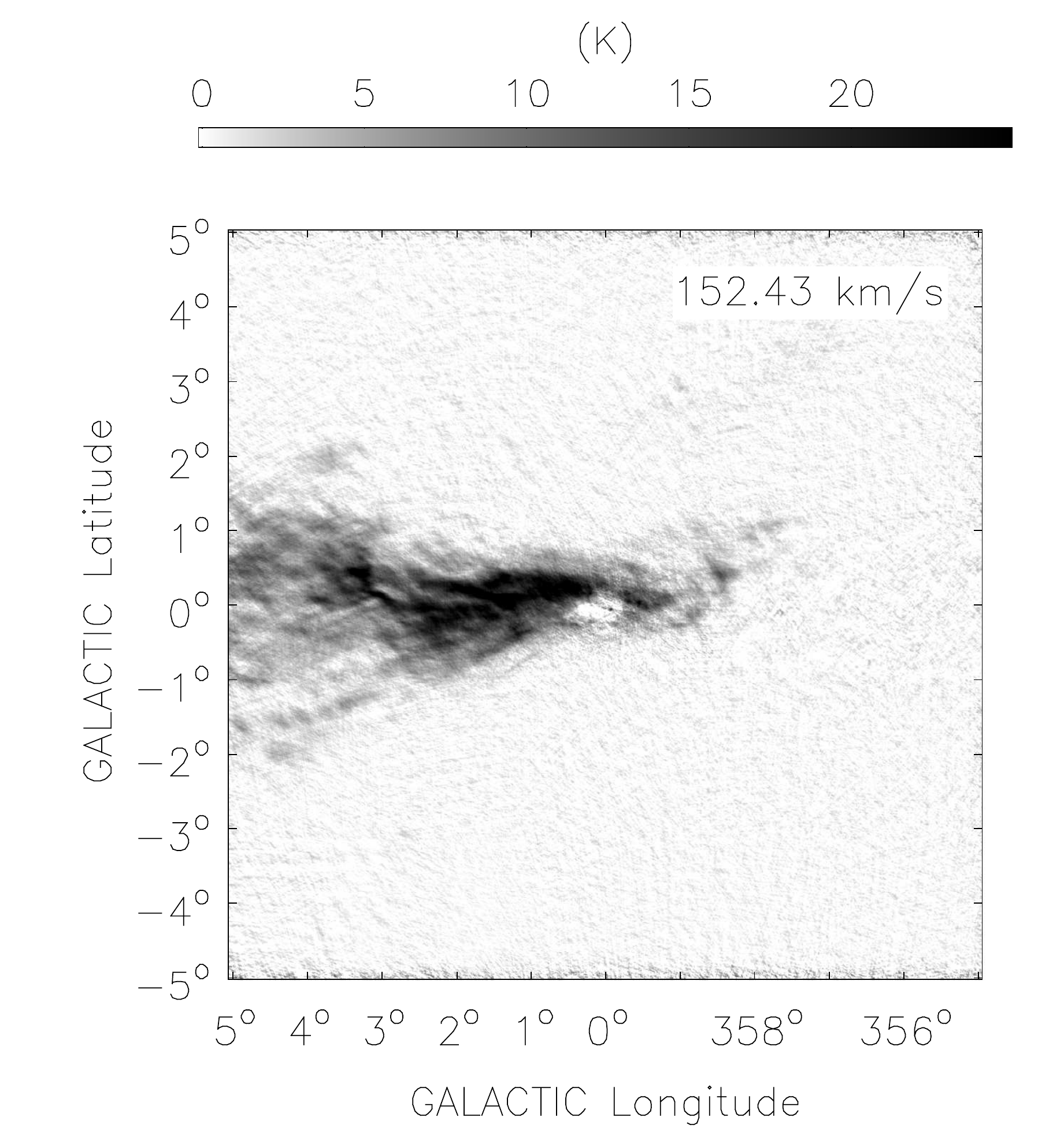}
\plottwo{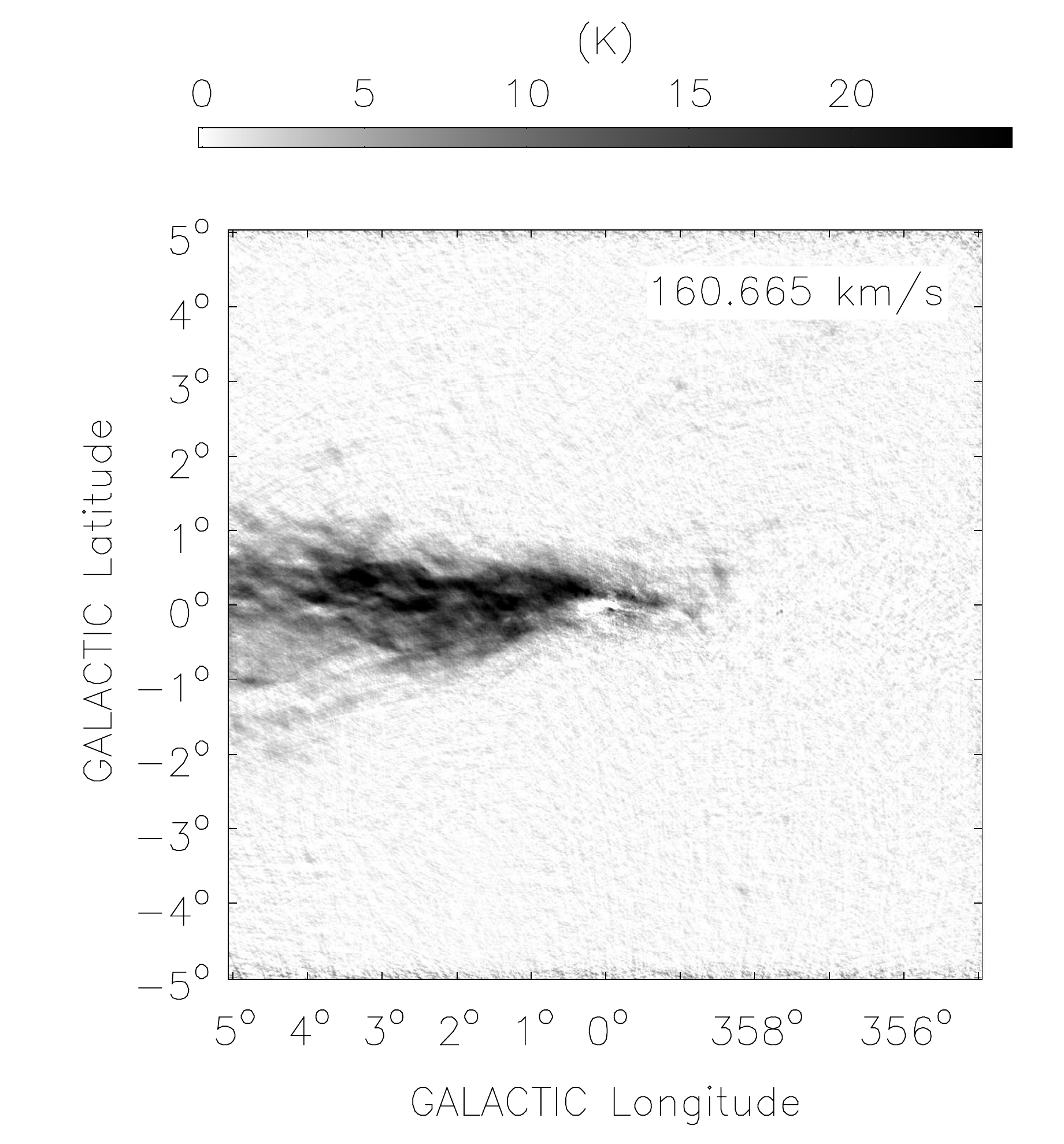}{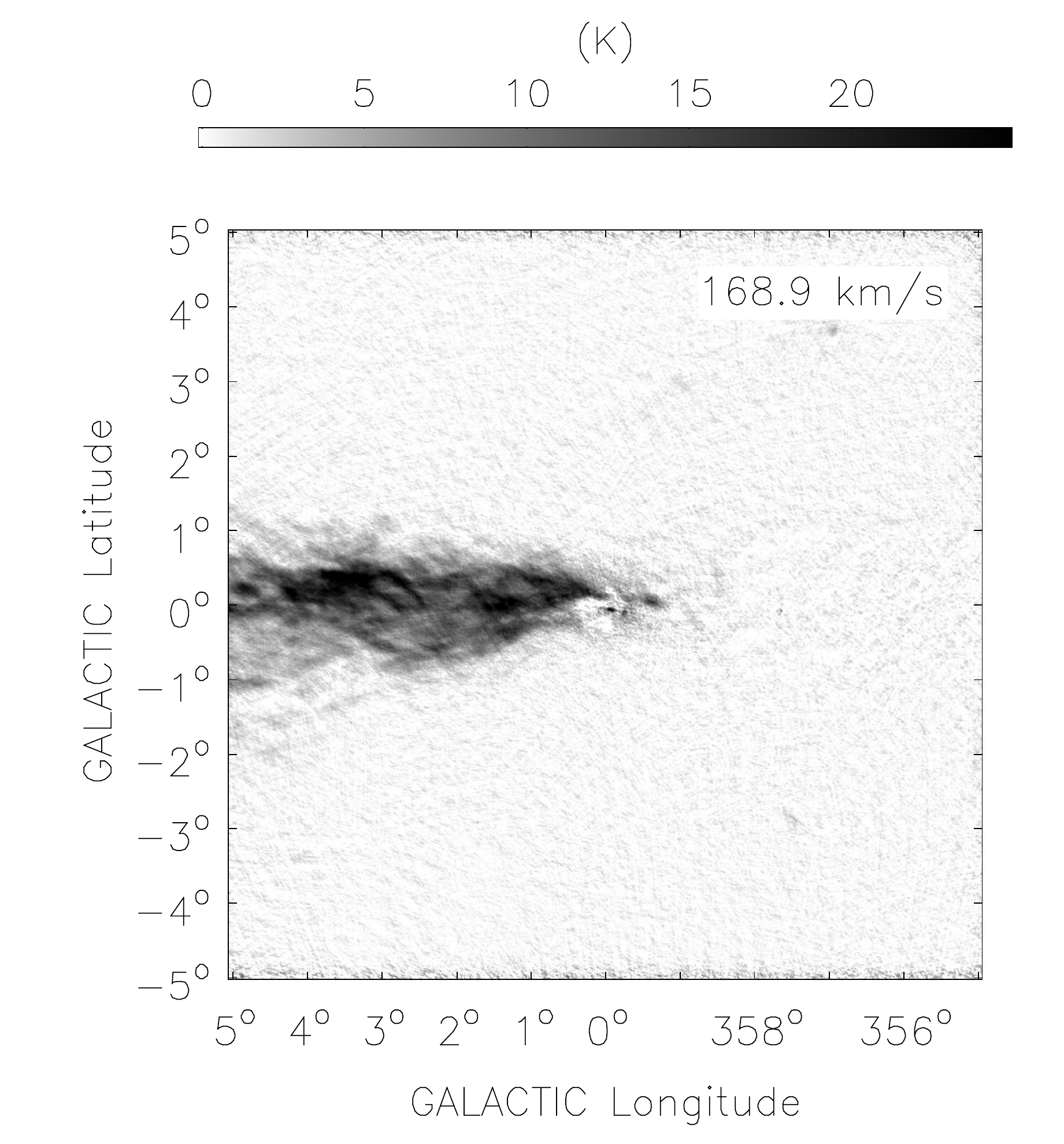}
\caption[]{}
\end{figure}
\begin{figure}
\figurenum{4i}
\centering
\plottwo{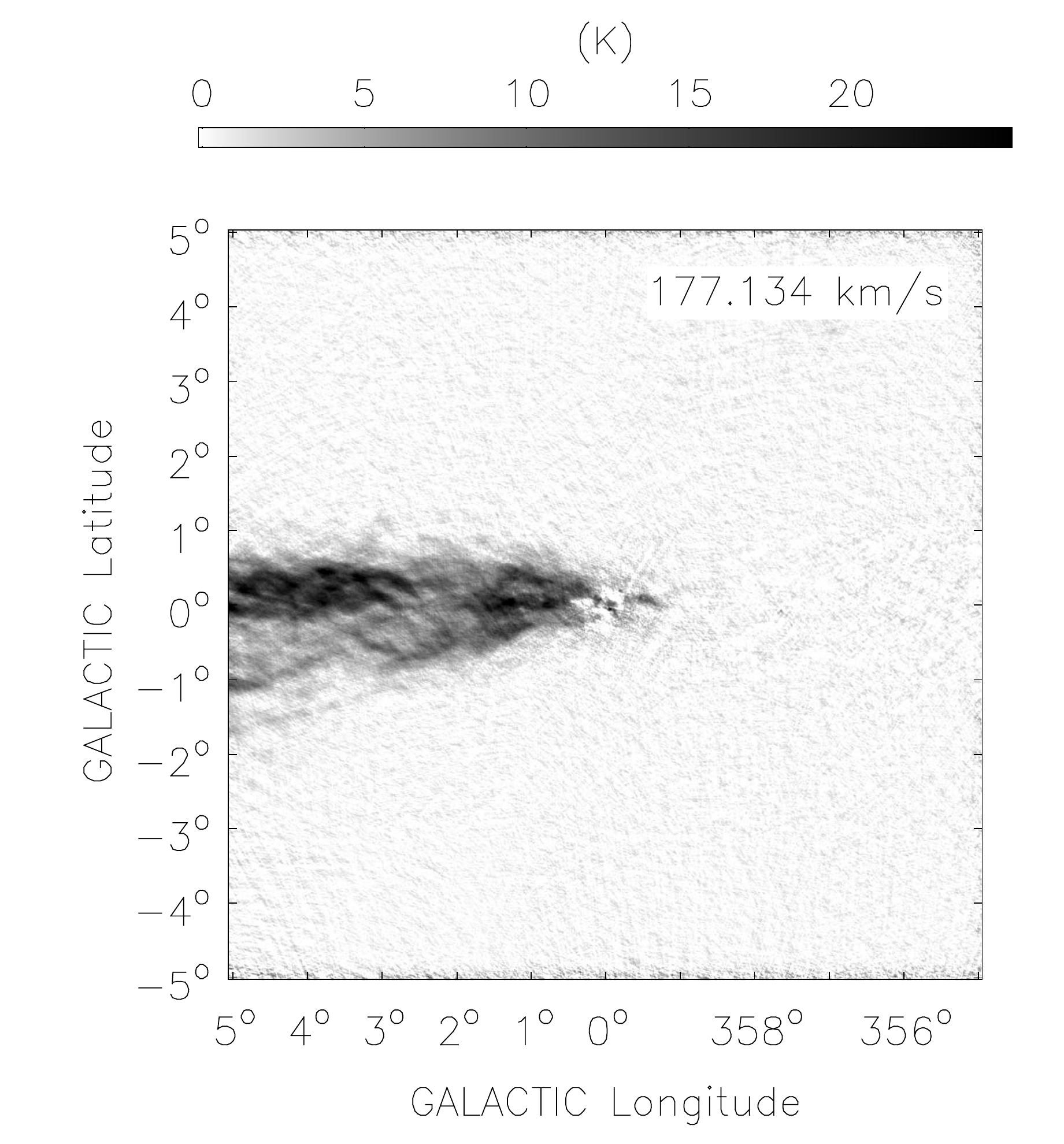}{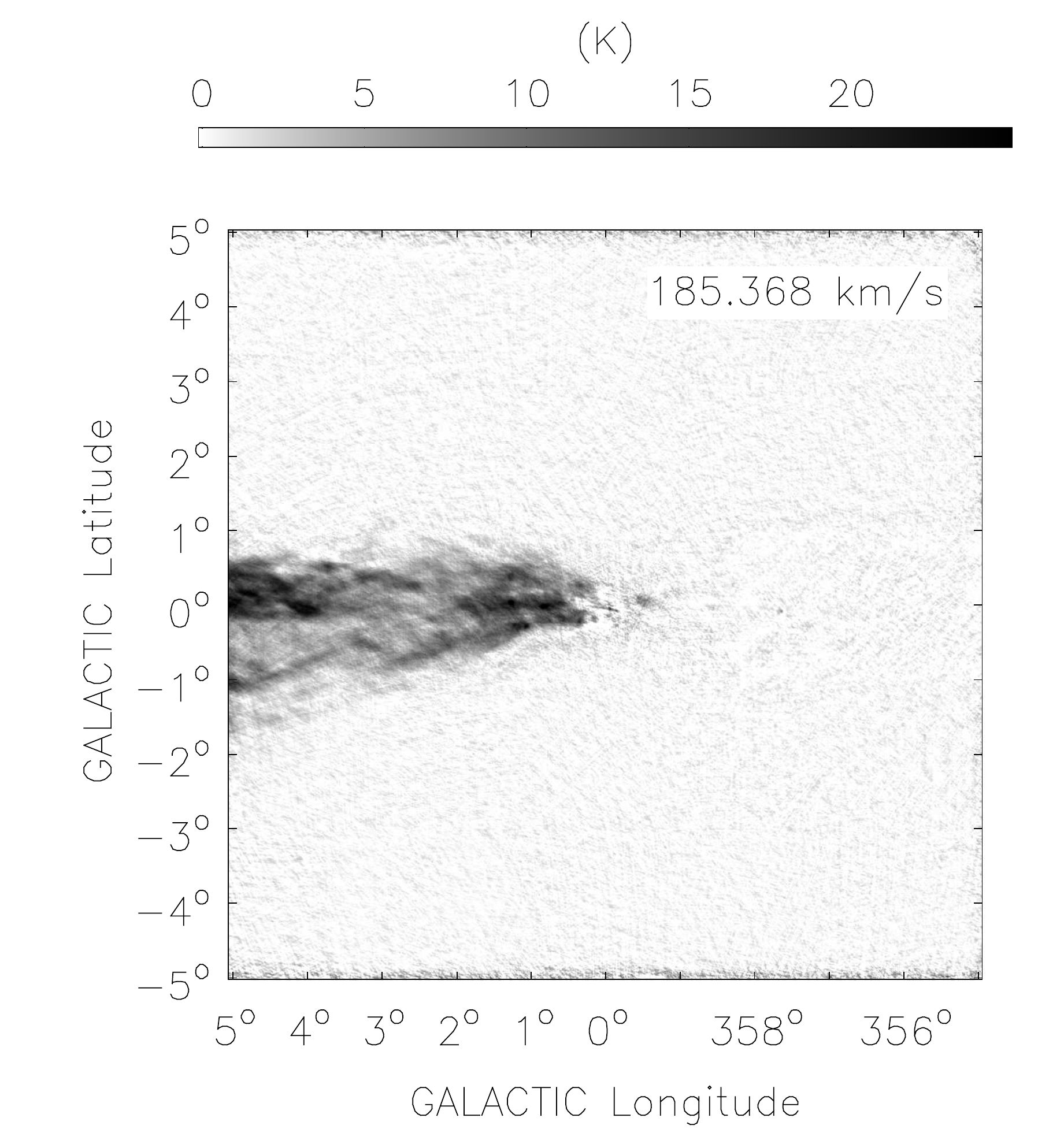}
\plottwo{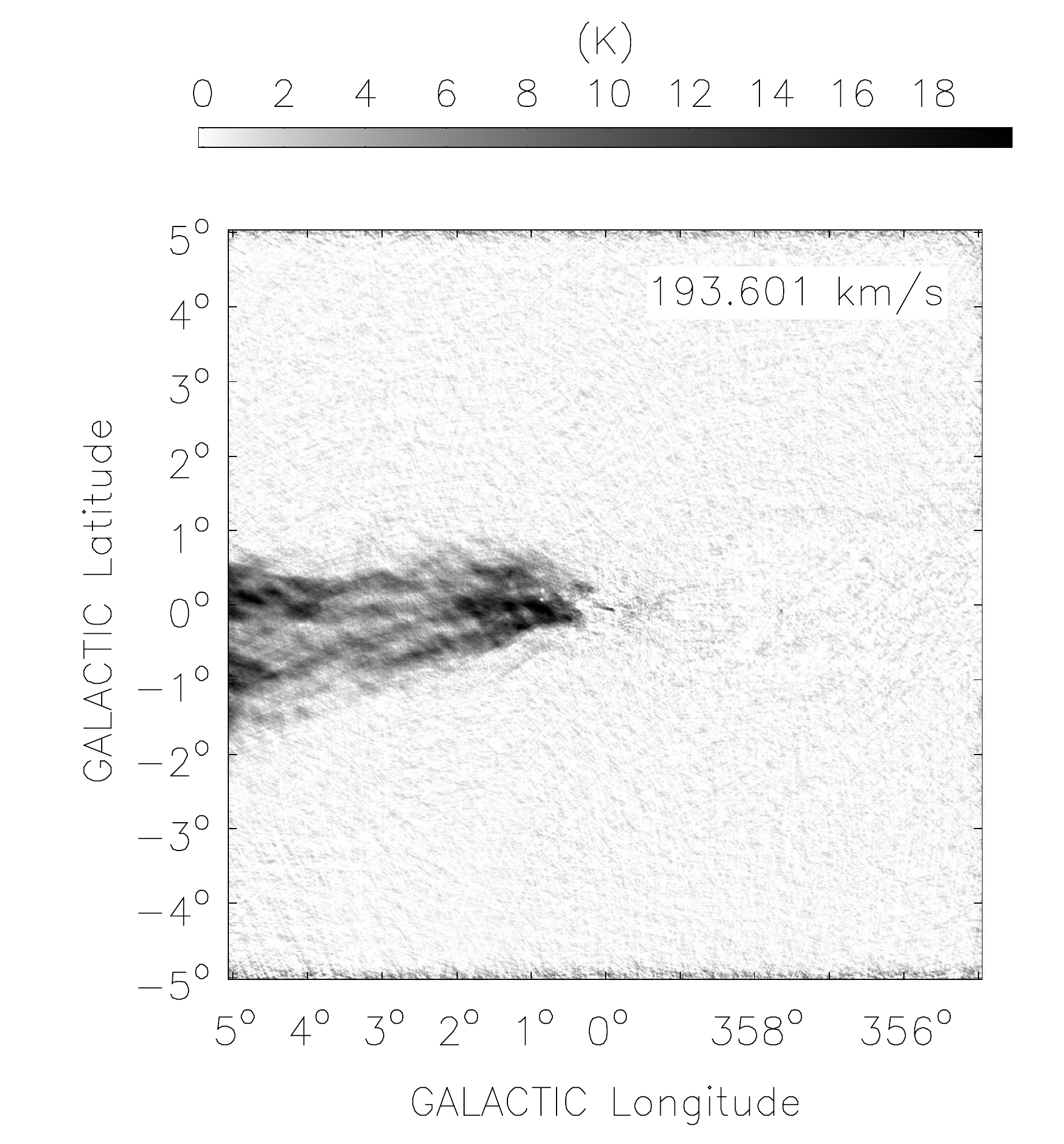}{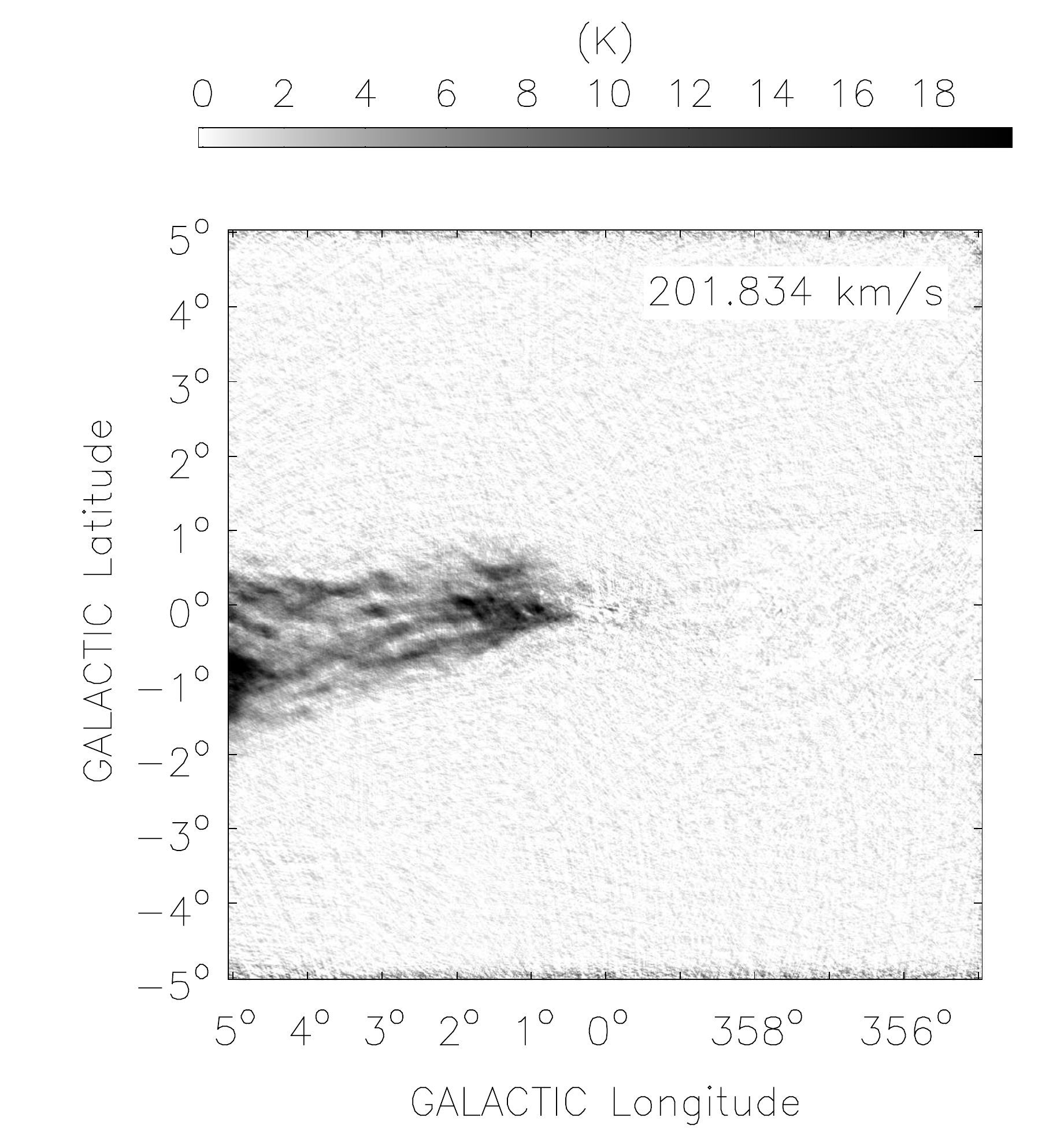}
\plottwo{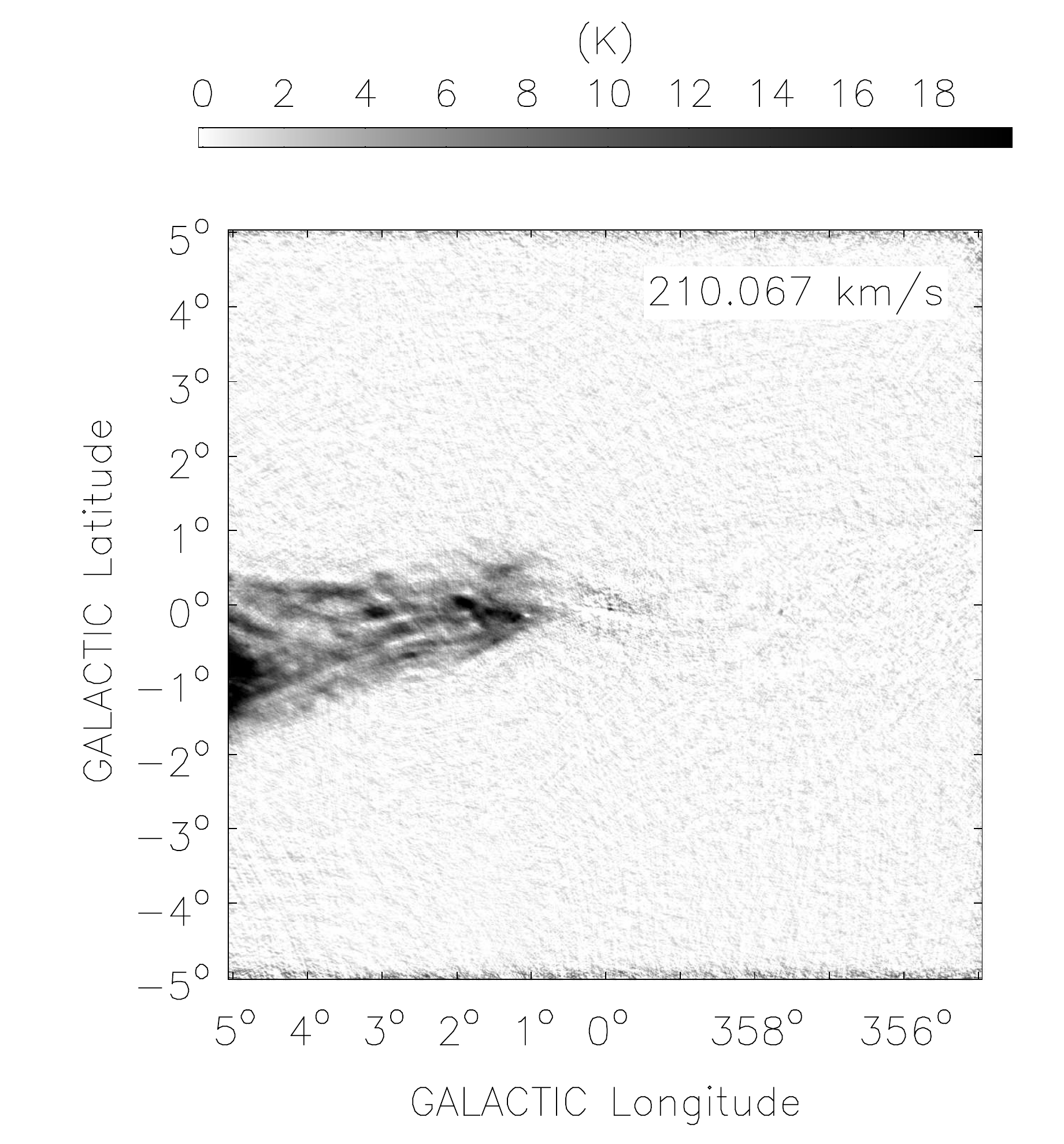}{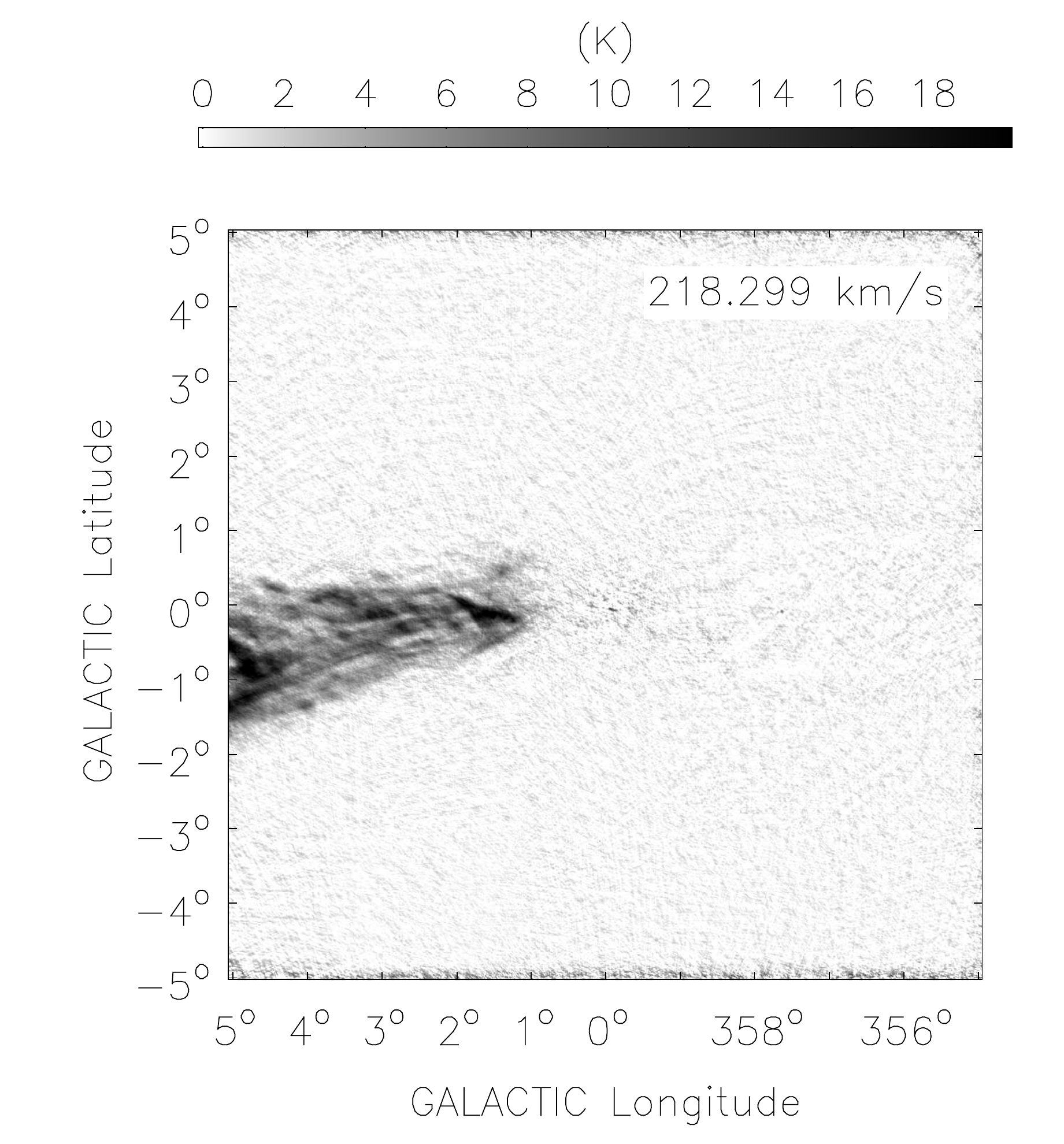}
\caption[]{}
\end{figure}
\clearpage

\begin{figure}
\centering
\epsscale{0.95}
\plottwo{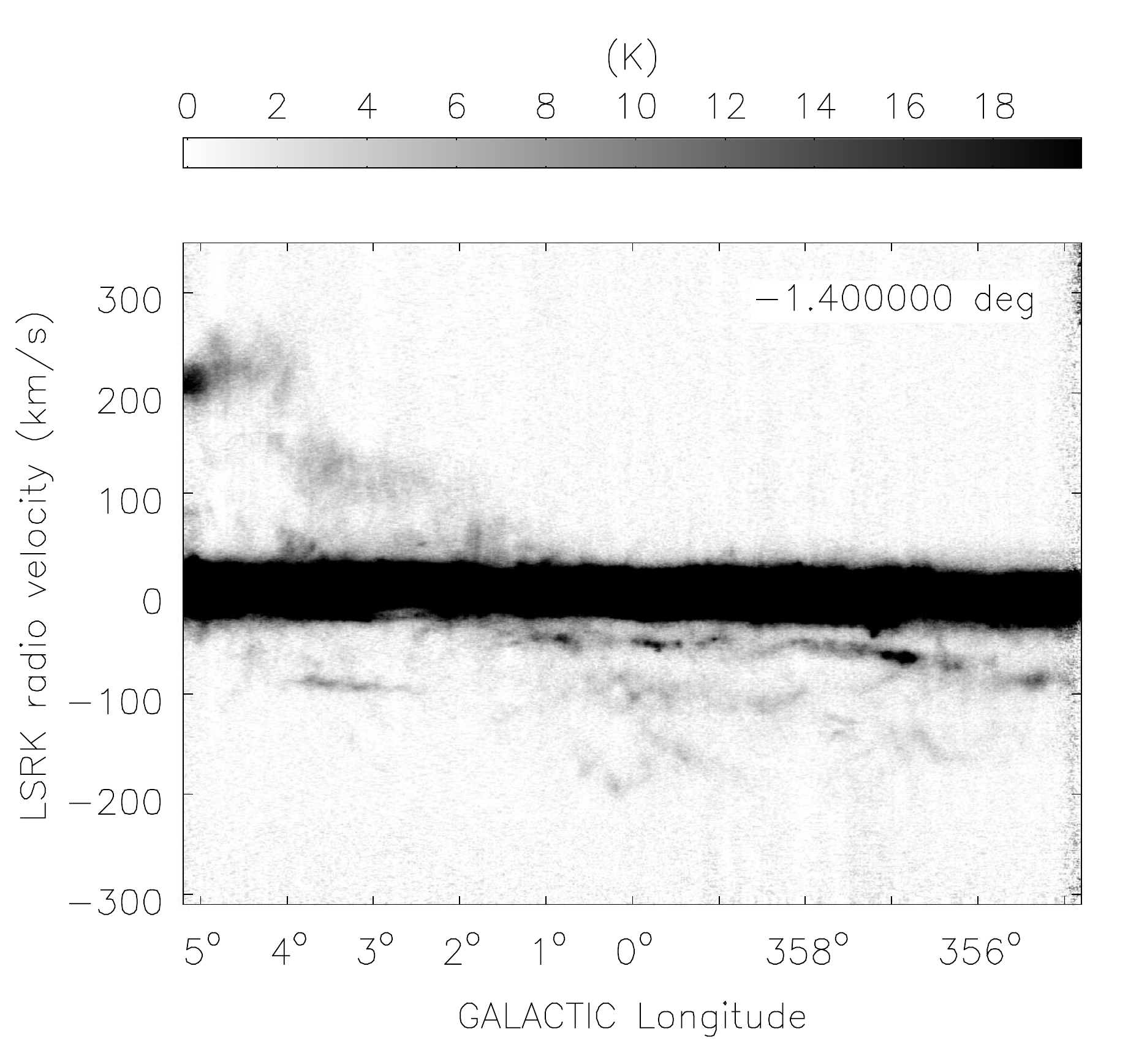}{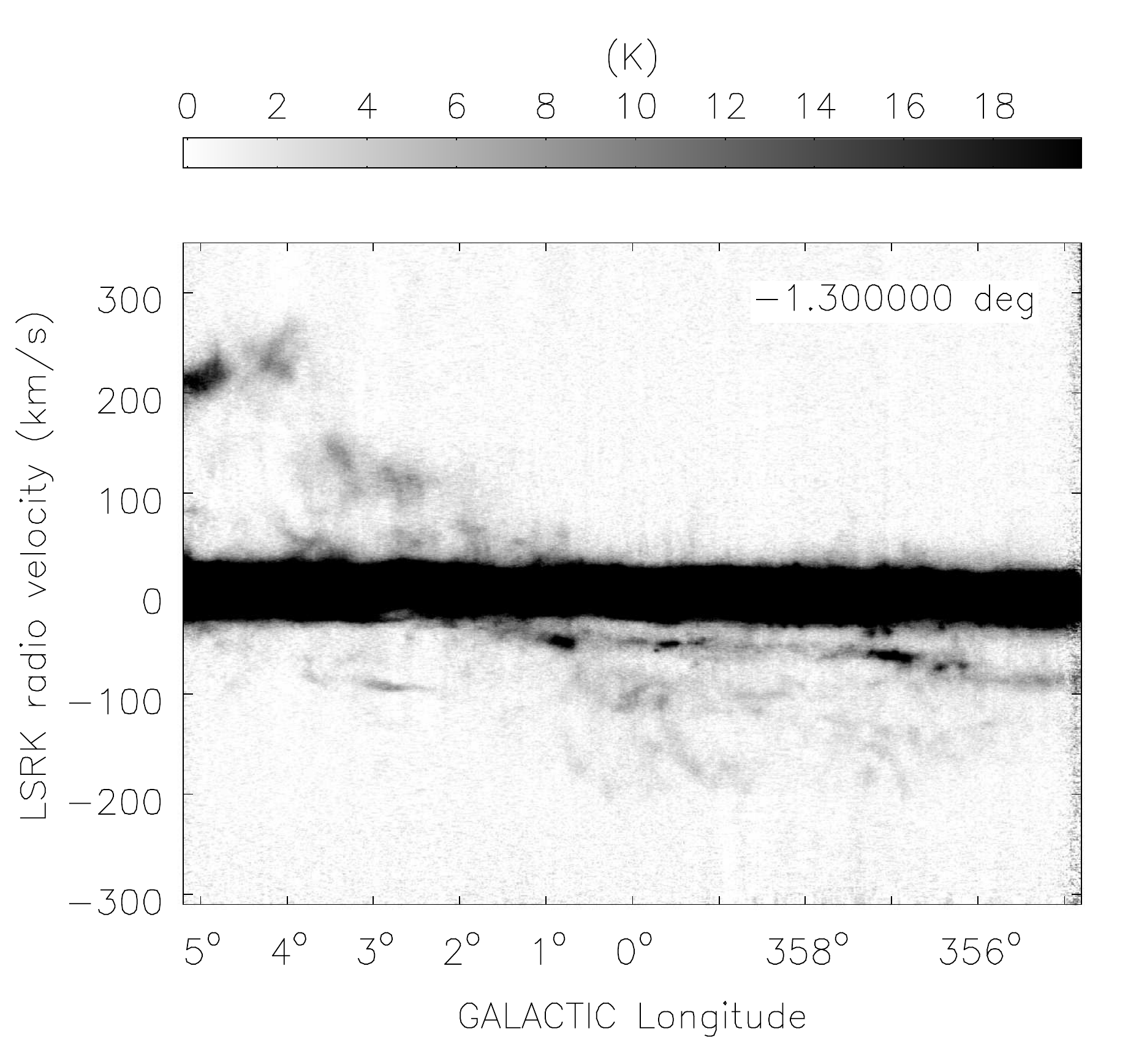}\\
\plottwo{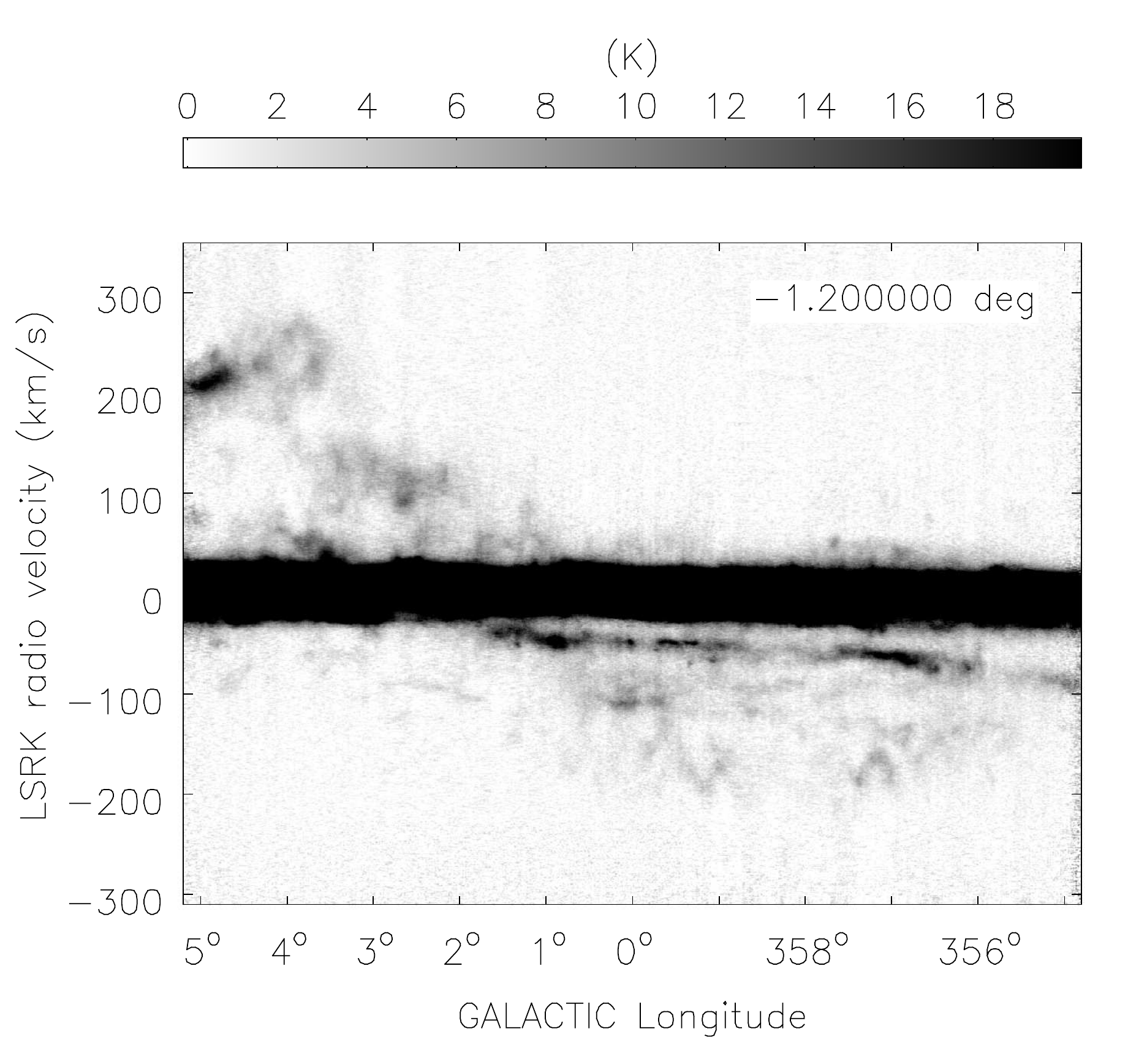}{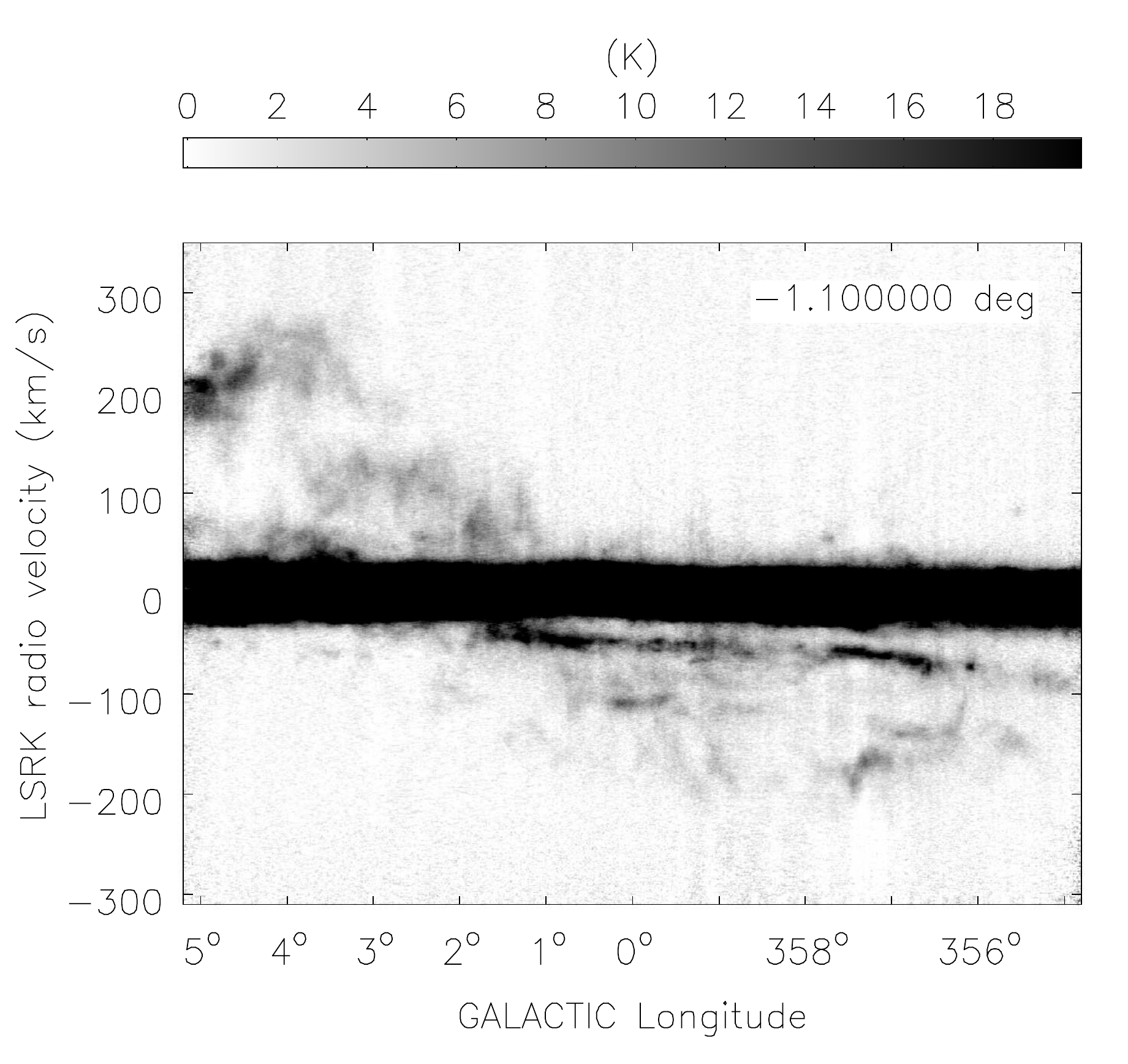}\\
\plottwo{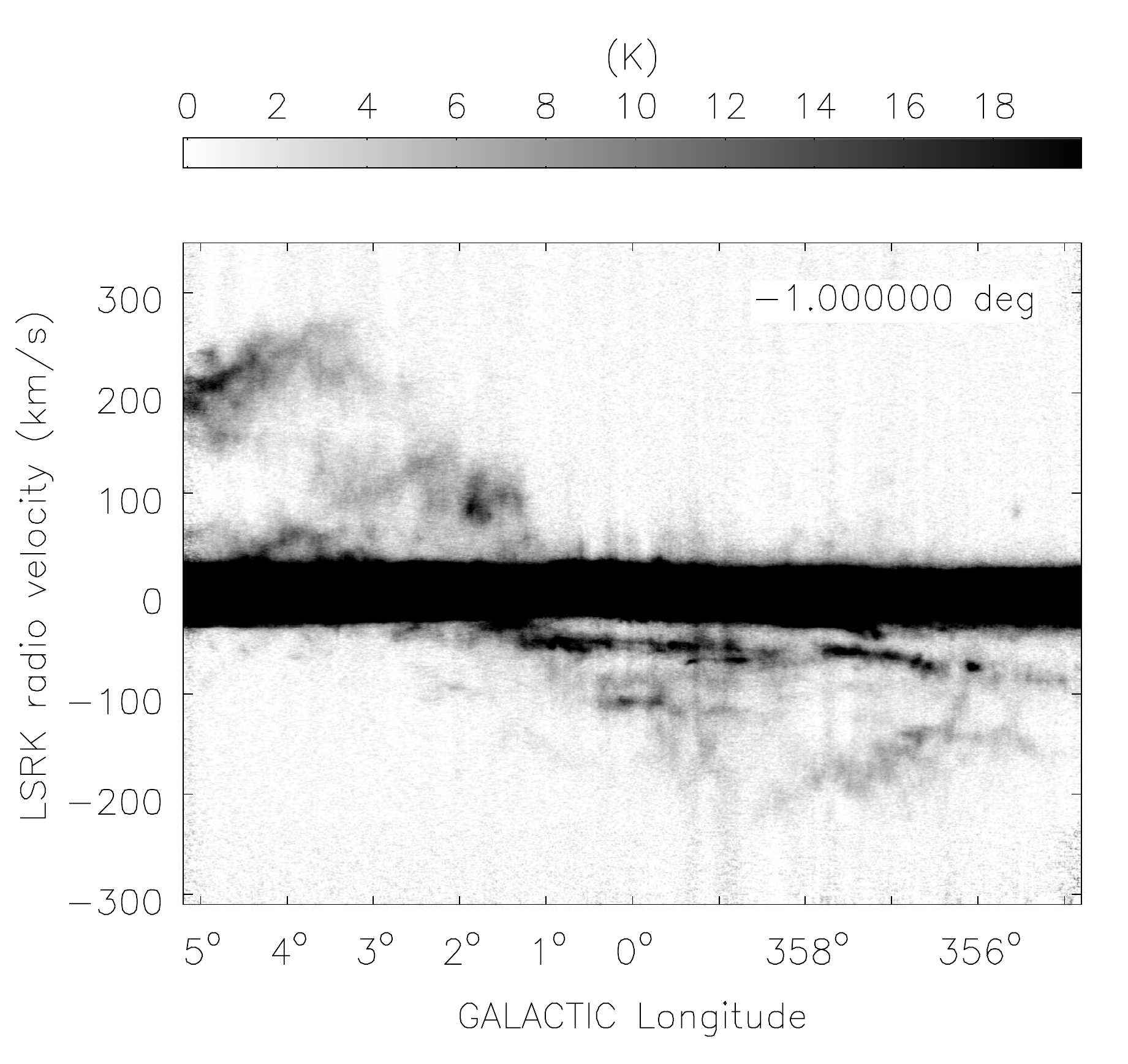}{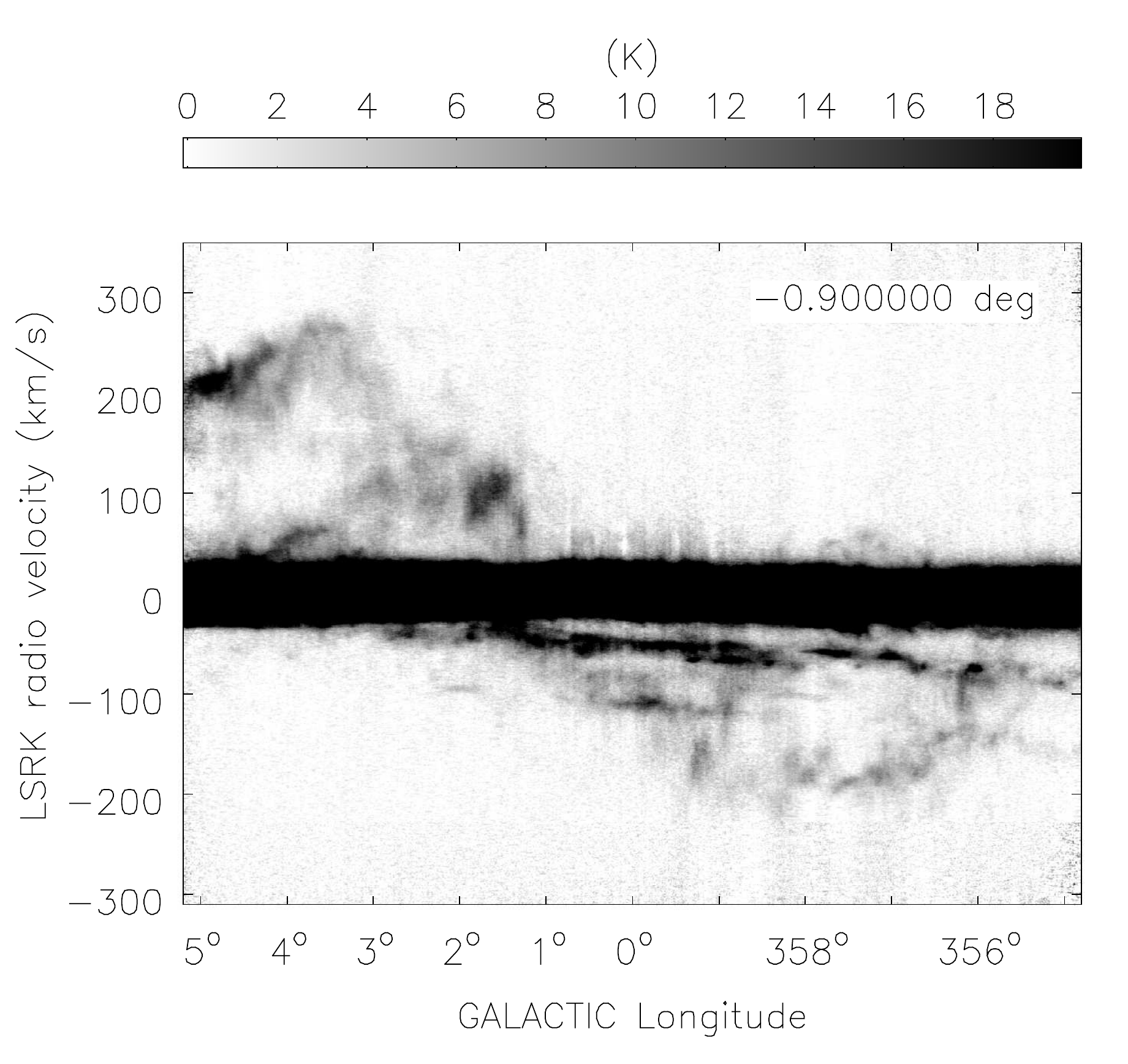}\\
\caption[]{Longitude-velocity images sampled at latitude intervals of
  $0.1^{\circ}$.  The greyscale for each panel is shown in the color
  wedges, each panel uses a different scaling.
\label{fig:lv}}
\end{figure}

\begin{figure}
\figurenum{5b}
\centering
\plottwo{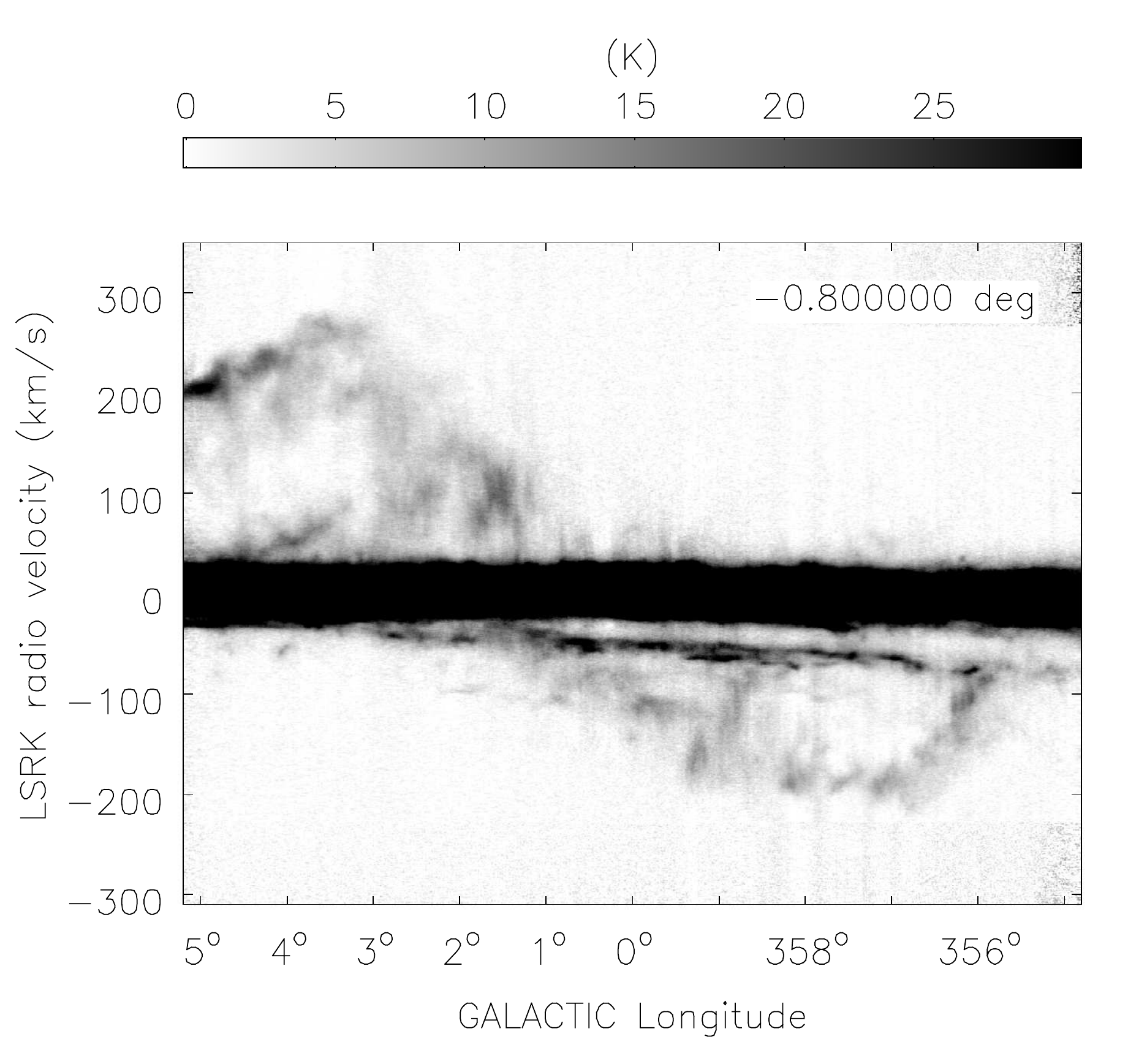}{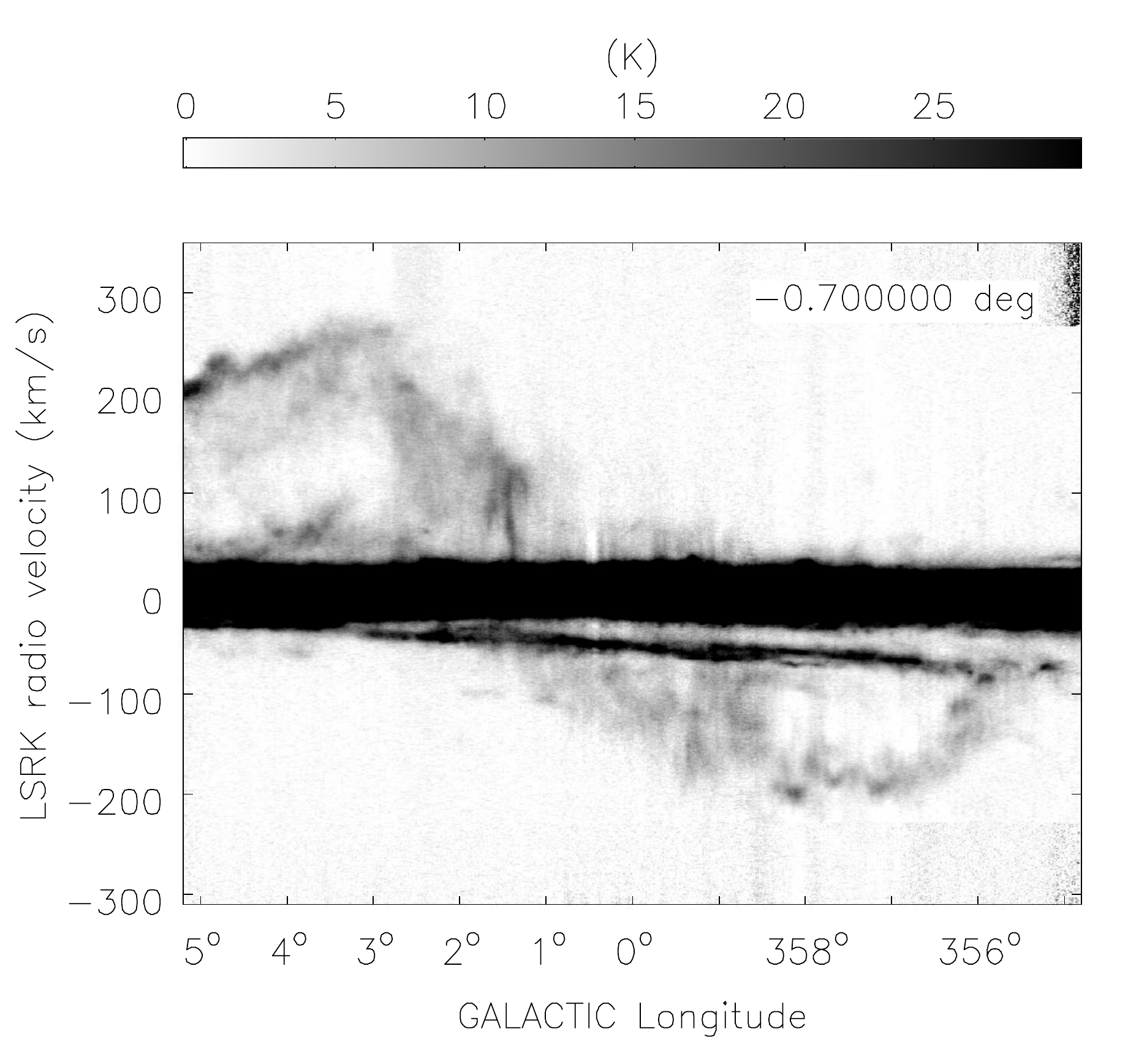}\\
\plottwo{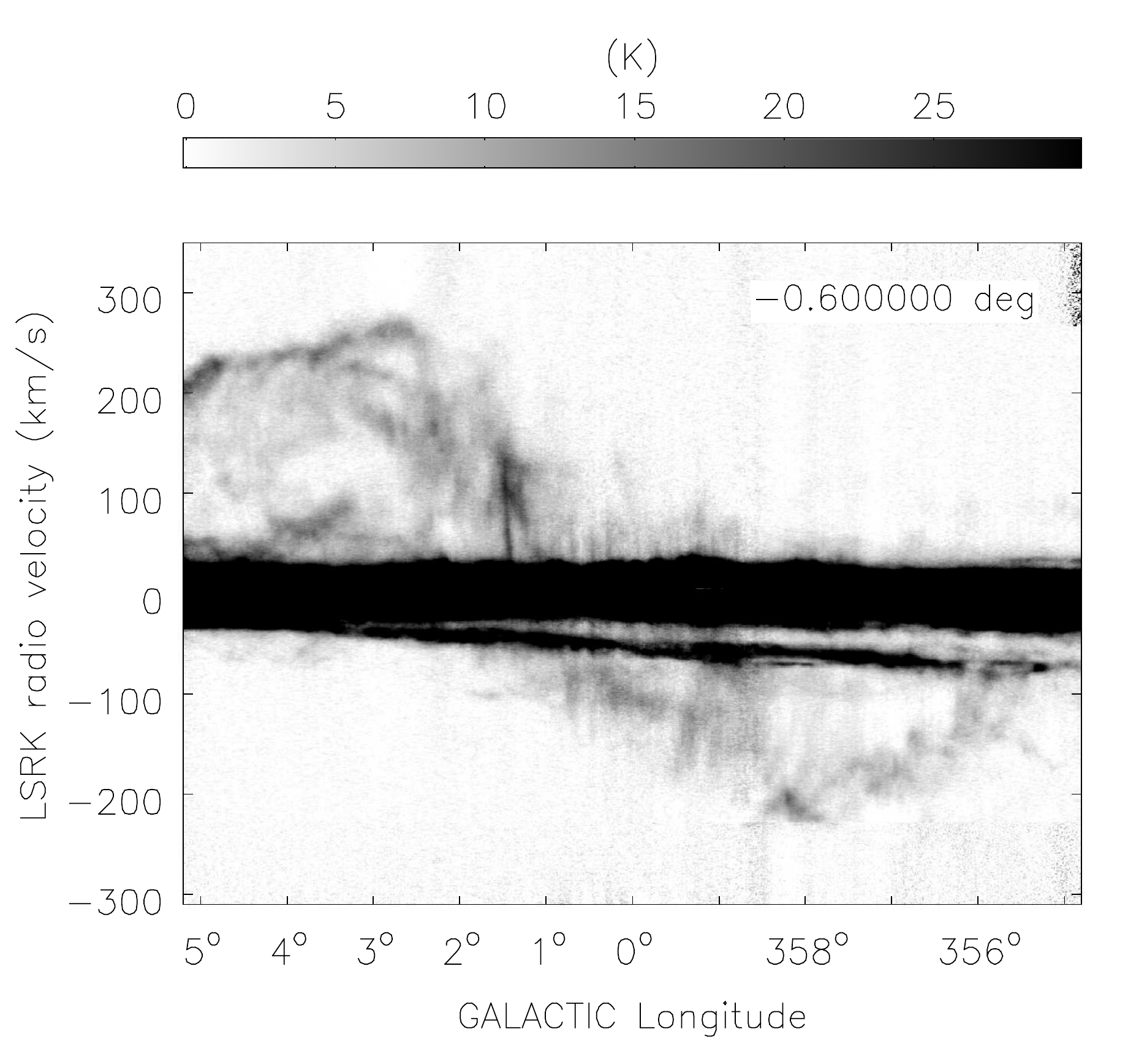}{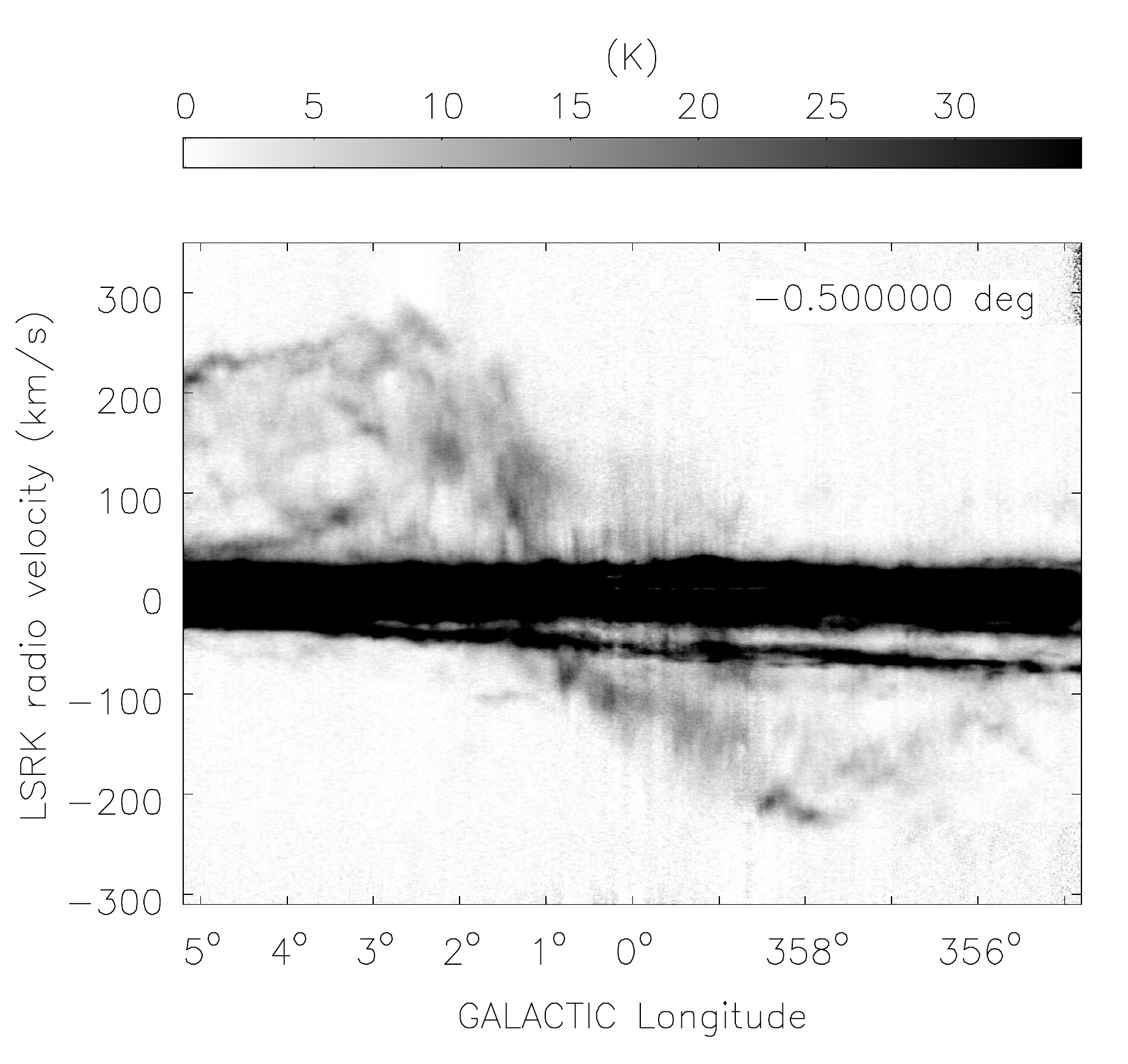}\\
\plottwo{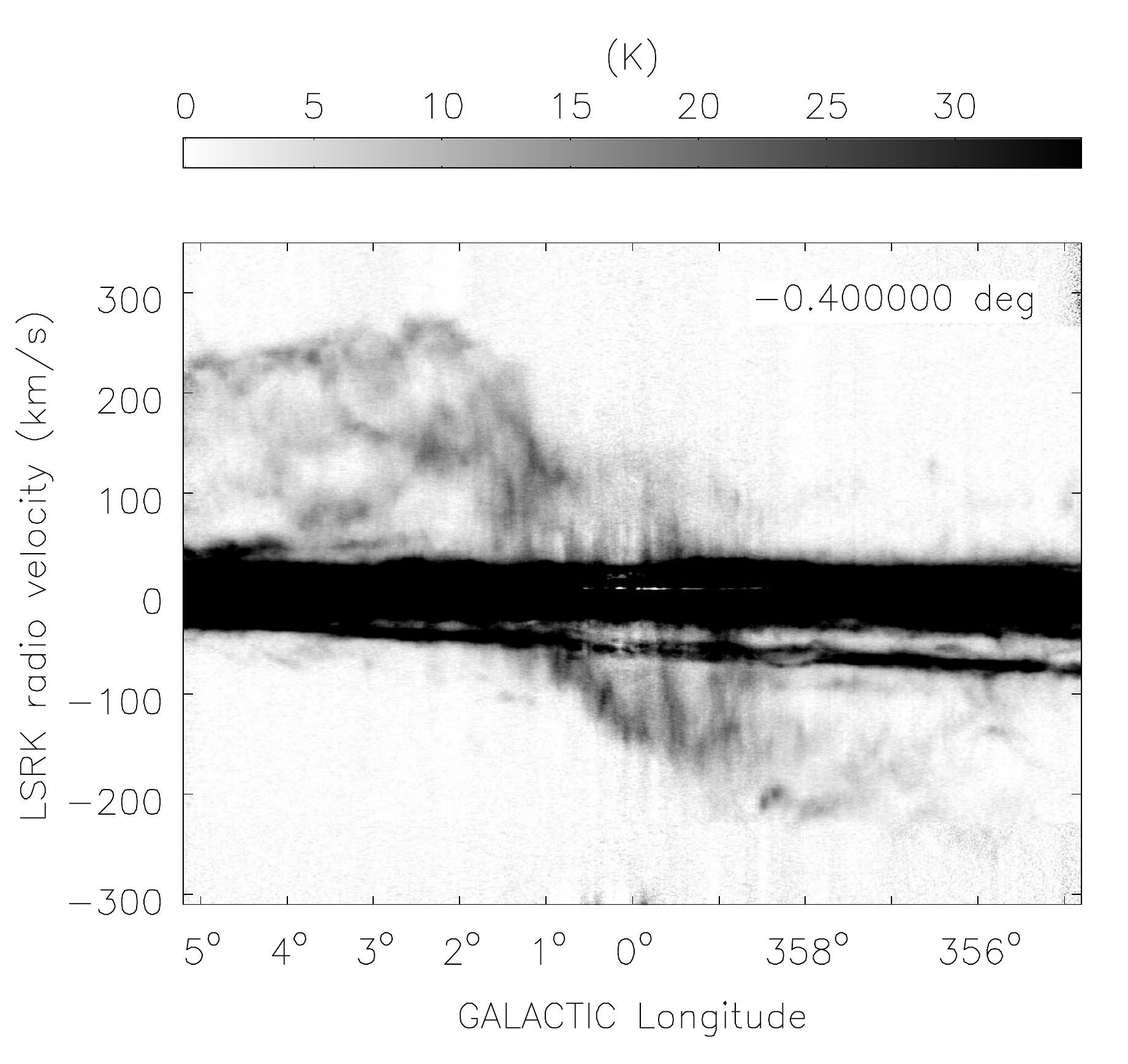}{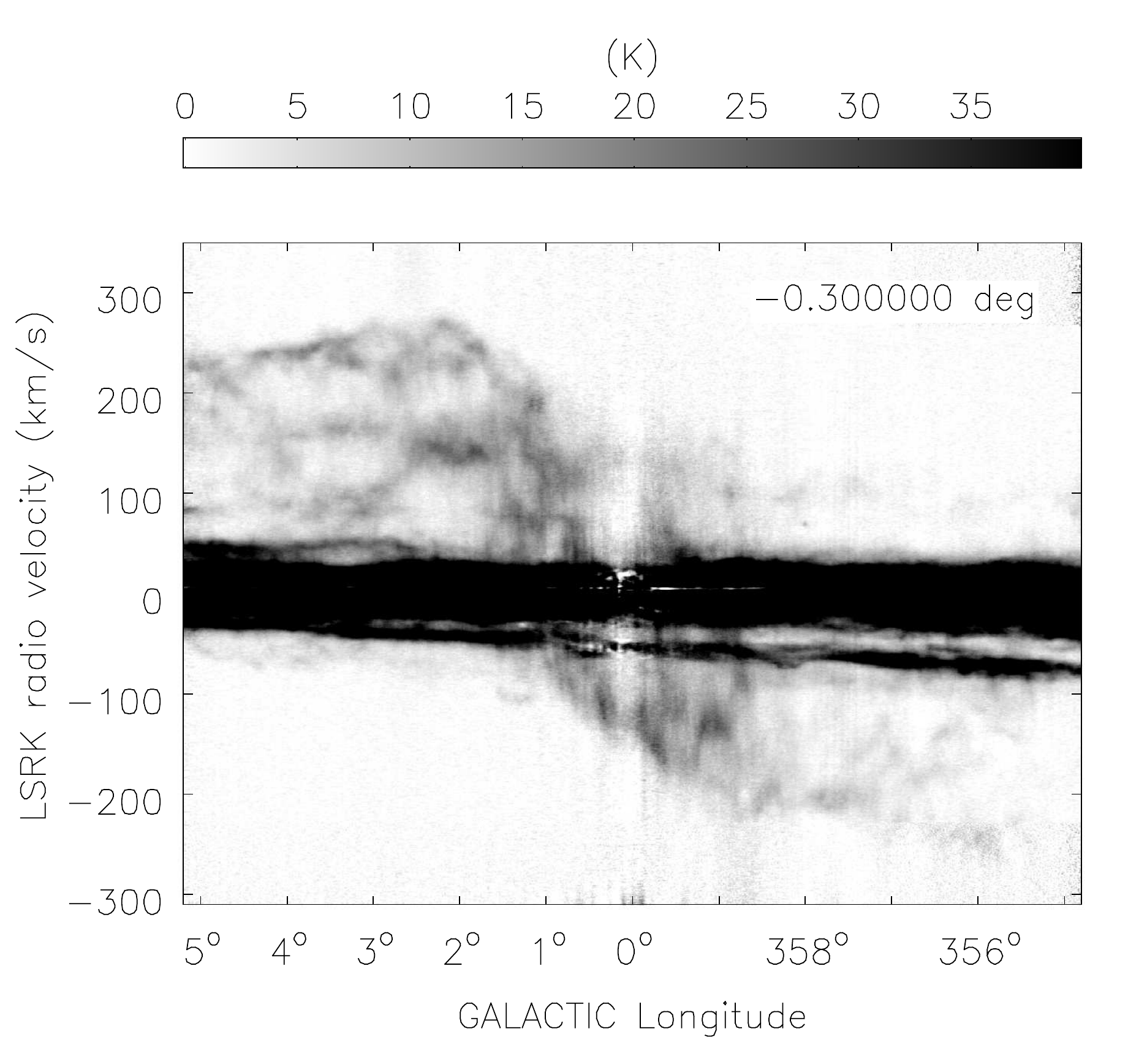}\\
 \caption[]{
}
\end{figure}

\begin{figure}
\figurenum{5c}
\centering
\plottwo{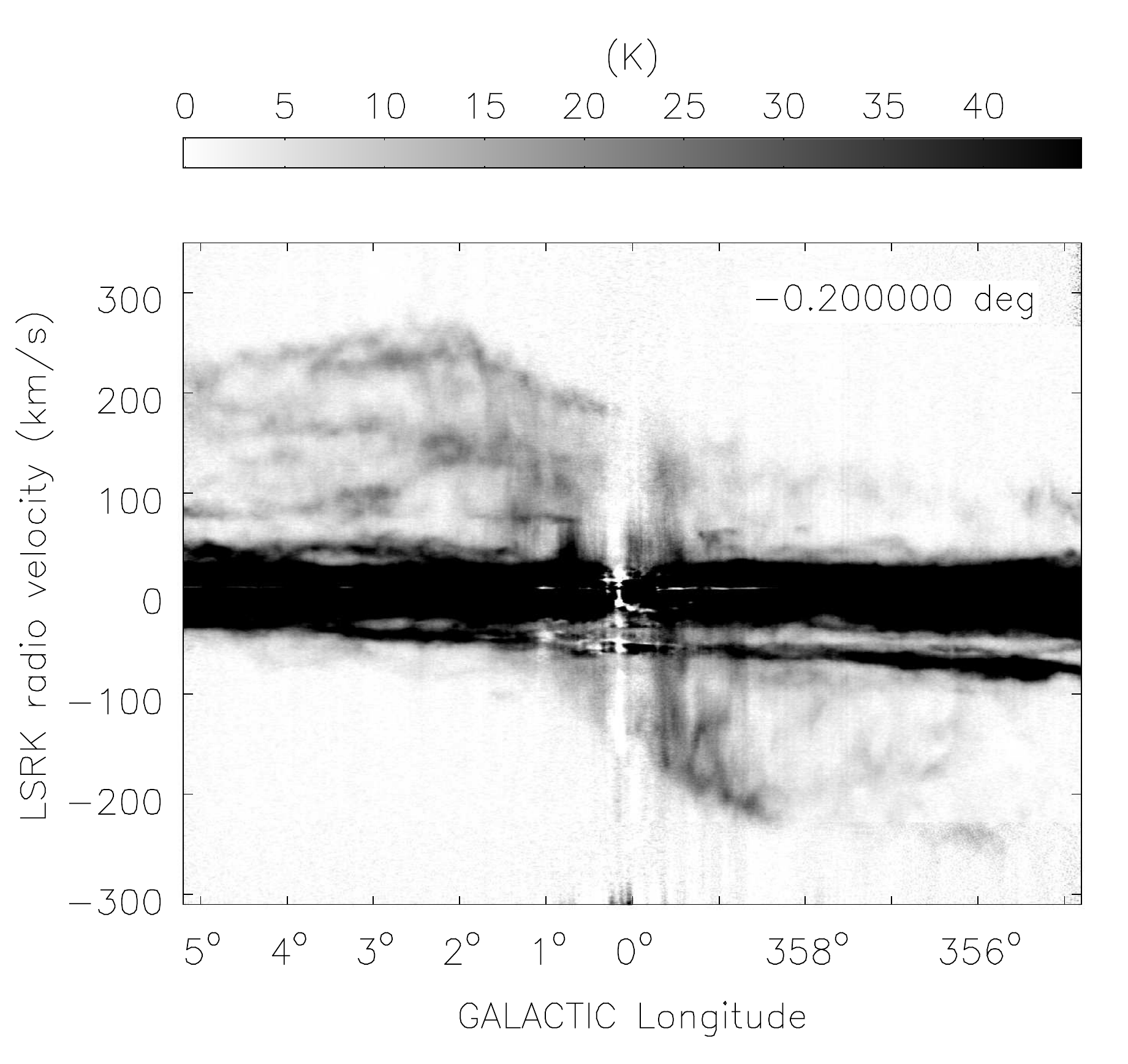}{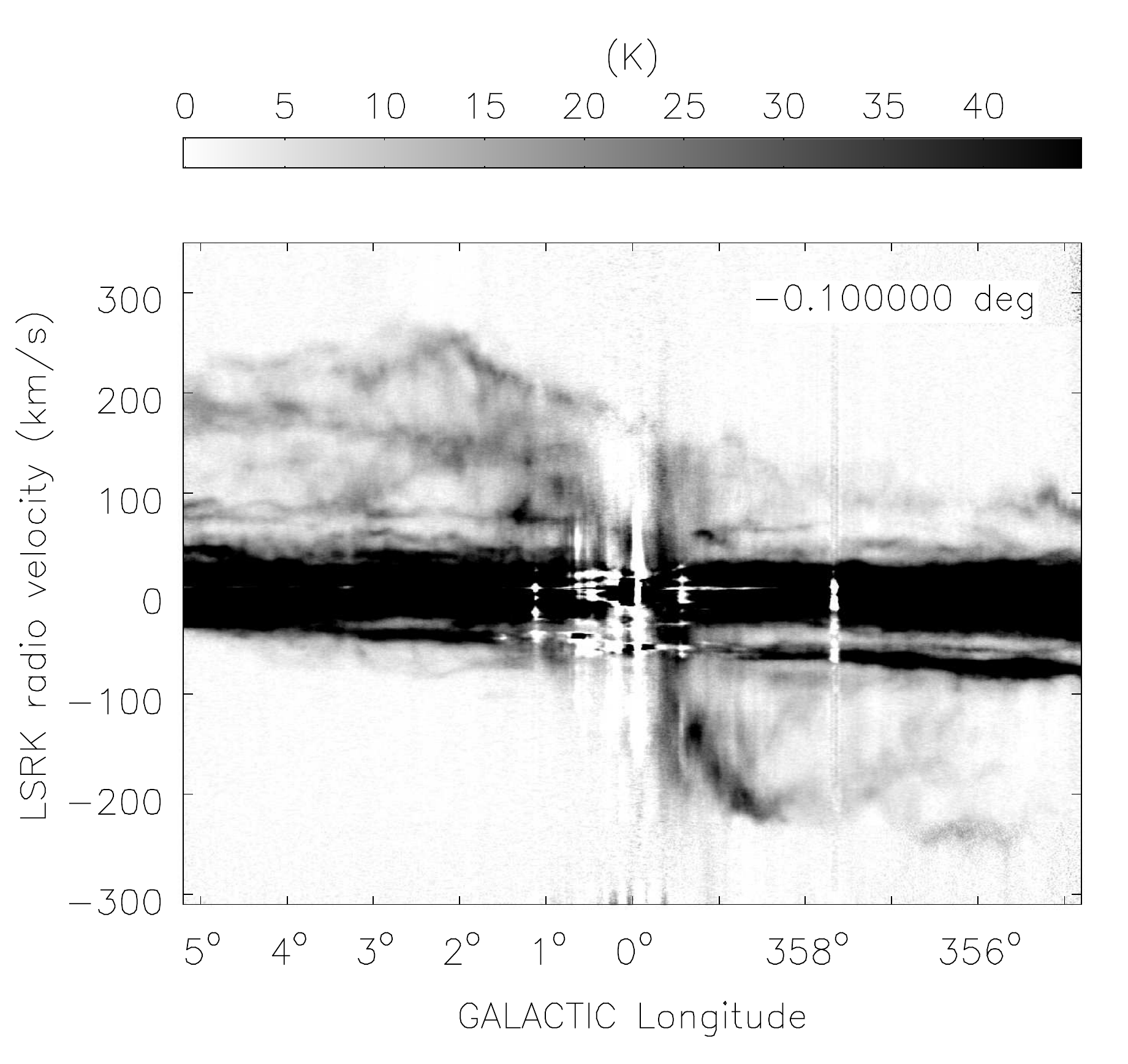}\\
\plottwo{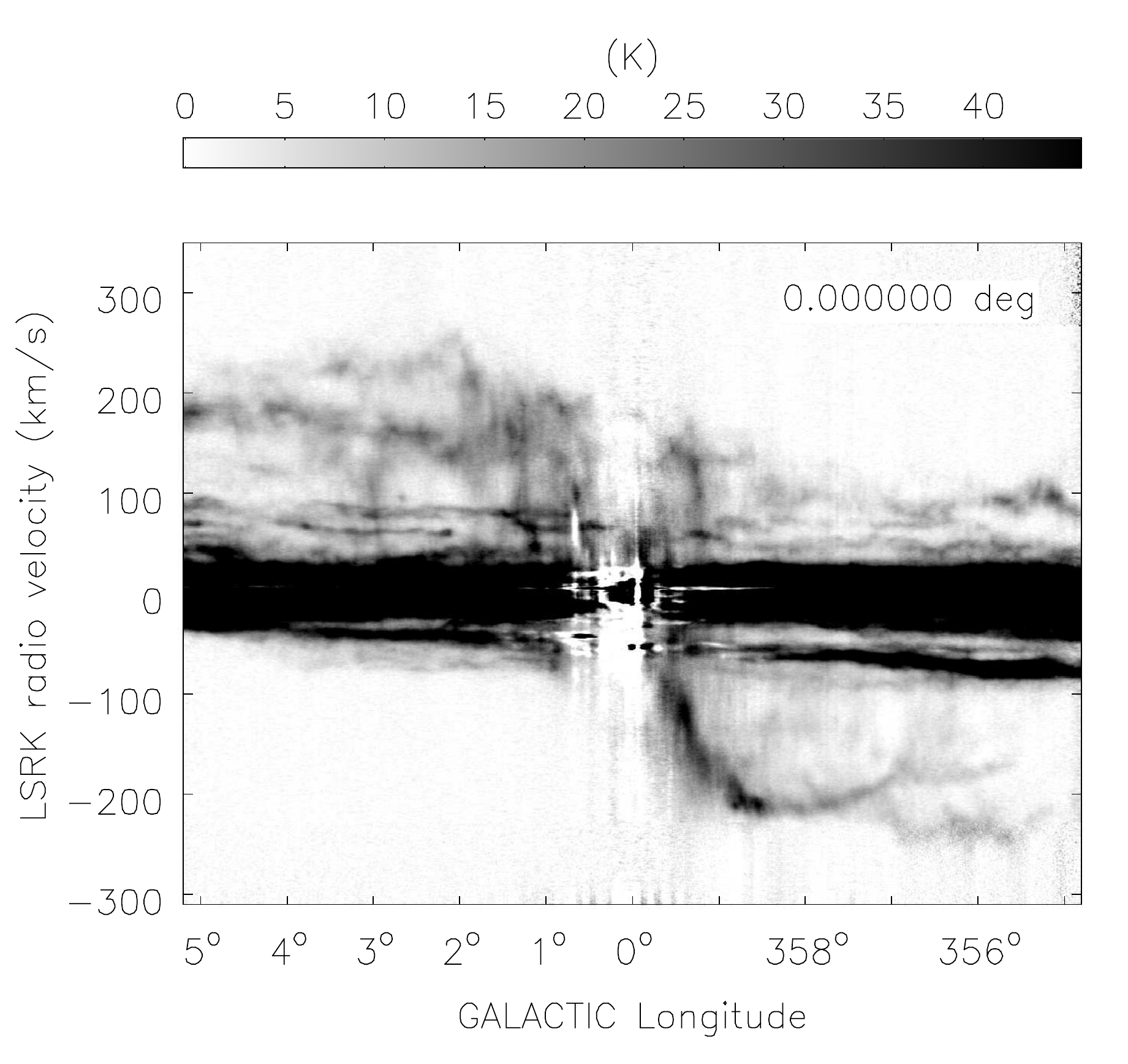}{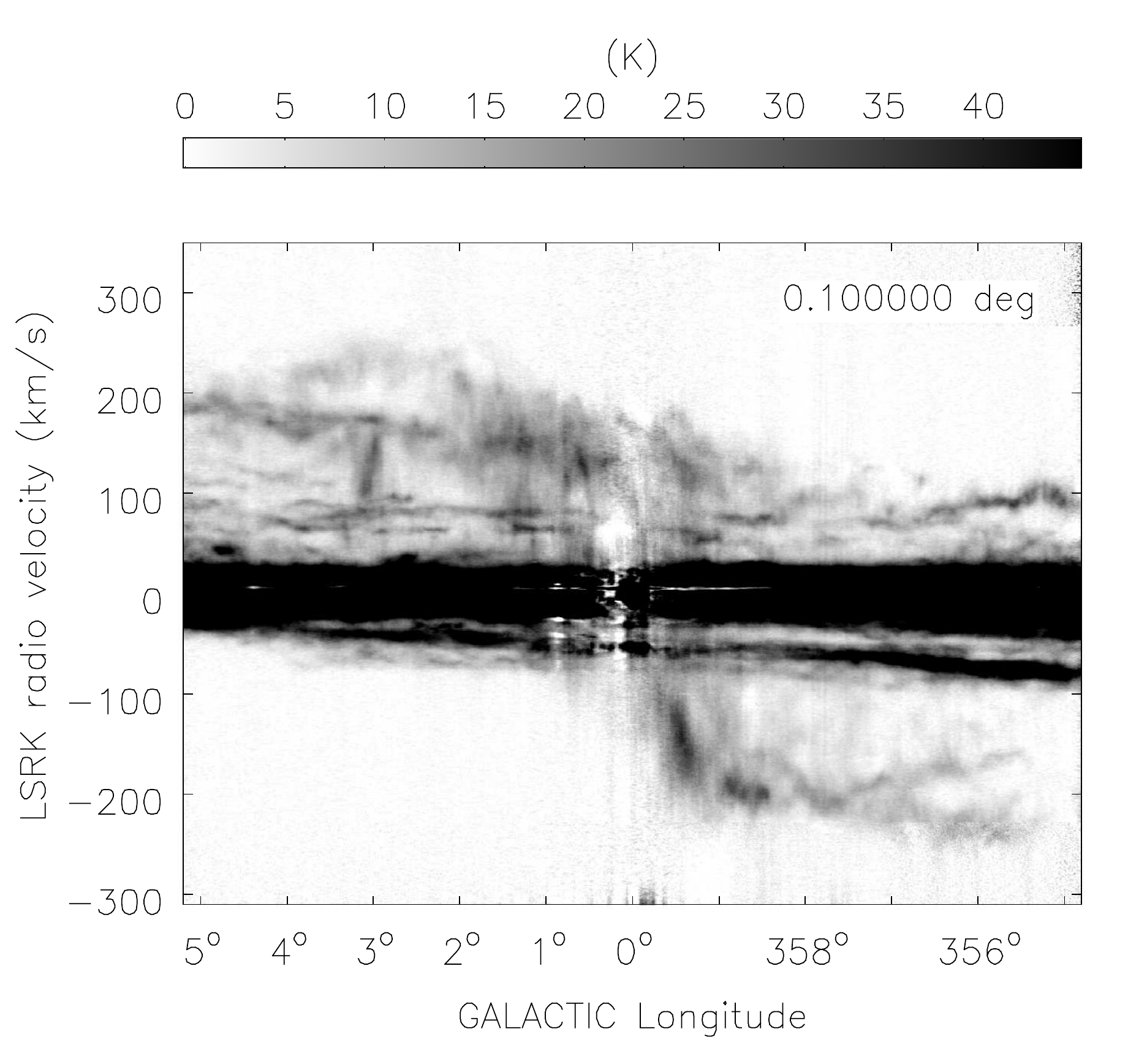}\\
\plottwo{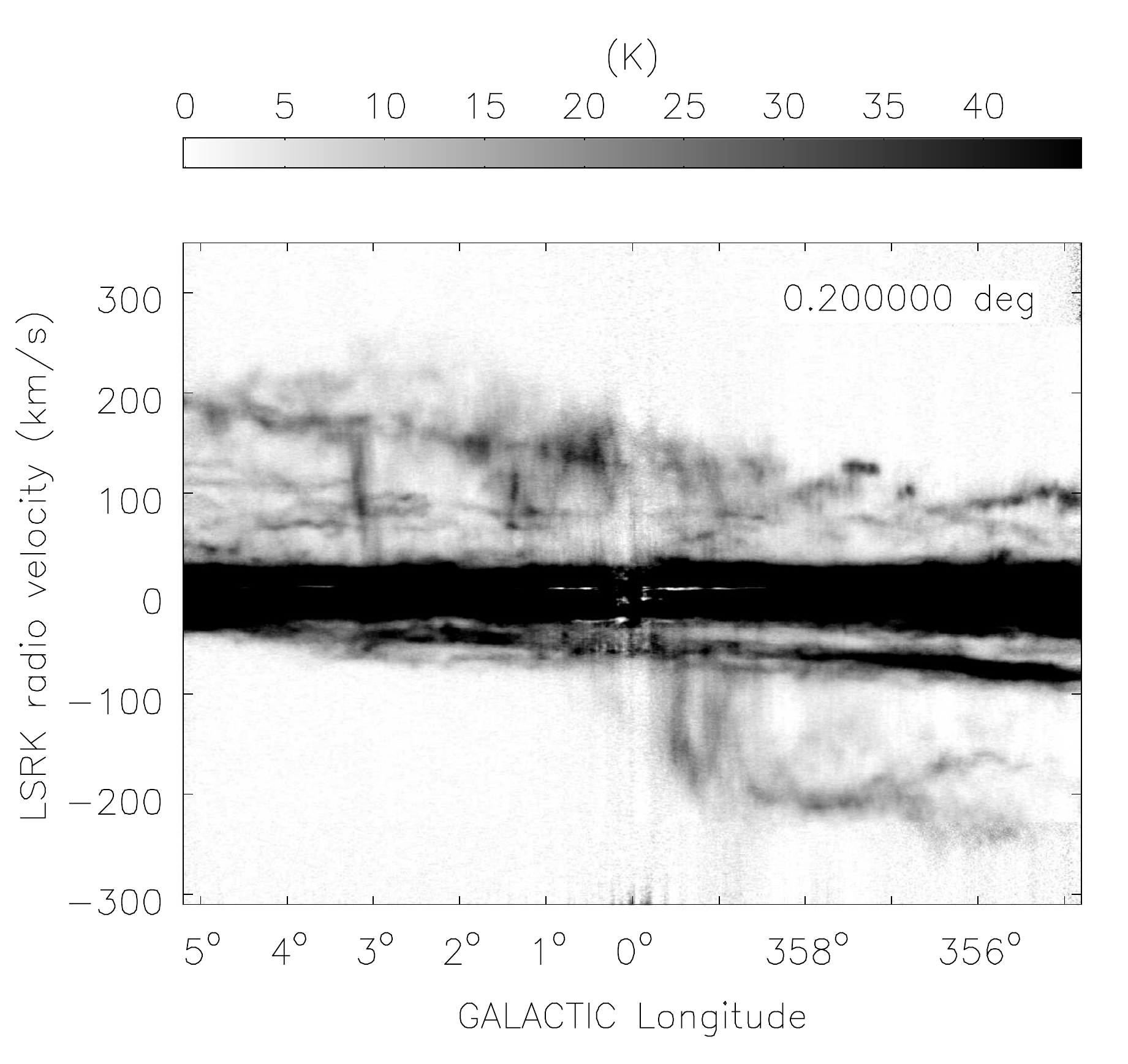}{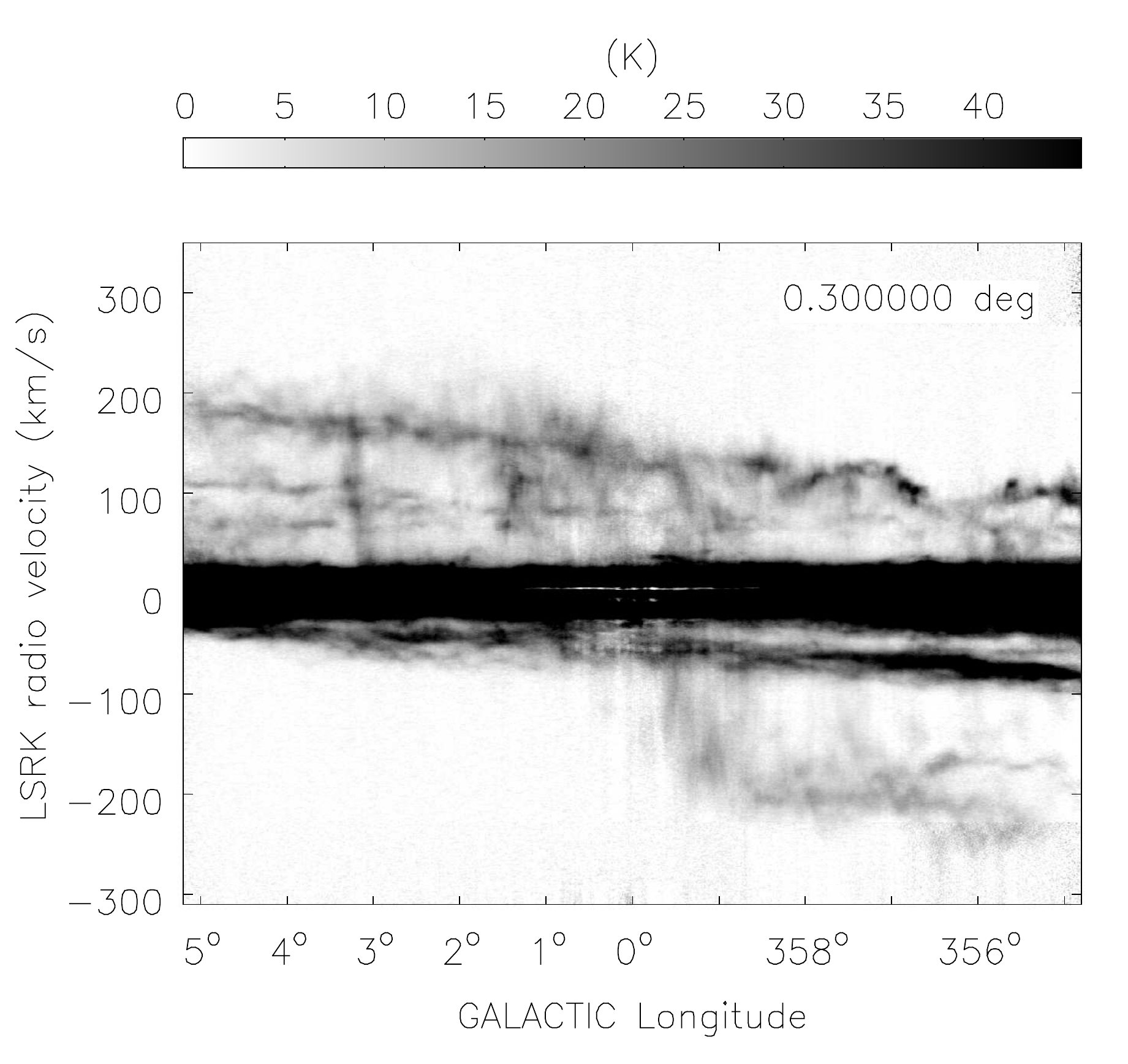}\\
 \caption[]{
}
\end{figure}

\begin{figure}
\figurenum{5d}
\centering
\plottwo{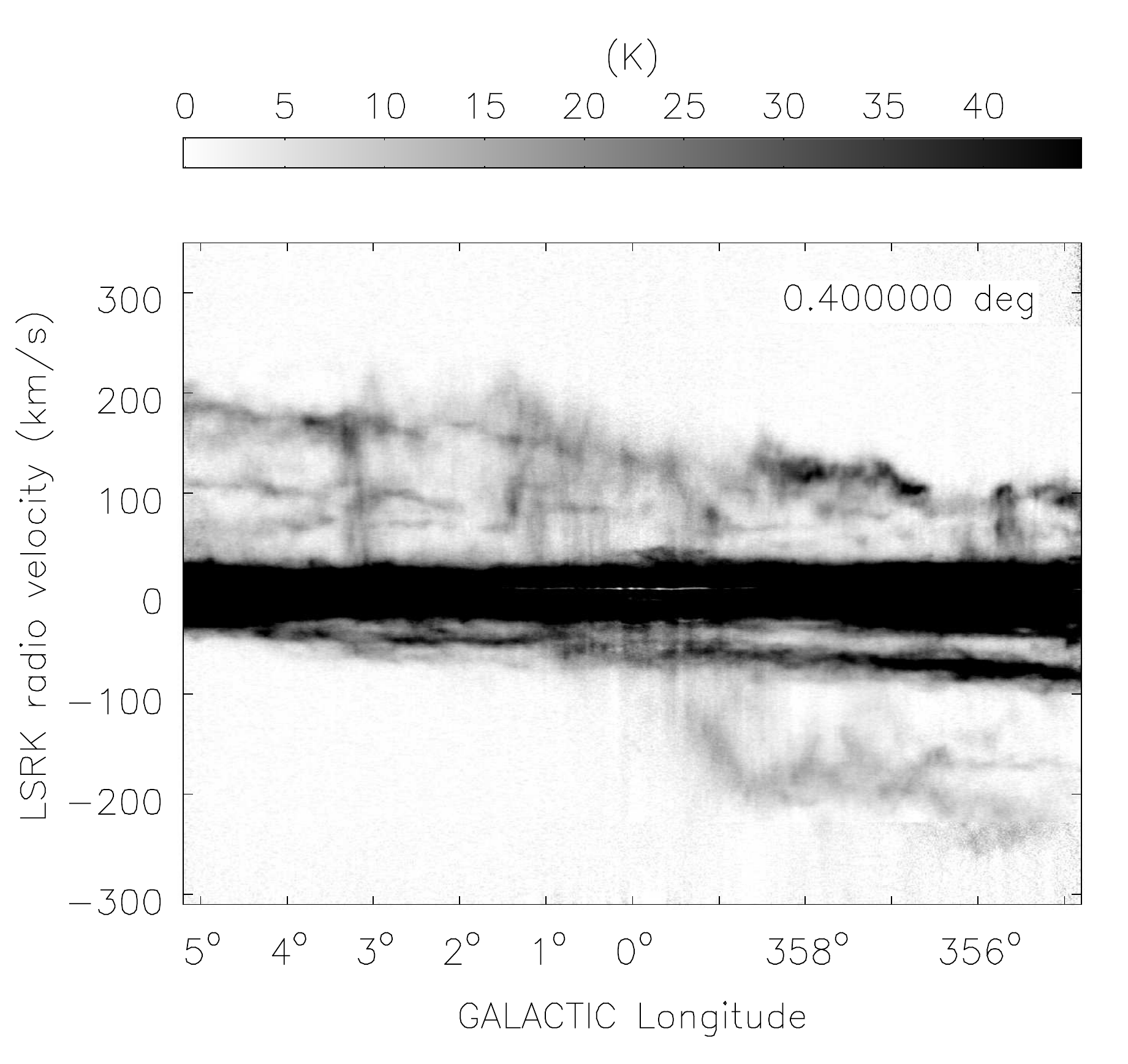}{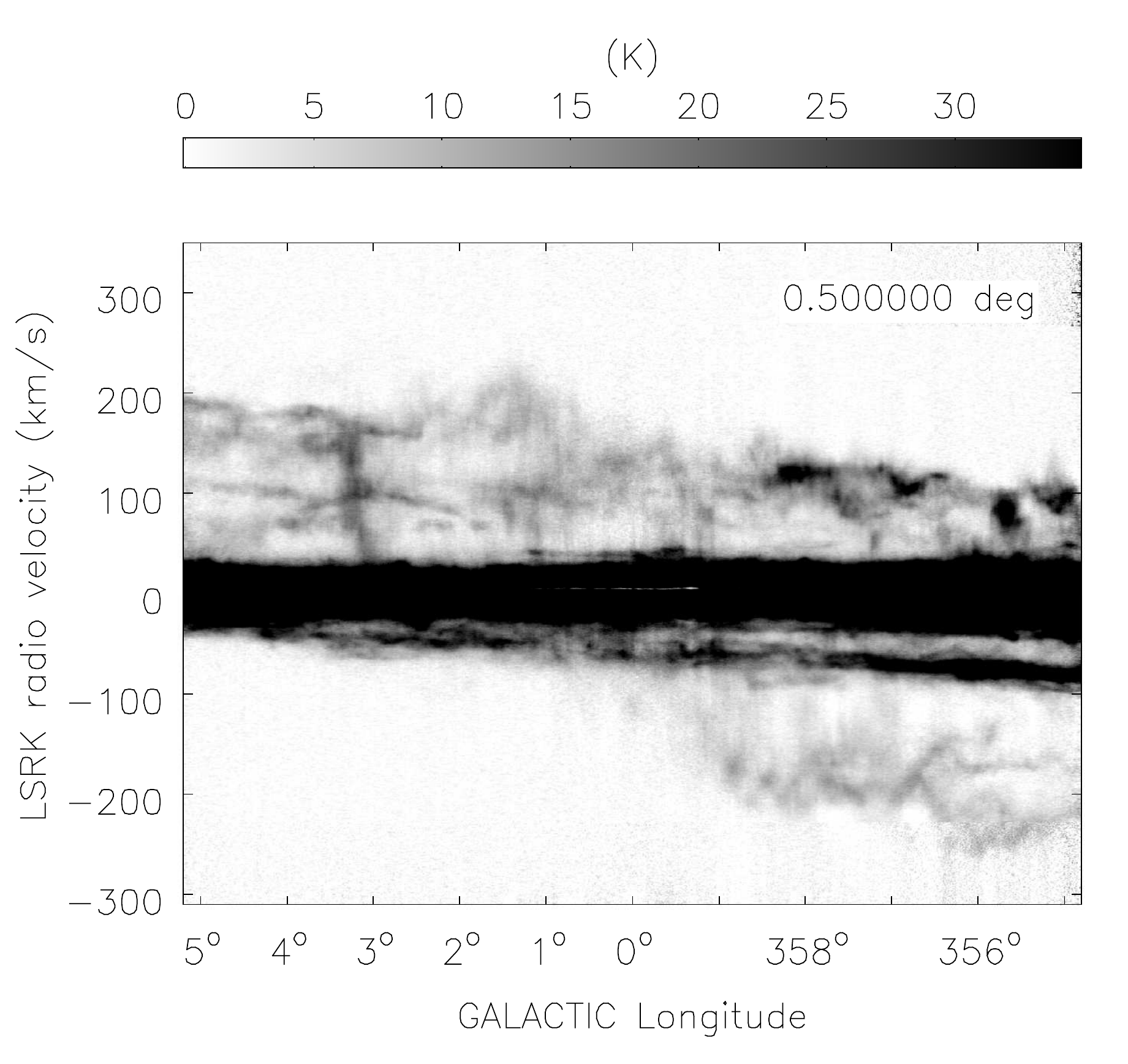}\\
\plottwo{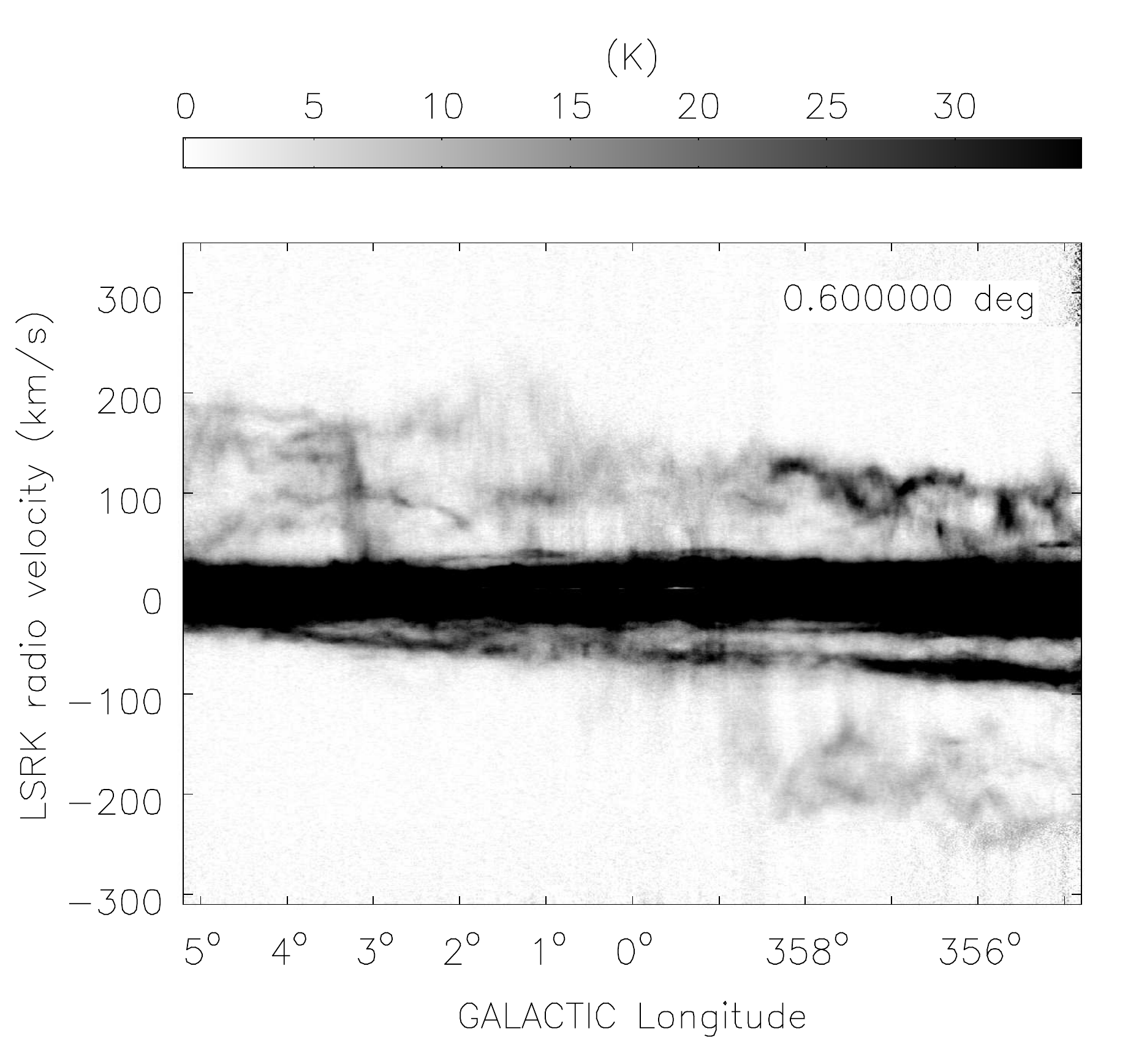}{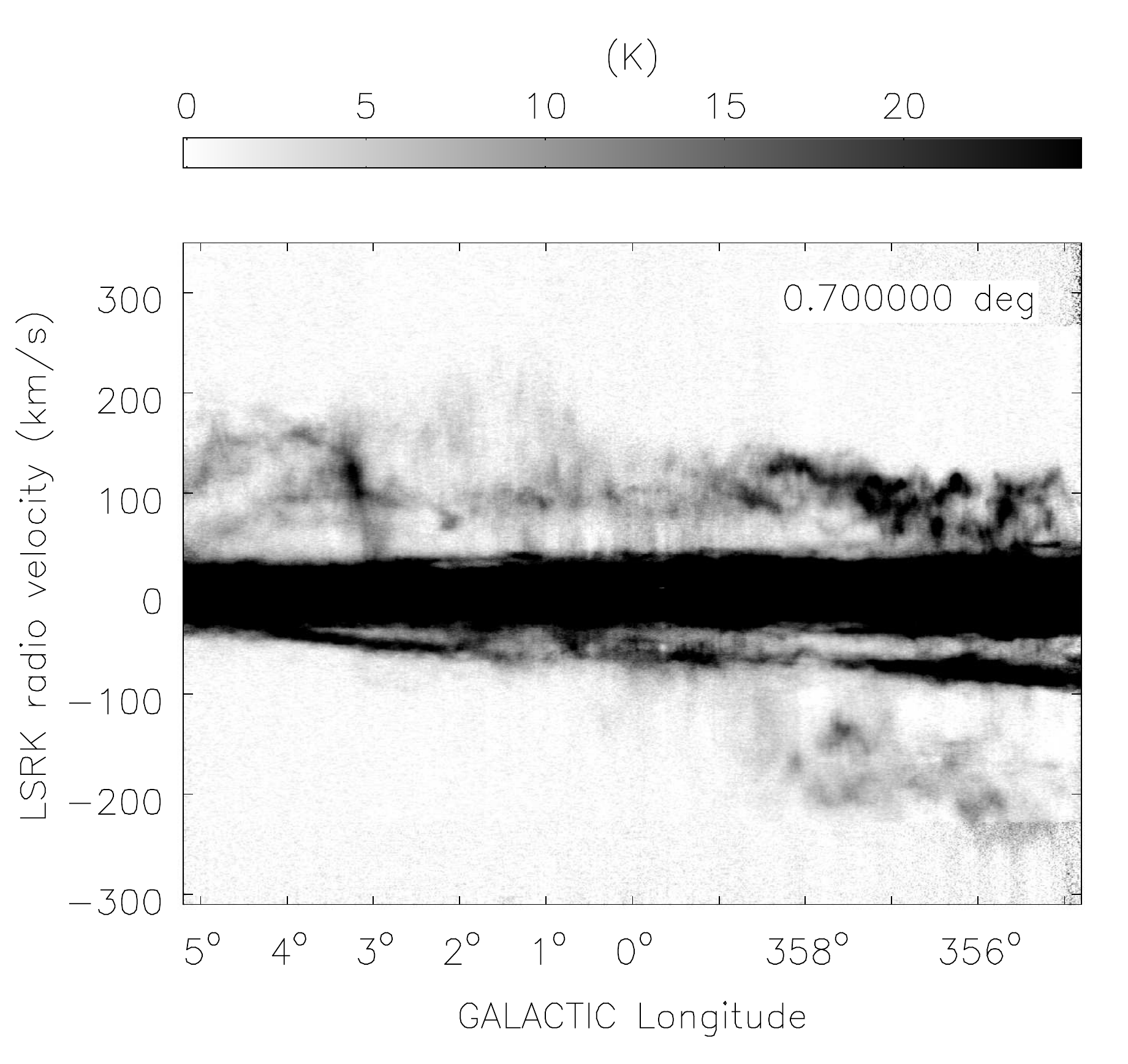}\\
\plottwo{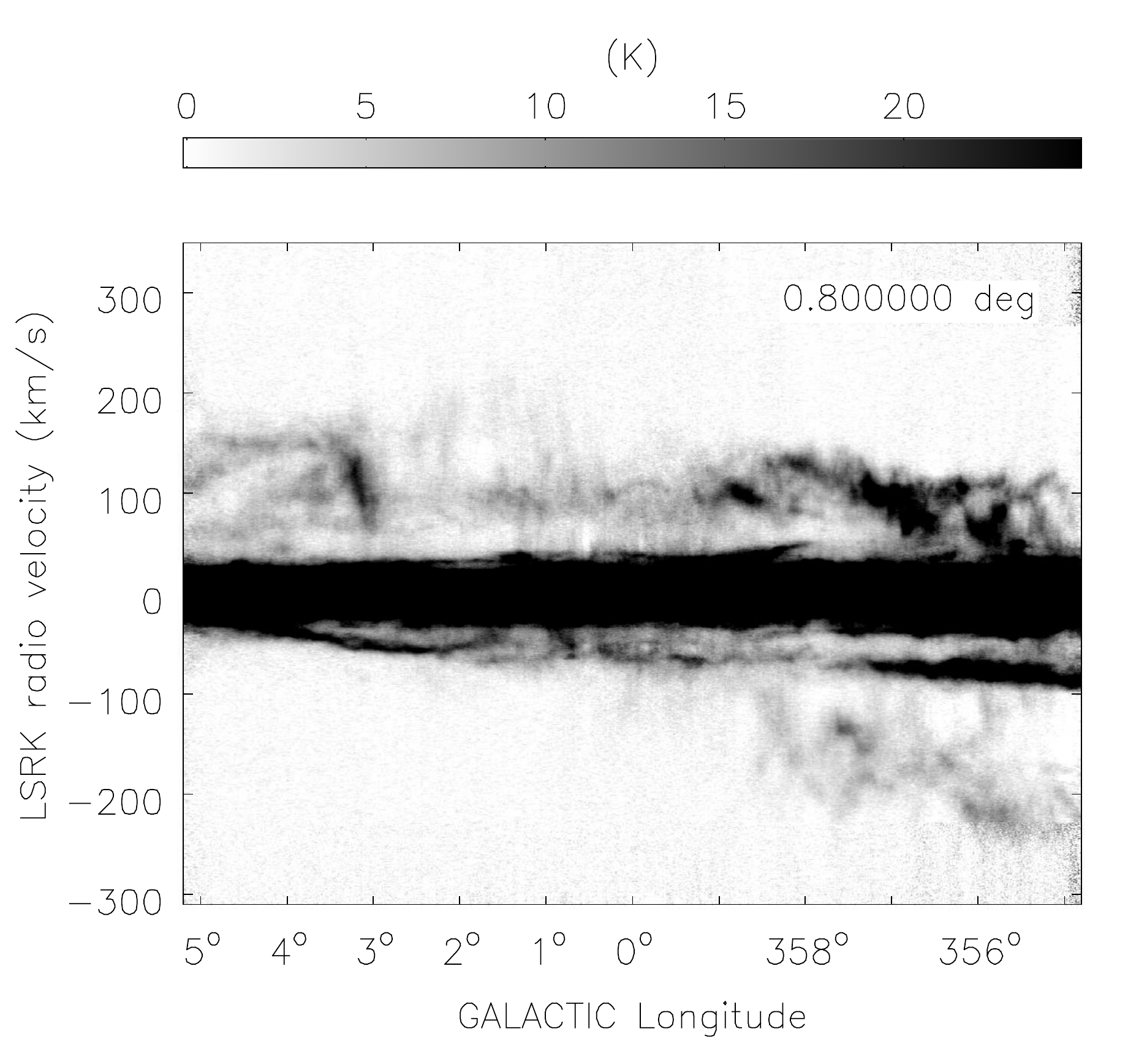}{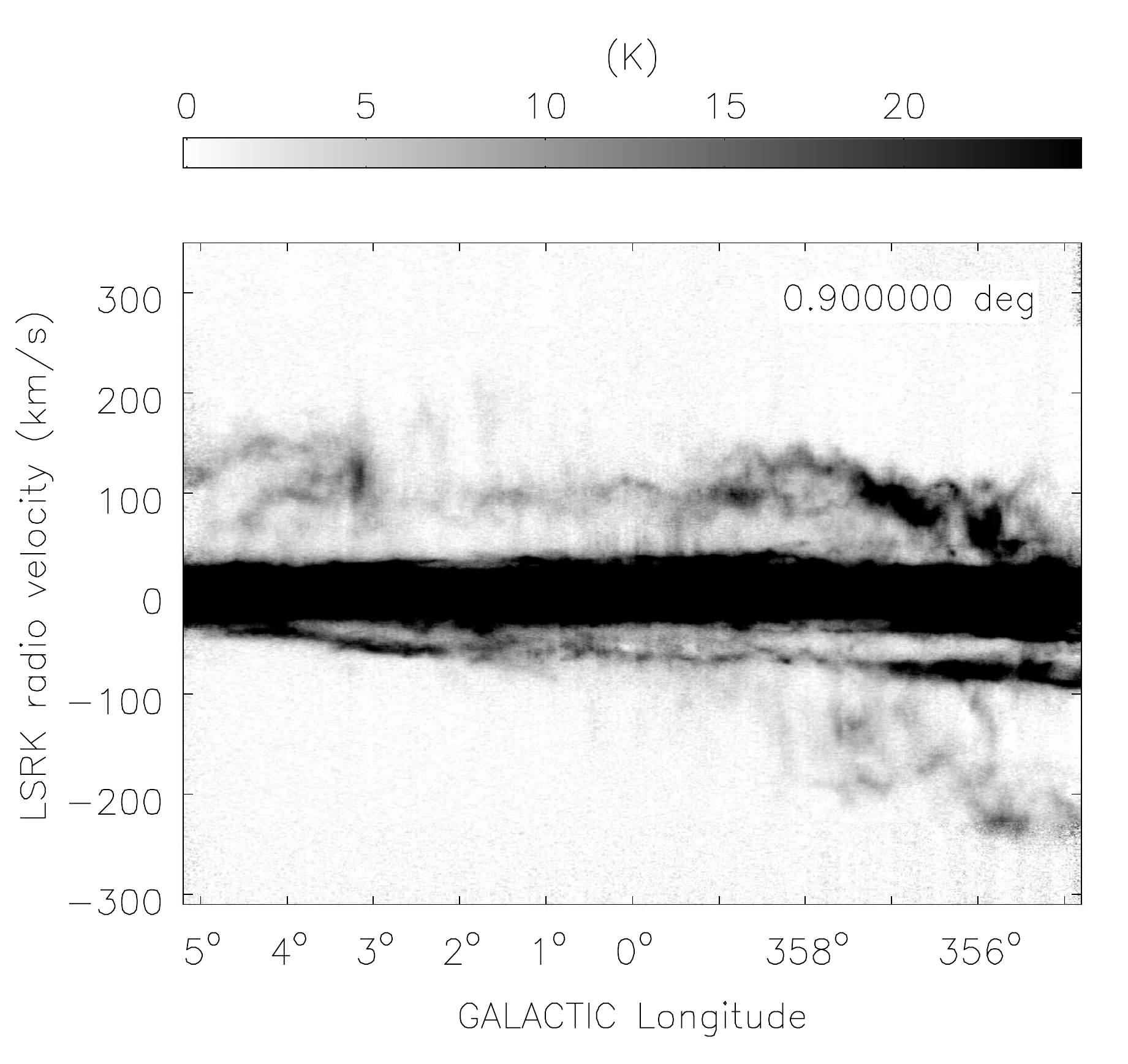}\\
 \caption[]{
}
\end{figure}

\begin{figure}
\figurenum{5e}
\centering
\plottwo{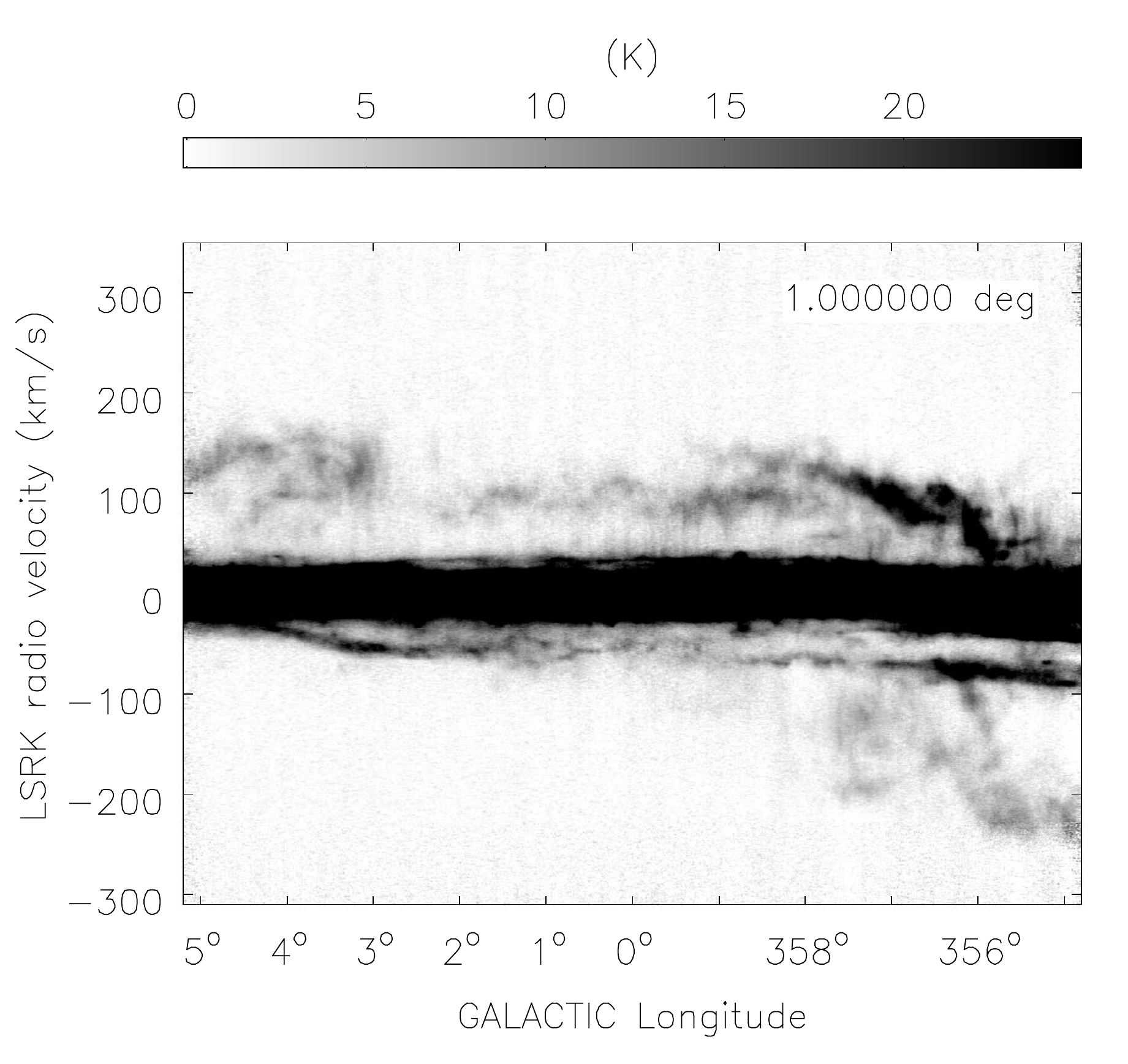}{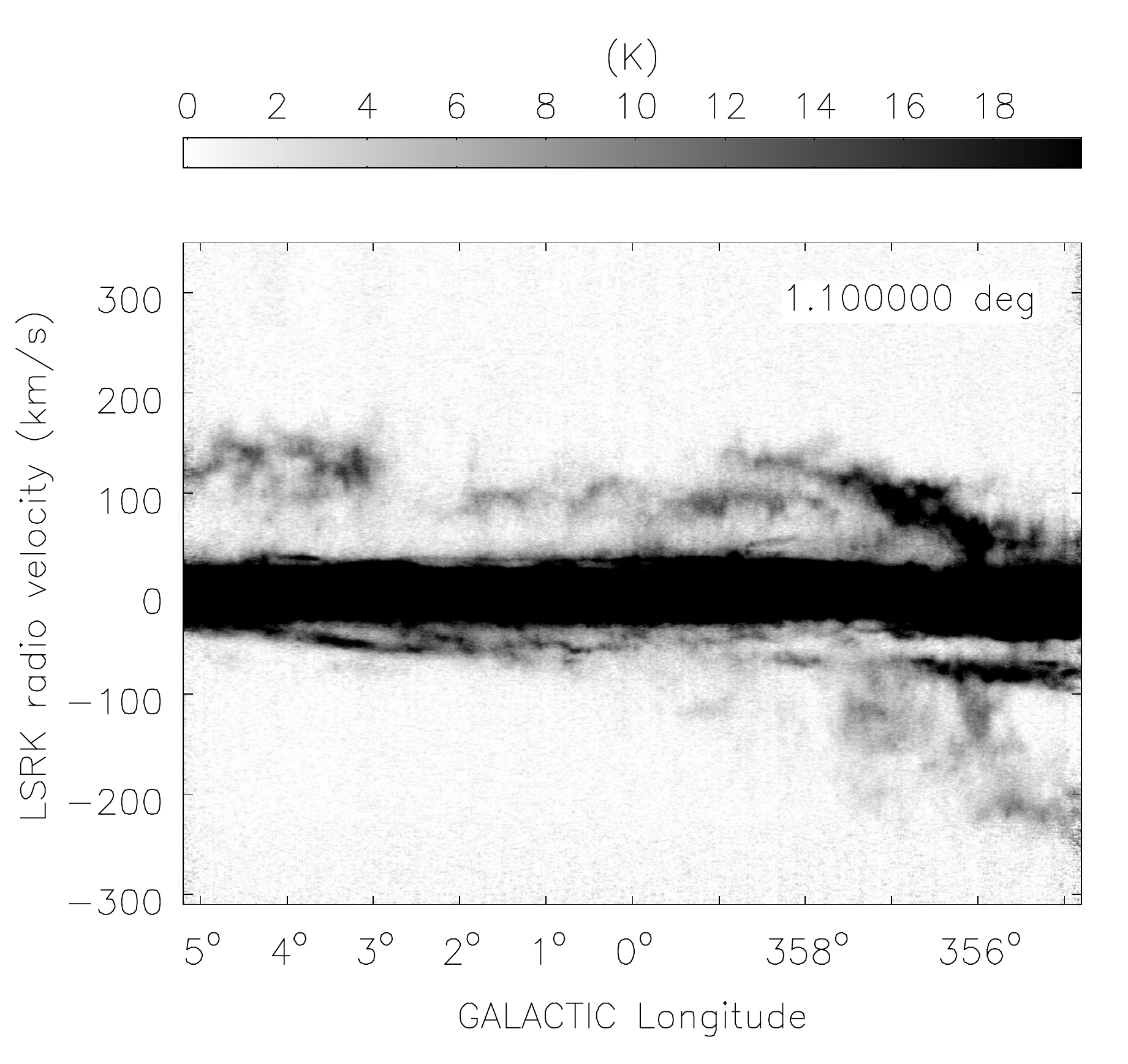}\\
\plottwo{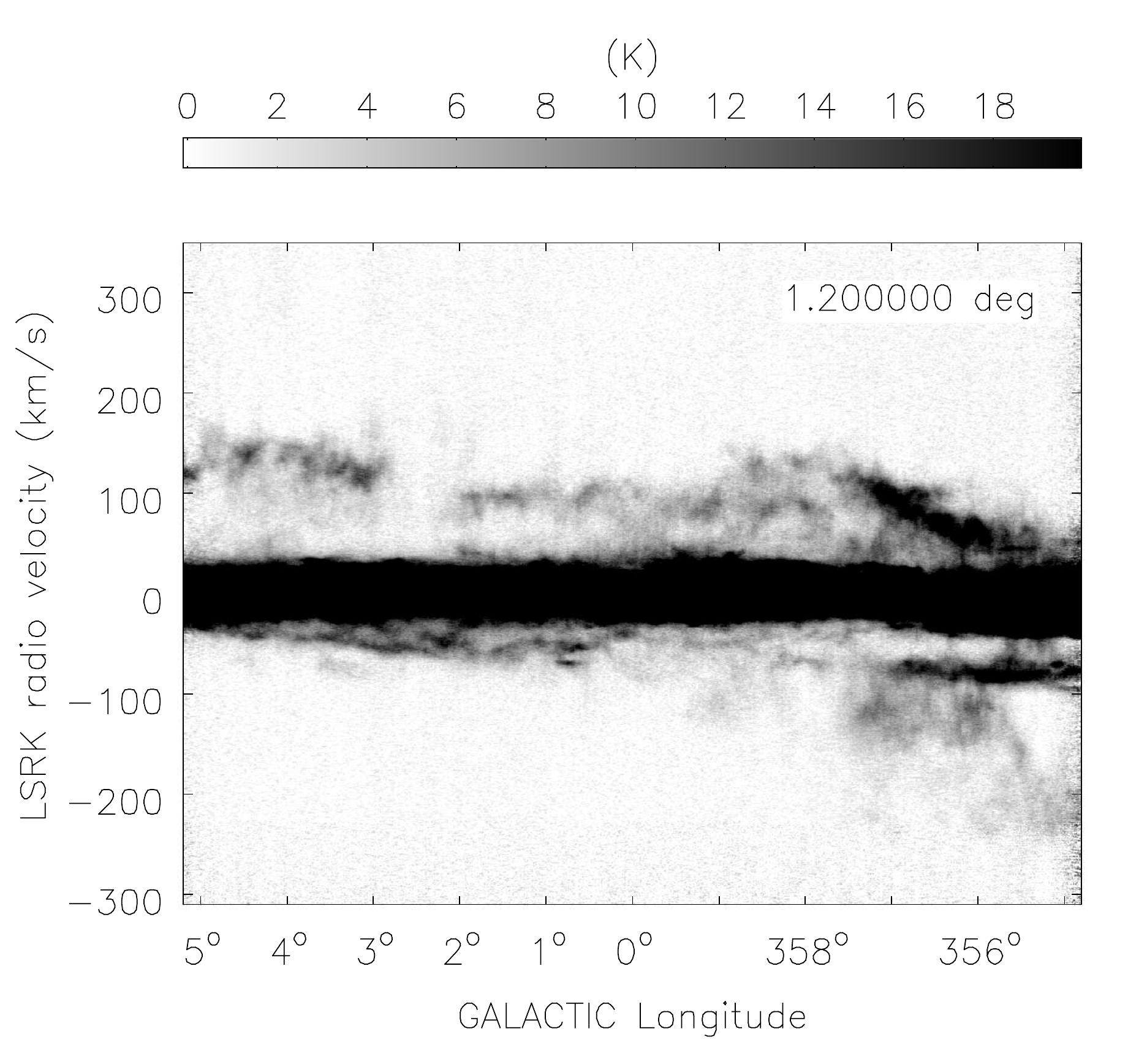}{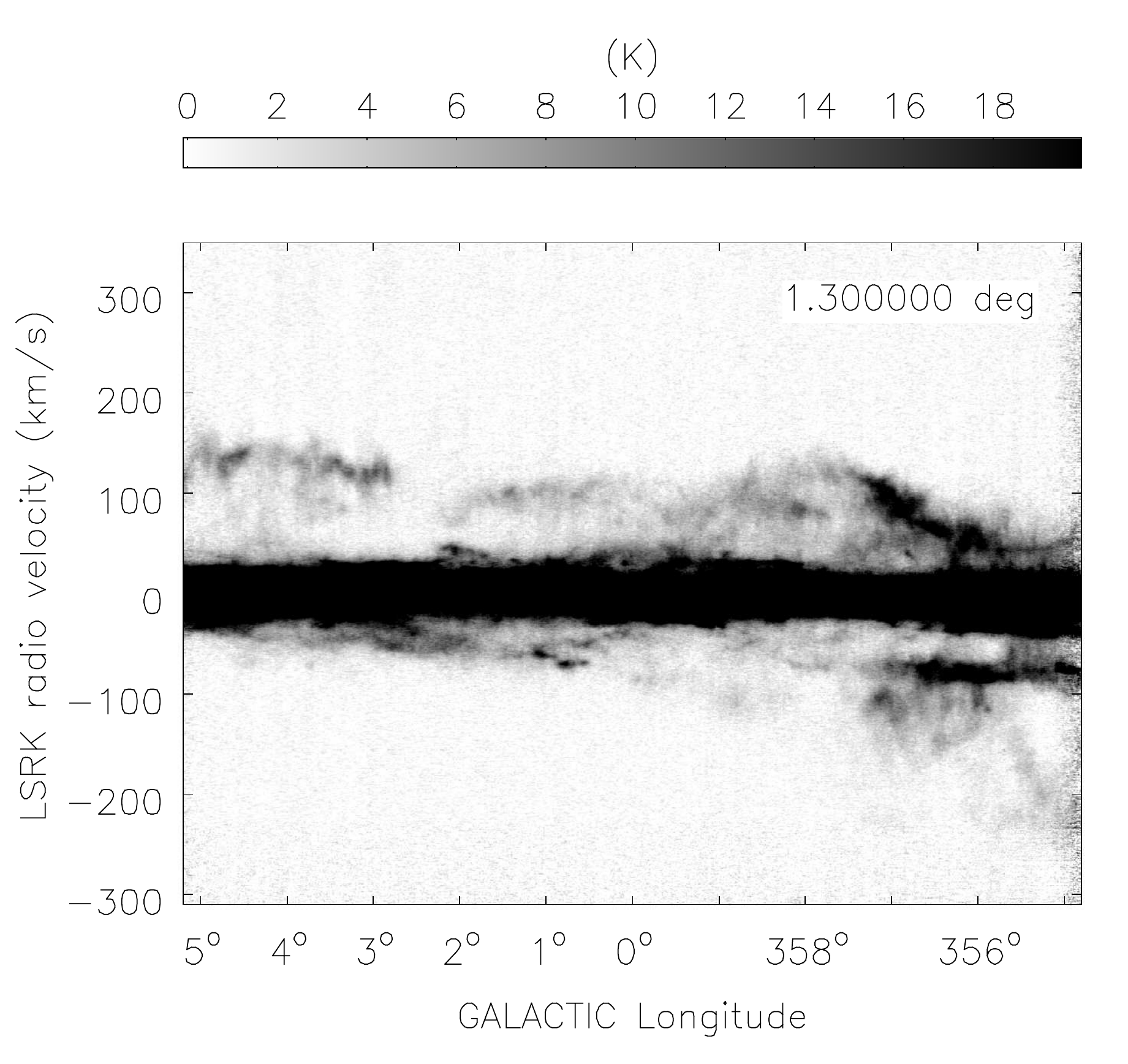}\\
\plottwo{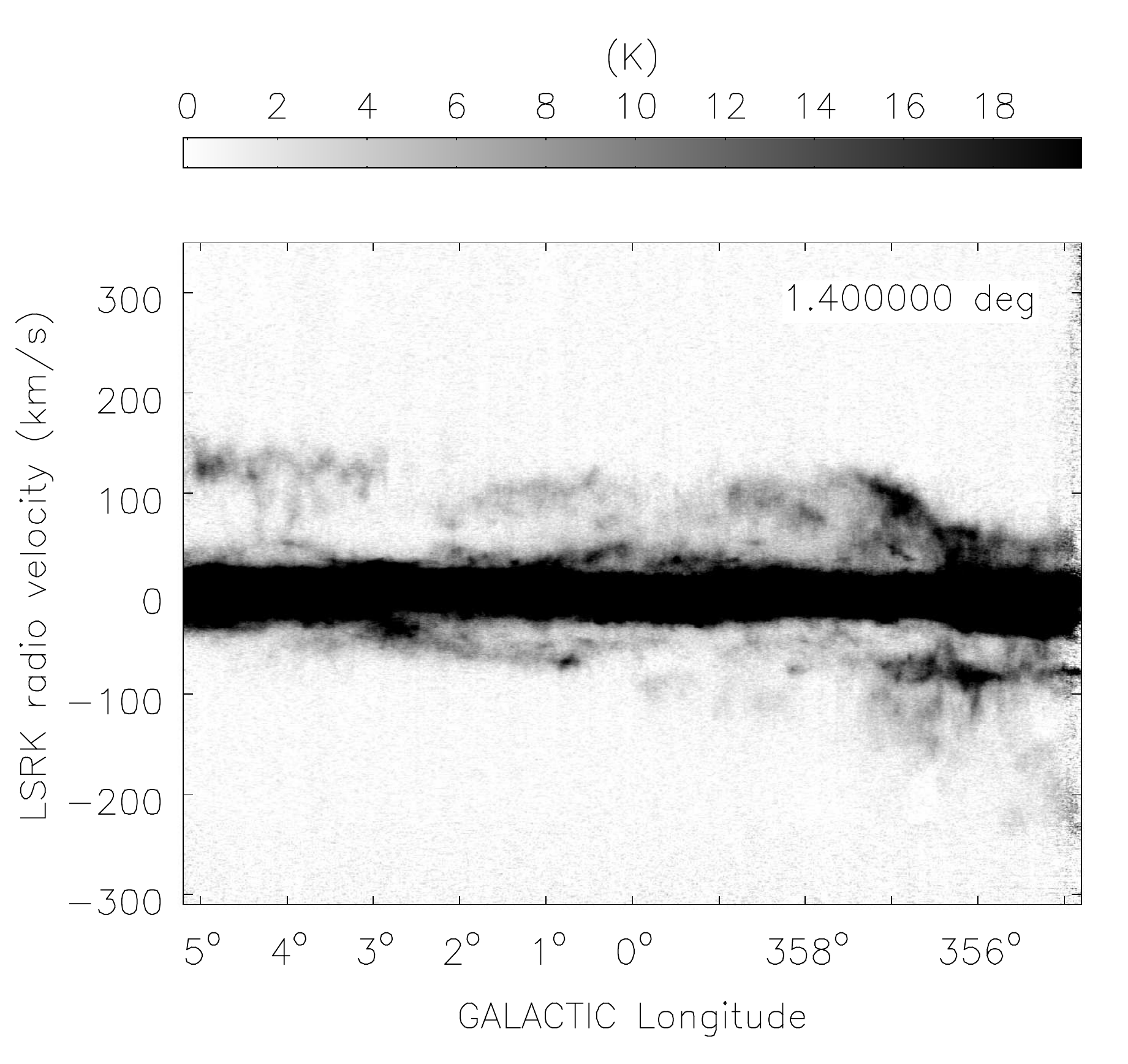}{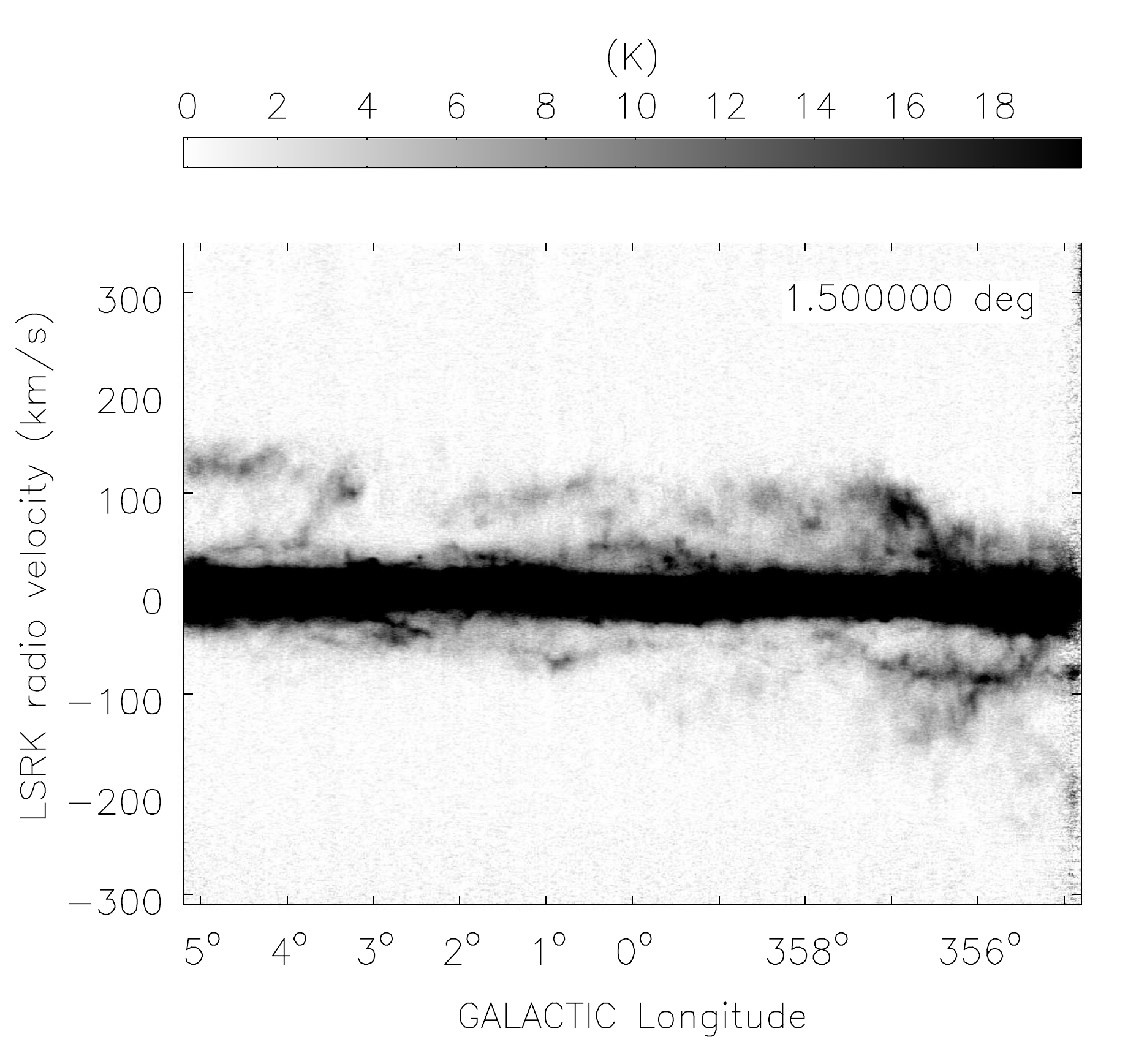}\\
 \caption[]{
}
\end{figure}
\clearpage

\begin{figure}
\centering
\includegraphics[width=6in]{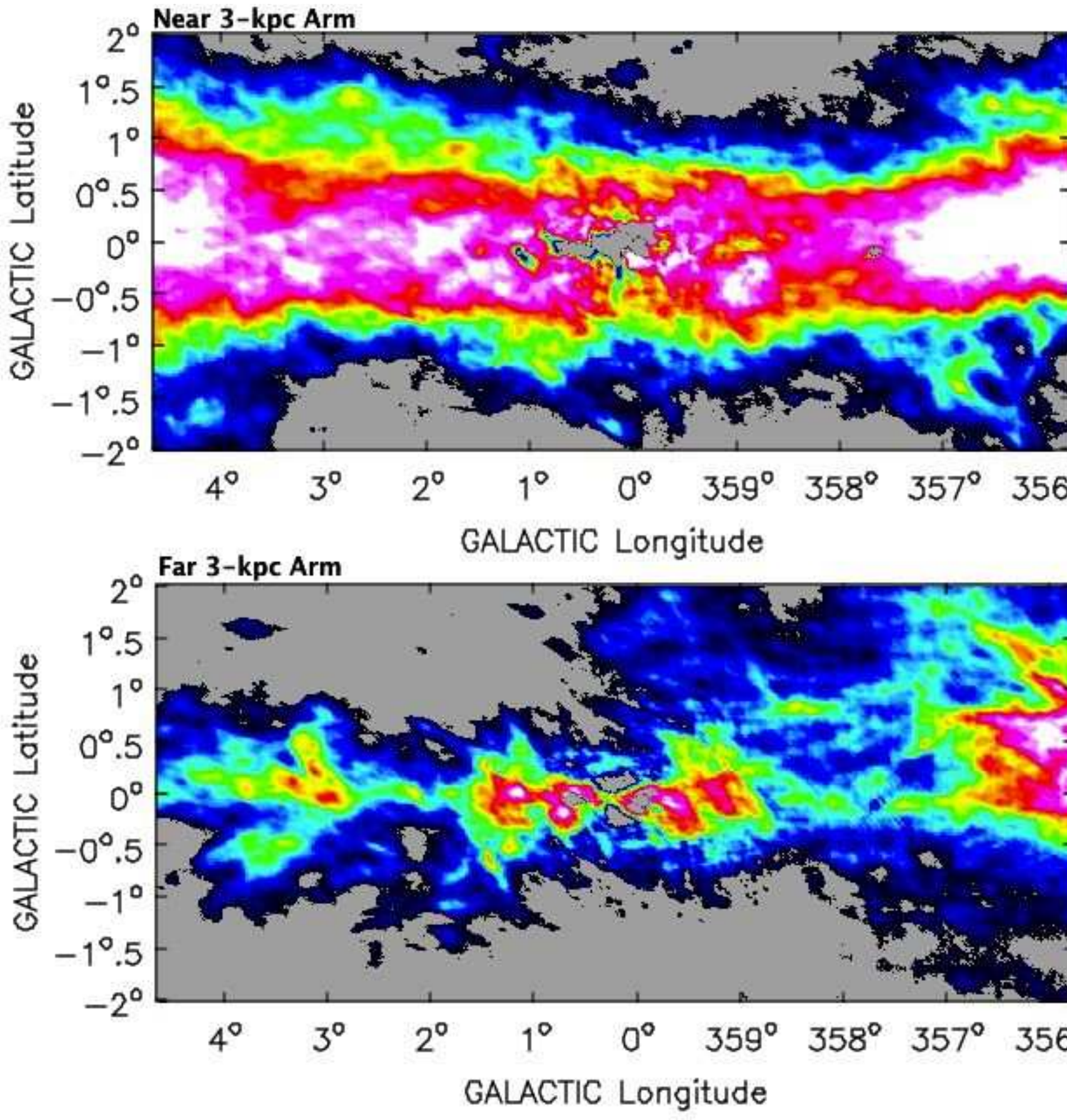}
\caption[]{Velocity integrated \HI\ intensity maps of the near (top) and
  far (bottom) 3-kpc arms following the definitions of
  \citet{dame08}. Emission was integrated over 26 km/s bins, centered
  at each longitude on the arm at a velocity as defined by the linear
  fits of \citet{dame08}: $v_{lsr}=-53.1 + 4.16l$ \kms\ for the near
  arm and $v_{lsr}=+56.0 + 4.08l$ \kms\ for the far arm.
  \label{fig:3kpcarms}}
\end{figure}

\begin{figure}
\centering
\includegraphics[width=6in]{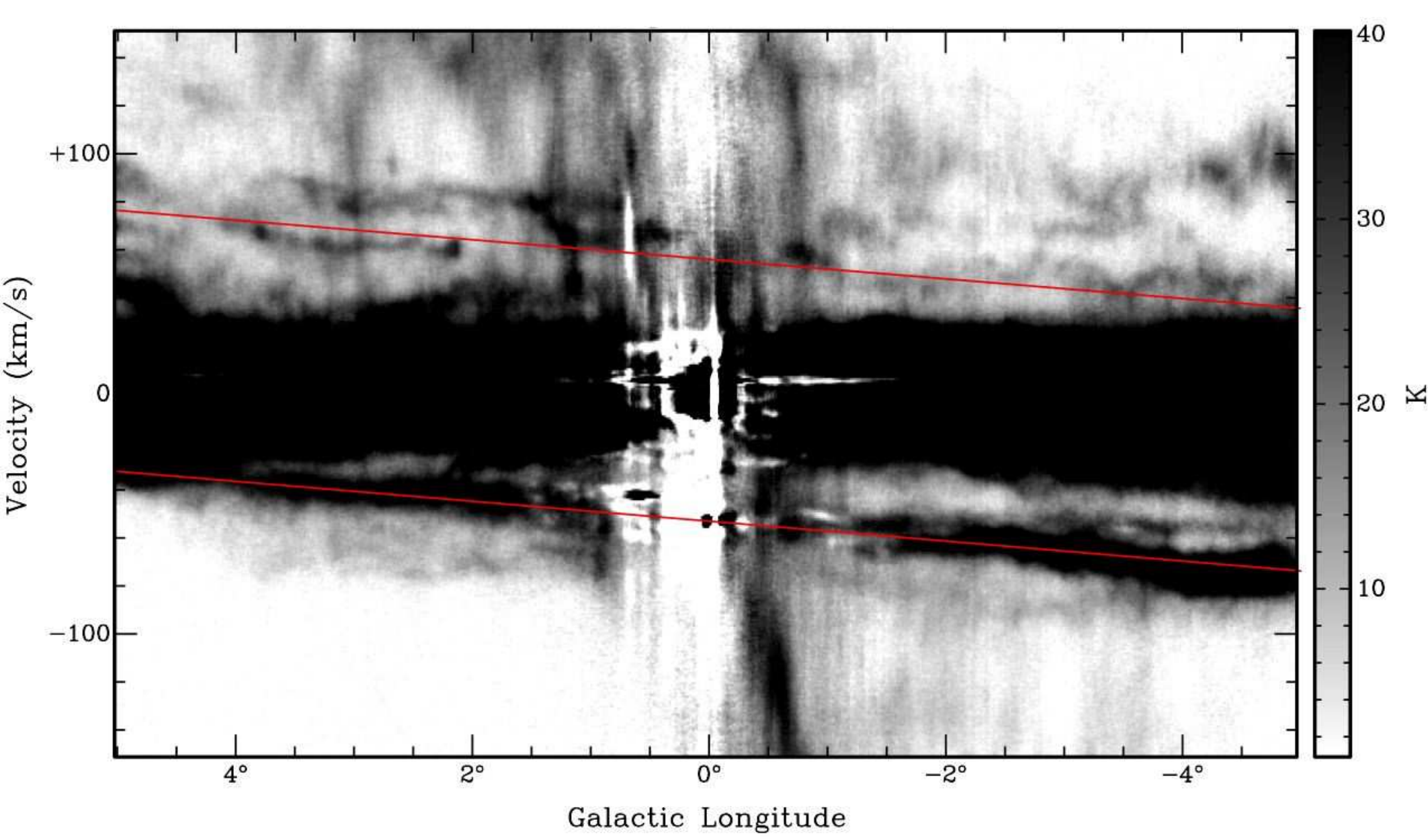}
\caption[]{Longitude-velocity at $b=0^{\circ}$ with red lines showing the linear fits of the 3-kpc arms from \citet{dame08}.
\label{fig:3kpc_lv}}
\end{figure}

\begin{figure}
\centering
\includegraphics[width=4in]{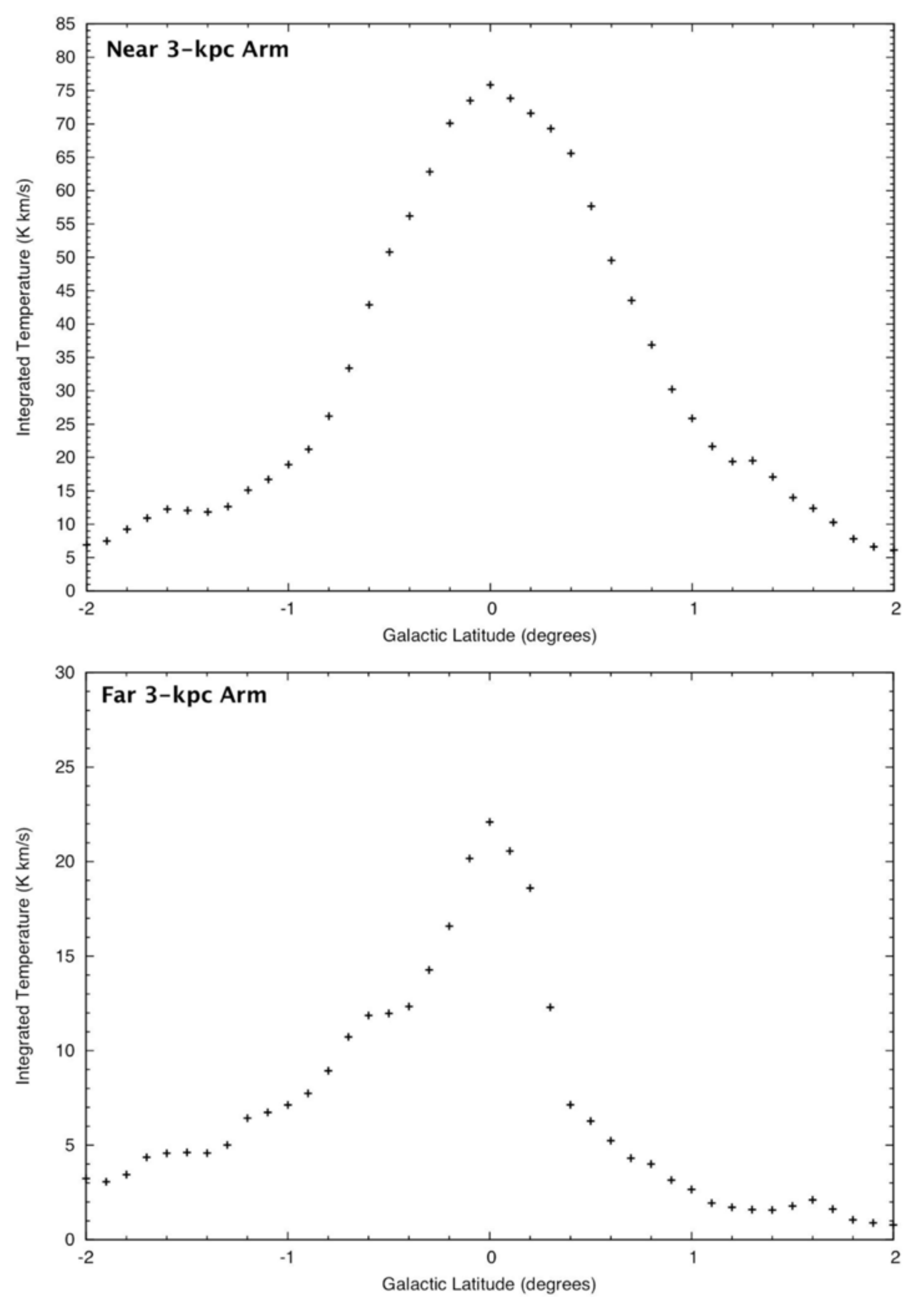}
\caption[]{Latitude profiles of the integrated \HI\ emission from the near (top) and far (bottom) 3-kpc arms. The unrelated blended and vertical sections discussed in \citet{dame08} have been excluded. 
\label{fig:3kpc_lat}}
\end{figure}

\begin{figure}
\centering
\includegraphics[width=4in]{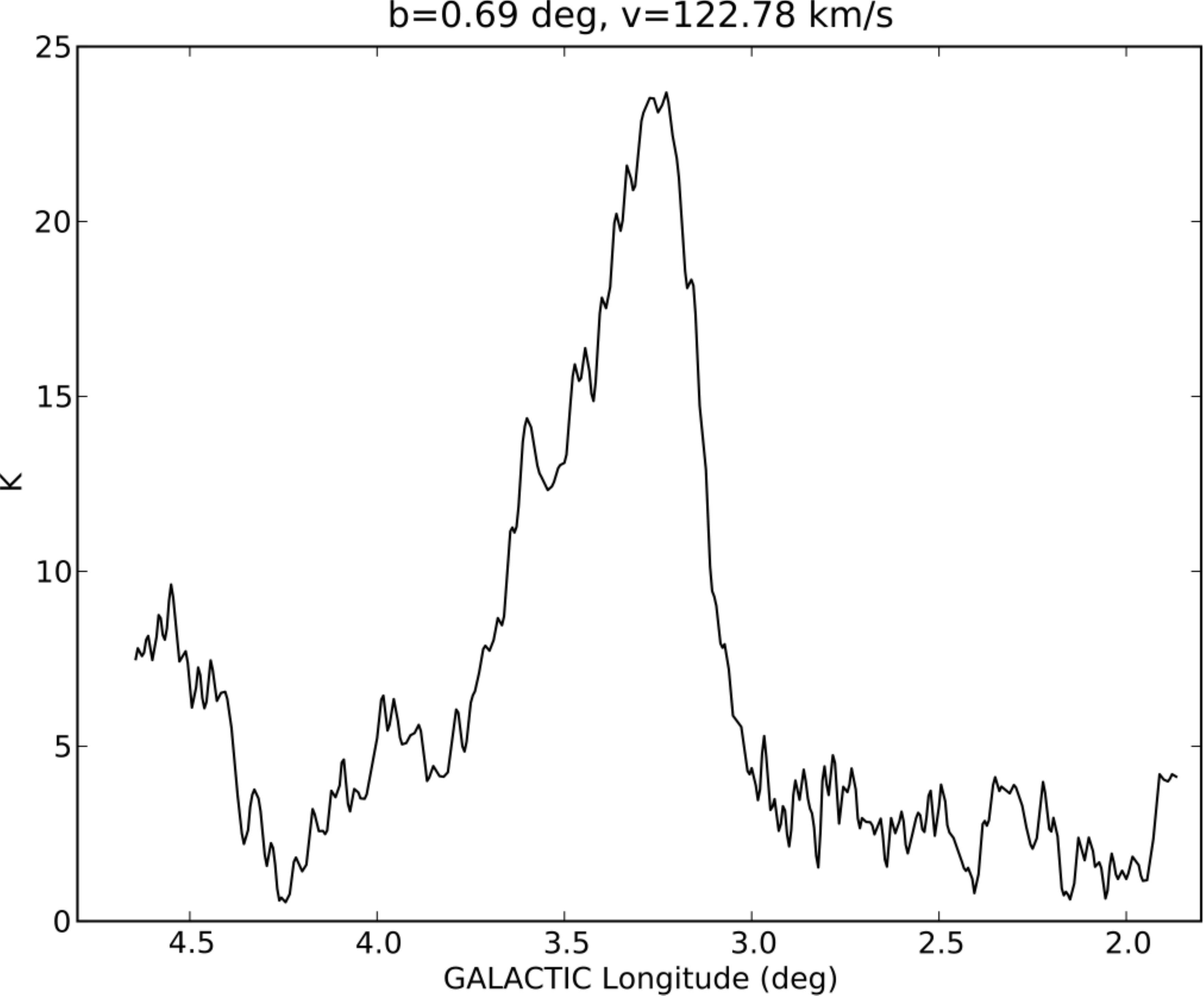}\\
\includegraphics[width=4in]{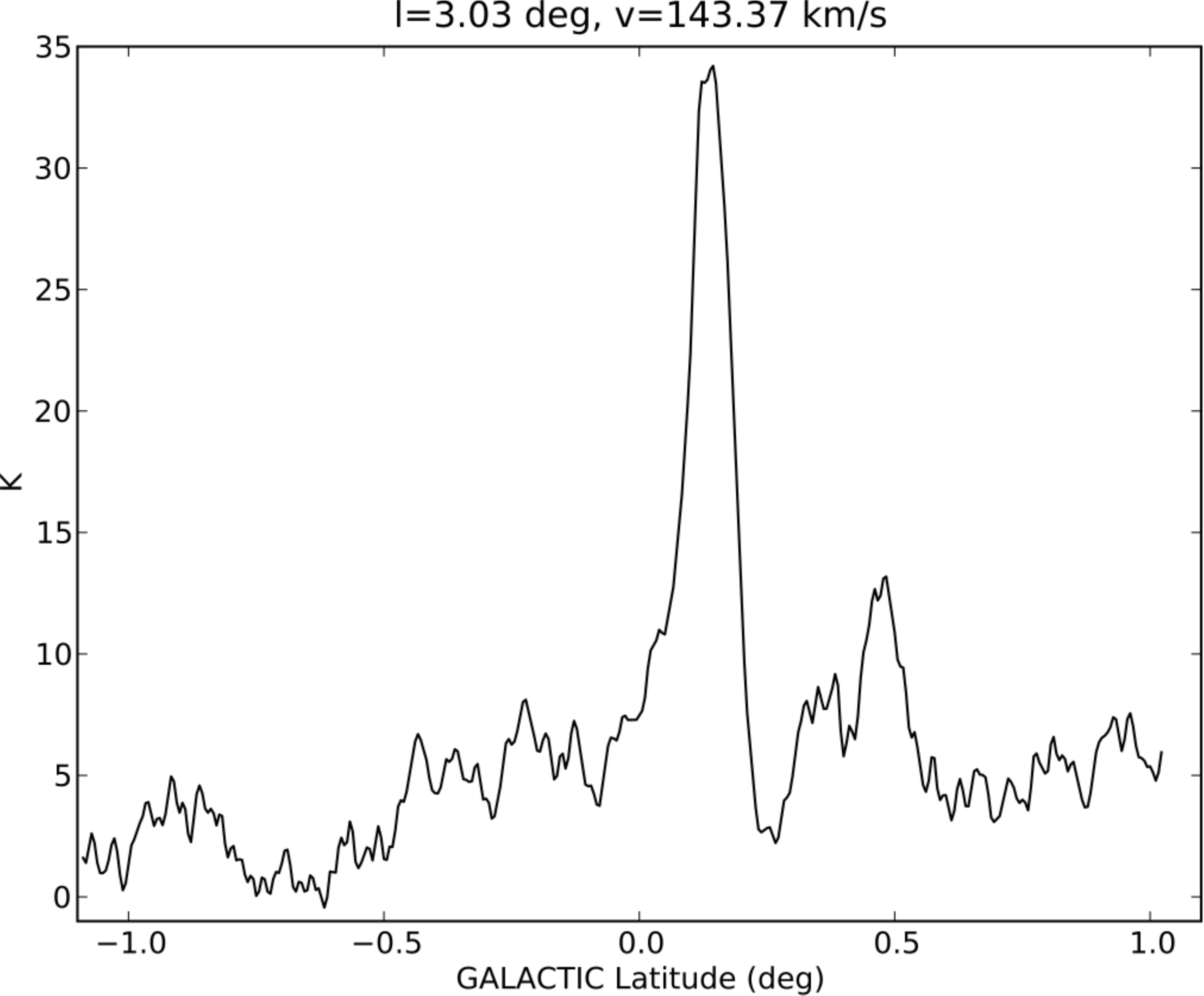}
\caption[]{Slices in longitude (latitude) across the Bania's Clump 2,
  showing its steep interior sides.  Top: Slice at $b=0.69\arcdeg$,
  $v=122.78$ \kms.  Bottom: Slice at $l=3.03\arcdeg$, $v=143.37$ \kms.
\label{fig:clump_slices}}
\end{figure}

\begin{figure}
\centering
\includegraphics[width=4in]{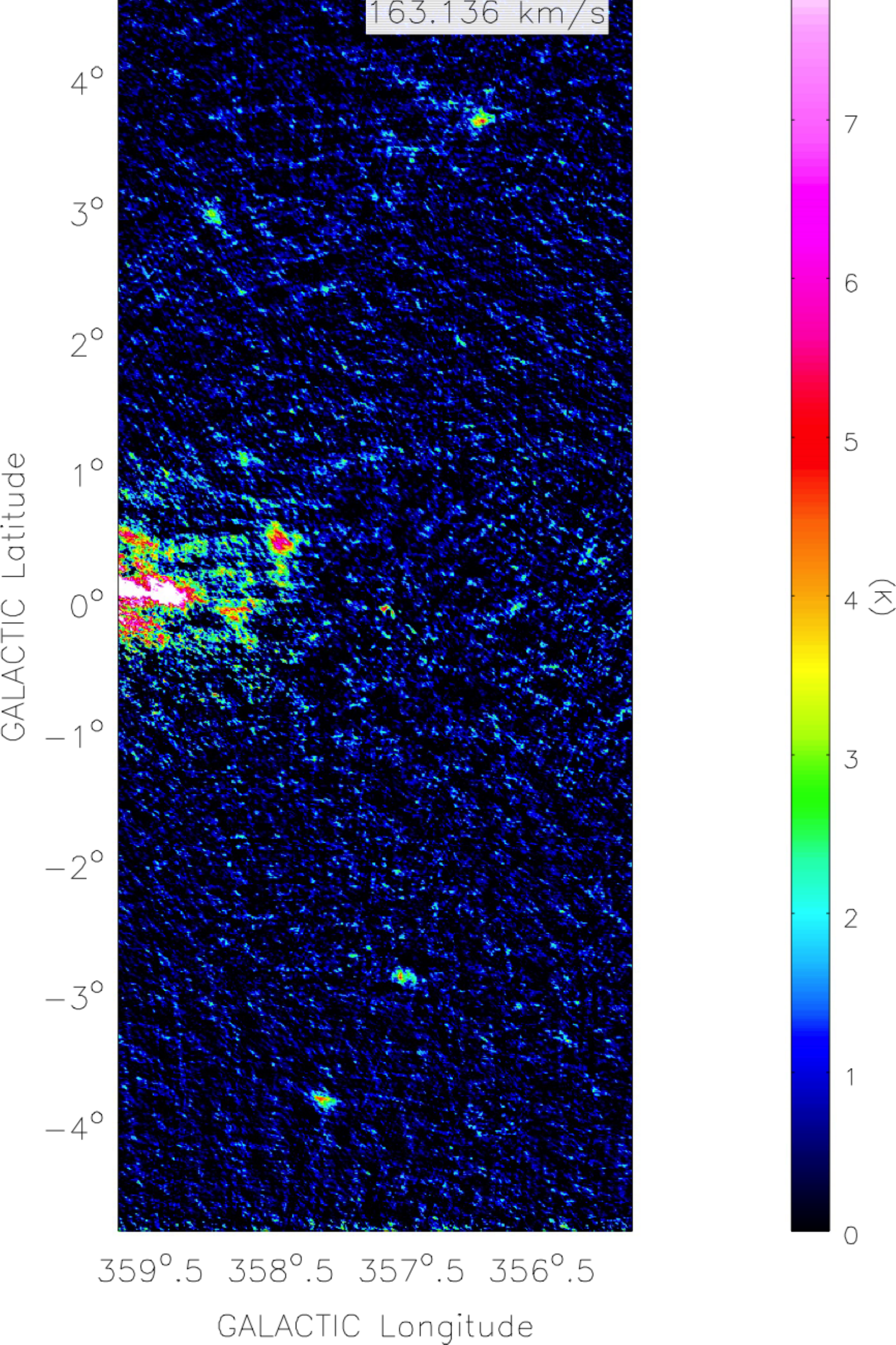}
\caption[]{Examples of compact, forbidden velocity clouds at
  $v=163~{\rm km~s^{-1}}$ observed high above the Galactic plane.  The
  color scale goes from $0 - 8$ K with a scaling power of $-0.6$, as
  shown in the wedge at the right.
  These four clouds at: $(l,b)=(358.2\arcdeg, -3.8\arcdeg)$,
  $(l,b)=(357.5\arcdeg, -2.9\arcdeg)$, $(l,b)=(357.0\arcdeg,
  +3.7\arcdeg)$ and $(l,b)=(359.0\arcdeg, +2.9\arcdeg)$ are
  representative examples out of the approximately 60 clouds
  detected in the survey area.  
  \label{fig:small_clouds}}
\end{figure}
\end{document}